\RequirePackage{fix-cm}
\documentclass[smallextended]{svjour3}       
\journalname{Foundations of Physics}
\usepackage{graphicx}
\usepackage{amsfonts}
\usepackage{amsmath} 
\usepackage{bm}
\usepackage{hyperref}

\begin{document}

\title{Quantum electrostatics, Gauss's law, and a product picture for quantum electrodynamics; or, the temporal gauge revised}

\titlerunning{Quantum electrostatics, Gauss's law and a product picture for QED}        

\author{Bernard~S.~Kay}

\institute{Department of Mathematics, University of York, York YO10 5DD, UK\hfil\break
             \email{bernard.kay@york.ac.uk}         
} 

\date{Received: date / Accepted: date}

\maketitle

\begin{abstract}
We provide a suitable theoretical foundation for the notion of the quantum coherent state which describes the electrostatic field due to a static external macroscopic charge distribution introduced by the author in 1998 and use it to rederive the formulae obtained in 1998 for the inner product of a pair of such states.  (We also correct an incorrect factor of $4\pi$ in some of those formulae.)  Contrary to what one might expect, this inner product is usually non-zero whenever the total charges of the two charge distributions are equal, even if the charge distributions themselves are different.  We actually display two different frameworks that lead to the same inner-product formulae, in the second of which Gauss's law only holds in expectation value.   We propose an experiment capable of ruling out the latter framework.   We then address the problem of finding a \textit{product picture} for QED -- i.e.\ a reformulation in which it has a total Hamiltonian, arising as a sum of a free electromagnetic Hamiltonian, a free charged-matter Hamiltonian and an interaction term, acting on a Hilbert space which is a subspace (the \textit{physical subspace}) of the full tensor product of a charged-matter Hilbert space and an electromagnetic-field Hilbert space. (The traditional Coulomb gauge formulation of QED isn't a product picture in this sense because, in it, the longitudinal part of the electric field is a function of the charged matter operators.)  Motivated by the first framework for our coherent-state construction, we find such a product picture and exhibit its equivalence with Coulomb gauge QED both for a charged Dirac field and also for a system of non-relativistic charged balls.  For each of these systems, in all states in the physical subspace (including the vacuum in the case of the Dirac field) the charged matter is entangled with longitudinal photons and Gauss's law holds as an operator equation; albeit the electric field operator (and therefore also the full Hamiltonian) while self-adjoint on the physical subspace, fails to be self-adjoint on the full tensor-product Hilbert space.  The inner products of our electrostatic coherent states and the product picture for QED are relevant as analogues to quantities that play a r\^ole in the author's matter-gravity entanglement hypothesis.  Also, the product picture provides a temporal gauge quantization of QED which appears to be free from the difficulties which plagued previous approaches to temporal-gauge quantization.    
\end{abstract}


\section{\label{Sect:Intro} Introduction}

\subsection{\label{Sect:Intro1} Electrostatic coherent states for external classical charges}

Imagine two static electric fields, which, in a classical description, have values, $\bm E^{\mathrm{class}}_1(\bm x)$ and $\bm E^{\mathrm{class}}_2(\bm x)$, which result from two distinct classically described static charge distributions, $\rho_1(\bm x)$ and  $\rho_2(\bm x)$.   The relevant Maxwell equations are Gauss's law, 
\begin{equation}
\label{Gauss}
\bm{\nabla\cdot E}^{\mathrm{class}}=\frac{\rho}{\epsilon_0},
\end{equation}
(where $\epsilon_0$ is the permittivity of vacuum\footnote{\label{ftntUnits}   We explicitly include $\epsilon_0$ and $\hbar$ and $c$ in our opening paragraphs but later set them all to 1.  We are of course free to choose $\epsilon_0$ to be whatever we like (provided the permeability of vacuum, $\mu_0$ is taken to be $1/\epsilon_0 c^2$) but whatever we choose of course affects our unit of charge.   We caution the reader that, in \cite{KayNewt}, $\epsilon_0$ was taken to be $1/4\pi$.   Let us also note here that we may restore $\epsilon_0$, $\hbar$ and $c$ in all our later equations by noting that $\pi = -\epsilon_0 E$ and by making the following insertions in our equations:  (a) a factor of $\epsilon_0^{-1}$ in front of the term $\frac{1}{2}\bm \pi^2$ and a factor of $\epsilon_0 c^2$ ($=1/\mu_0$) in front of the term $\frac{1}{2}(\bm\nabla\bm\times\bm A)^2$ in the Hamiltonian ((\ref{Ham}) and subsequent equations) for the free electromagnetic field; (b) a factor of $c$ in front of each factor of $k$ in Equation (\ref{tradphipi}) which relates the $\phi$ and $\pi$ of a scalar field to creation and annihilation operators and a factor of $\epsilon_0 c$ in front of each factor of $k$ in all equations ((\ref{tradApi}), (\ref{Ahat}) etc.) which relate creation and annihilation operators to electromagnetic $\bm\pi$ and $\bm A$ etc.; (c) (in consequence of (b)) the $\chi(\bm k)$ of Equation (\ref{chiscalar}) becomes $|ck|^{1/2}\phi_{\mathrm{sc}}(\bm k)/\sqrt 2$, while the $\chi^i(\bm k)$ of (\ref{phichi}) becomes $k^i\phi(\bm k)/\sqrt{2\epsilon_0 c}|k|^{1/2}$; (d) a factor of $\hbar$ on the right hand side of all commutation and anticommutation relations; (e) a factor of $1/\hbar$ in the (complex) exponents in all expressions (\ref{coh}), (\ref{Uext}), (\ref{U}) etc.\ for the operator(s) we call $U$ and other related exponentials as well as a factor of $1/\hbar$ in the exponents in Equations (\ref{scalarchiUomega}), (\ref{UOmegDeriv}) etc.\ and in front of the $D_0$ of (\ref{D0formula}) and similarly for Equation (\ref{spin1chiUomega}) etc.\ and in front of the $D_1$ of (\ref{D1formula}), as well, of course, as in the exponent of any expression of form $\exp(-iHt)$ where $H$ is a Hamiltonian; (f) a factor of $\hbar$ in front of $\bm k\bm \cdot \bm a$ in Equation (\ref{IIalt}) and subsequent related equations including in Footnote \ref{ftntDetails}; (g) A factor of $c$ in front of the free electromagnetic Hamiltonian when expressed in terms of creation and annihilation operators -- i.e.\ in Equations (\ref{HQEDHam}) and (\ref{PPSchroHamQEDApprox}).  

\smallskip

Note that with these conventions, we have $[a(\bm x), a^+(\bm x')] = \hbar\delta^{(3)}(\bm x - \bm x')$, $[a(\bm k), a^+(\bm k')] = \hbar \delta^{(3)}(\bm k + \bm k')$ etc. })
and, assuming the absence of any time-dependent magnetic fields, $\bm\nabla \bm\times \bm E = \bm 0$.   The latter equation is solved, as usual, by the introduction of an electrical potential, $\phi$, such that 
\begin{equation}
\label{Epot}
\bm E^{\mathrm{class}}=-{\bm\nabla}\phi.
\end{equation}
Combining (\ref{Gauss}) and (\ref{Epot}), we have
\begin{equation}
\label{Poisson}
\nabla^2\phi=-\frac{\rho}{\epsilon_0}
\end{equation}
with solution
\begin{equation}
\label{phirho}
\phi(\bm x)=\int \frac{\rho(\bm y)}{4\pi\epsilon_0|\bm x-\bm y|}\, d^3y.
\end{equation}
$\bm E^{\mathrm{class}}_1$ and $\bm E^{\mathrm{class}}_2$ are then given by (\ref{Epot}) for $\phi_1, \phi_2$ given by (\ref{phirho}) with $\rho=\rho_1, \rho_2$ respectively.

\smallskip

To give an example, our charge distributions, $\rho_1$  and $\rho_2$, might be two different possible charge distributions within a macroscopic  glass ball or, indeed (see Section \ref{Sect:Intro2} below), two states involving a single such glass ball, with a single such charge distribution (perhaps uniform), centred at two different locations.  Even though they are macroscopic and thus, for many purposes, treatable as classical, like everything in nature, our charge distributions and their electric fields must surely each ultimately be describable in quantum theory.   We shall first be interested in the question of what is the best description of their electric fields in terms of normalized quantum state vectors, say $\Psi_1$ and $\Psi_2$ (each subject to the usual phase ambiguity) in an appropriate Hilbert space when we continue to model our (static) charge distributions, $\rho_1$ and $\rho_2$, as classical (i.e.\ as $c$ numbers [multiplied by the identity]).   In Section \ref{Sect:extsource}, we shall clarify the status of a certain extension of the standard theory of the free electromagnetic field which was first proposed in \cite{KayNewt} and which provides such a description in terms of a notion, also introduced in \cite{KayNewt}, of (what we shall call here) \textit{electrostatic coherent state}  -- a notion which involves (non-dynamical) longitudinal photons.  (We remark that in \cite{KayNewt}, electrostatic coherent states  were discussed as a useful mathematical analogy to gravitostatic coherent states, which were the principle focus of that paper.   See the further remarks and references towards the end of Section \ref{Sect:Intro2}.) 

\cite{KayNewt} also gave an answer to the question:  

\smallskip

\noindent
\textit{What is the transition amplitude $\langle \Psi_1|\Psi_2\rangle$ betwen two such electrostatic coherent states?} 

\smallskip

(Strictly, in view of the phase ambiguities mentioned above, we should ask what is $|\langle \Psi_1|\Psi_2\rangle|$.)   One might think that this is the sort of question on which our existing understanding of quantum electrodynamics (\textit{QED}) will easily give an uncontroversial answer.  However, this seems not to be the case. 
If $\rho_1$ and $\rho_2$ are different, $\phi_1$ and $\phi_2$ will be different and so will their electric fields.  Therefore we expect that $\Psi_1 \ne \Psi_2$ (however we adjust their phases) and thus the magnitude,  $|\langle \Psi_1|\Psi_2\rangle|$, of their inner product cannot be 1.   

The following orthogonality theorem might seem to suggest that unless the charge distributions, $\rho_1$ and $\rho_2$, are identical, $\langle\Psi_1|\Psi_2\rangle$ will be zero.  
\smallskip

\noindent
{\bf Orthogonality Theorem:} In its quantum description, let the electric field be represented mathematically by a self-adjoint operator-valued function, $\bm E$, on ${\mathbb R}^3$ with values in a Hilbert space, ${\cal H}_{\mathrm{electromag}}$, and assume that Gauss's law holds in the sense that $(\bm{\nabla\cdot E})(\bm x)\Psi_1 =\rho_1(\bm x)\Psi_1$ and $(\bm{\nabla\cdot E})(\bm x)\Psi_2 =\rho_2(\bm x)\Psi_2$ for some $\Psi_1, \Psi_2 \in {\cal H}_{\mathrm{electromag}}$.   Then, unless $\rho_1(\bm x)=\rho_2(\bm x) \ \forall \ {\bm x}\in {\mathbb R}^3$, $\langle\Psi_1|\Psi_2\rangle = 0$.

\smallskip

\noindent
(The proof follows immediately from the elementary calculation: $\rho_2(\bm x)\langle\Psi_1|\Psi_2\rangle = \langle\Psi_1|(\bm{\nabla\cdot E})(\bm x)\Psi_2\rangle
=\langle(\bm{\nabla\cdot E})(\bm x)\Psi_1|\Psi_2\rangle =
\rho_1(\bm x)\langle\Psi_1|\Psi_2\rangle$.)

However, the results we reported on in \cite{KayNewt} gave a value which is usually\footnote{\label{ftntUsually} We write ``usually'' because we are unaware of any example for which we have two different charge distributions whose total charges are the same, for which the two corresponding electrostatic coherent states are orthogonal, except for the simple capacitor example of Section \ref{Sect:capac} (where the total charge of the two distributions is zero).   And note that that example is on a (flat) space with a different global topology ($\mathbb{R}$ times the 2-torus) from $\mathbb{R}^3$.  See Footnote \ref{ftntCap}.} non-zero provided only the total charges
\[
Q_1=\int\rho_1(\bm x) \, d^3 x \quad\hbox{and}\ \ Q_2=\int\rho_1(\bm x) \, d^3 x
\]
are equal. 

We shall show in  Section \ref{Sect:extsource} here, that those results of \cite{KayNewt} can be derived in either one of two alternative theoretical frameworks, each of which sidesteps the above theorem.

In one of these frameworks, the electric field (the $-\hat{\bm\pi}$ of Section \ref{Sect:electrocoh}) is self adjoint but Gauss's law only holds in expectation value -- i.e.\break
$\langle\Psi_1|(\bm{\nabla\cdot E})(\bm x)\Psi_1\rangle =\rho_1(\bm x)$ etc.; in the other, the electric field (the $-\tilde{\bm\pi}$ of Section \ref{Sect:electrocoh})) satisfies Gauss's law in the sense of the above Orthogonality Theorem, but it (and  --  as we shall see in Section \ref{Sect:electrocoh} -- also the Hamiltonian) fails to be self-adjoint!   Both of the frameworks lead to identical formulae for our inner products, $|\langle\Psi_1|\Psi_2\rangle|$ but, as we shall discuss in Sections \ref{Sect:electrocoh} and \ref{Sect:capac}, the physical interpretation of these inner products is different in each framework.  

Similarly to, and related to, the different versions of Gauss's law in the $\tilde{\bm\pi}$ and $\hat{\bm\pi}$ frameworks, there is a different sense in which the energy of our electrostatic coherent states equals the classical energy in the two frameworks.   Thus in the $\tilde{\bm\pi}$ framework, we find that the coherent states are eigenstates of the (non-self-adjoint) version of the Hamiltonian relevant to that framework with the classical energy as the eigenvalue, while in the $\hat{\bm\pi}$ framework they are not eigenstates; instead the classical energy is just the expectation value in the coherent state of the relevant Hamiltonian. 

It should be possible to decide between the two frameworks in view of this latter difference.  As we discuss in Section \ref{Sect:capac}, one way to do this would be to repeatedly charge up (by a suitable equal amount each time) and then discharge a capacitor and to observe whether the amount of work that can be done each time (e.g.\ to heat up a resistor) on discharge, with the energy that had been stored in the capacitor fluctuates about some mean value as is (on certain assumptions and simplifications) predicted to happen in the $\hat{\bm\pi}$ framework, or is the same each time, as (on the same assumptions and simplifications) is predicted to happen in the $\tilde{\bm\pi}$ framework and, importantly, is also (on similar assumptions) predicted to happen on an analysis based on standard Coulomb gauge thinking.    In fact the experiment we propose is designed so that, in the $\hat{\bm\pi}$ framework, it is predicted that no work is done at all in approximately half of the runs of the experiment. 

Despite the fact that the $\tilde{\bm\pi}$ framework may seem, at first site, to be rather strange, we expect, for several reasons, that the experiment will rule out the $\hat{\bm\pi}$ framework and be consistent with the $\tilde{\bm\pi}$ framework.  The main reason for this expectation is simply the point which we made above that the predictions of the $\tilde{\bm\pi}$ framework agree with those of a standard Coulomb gauge analysis.  (We should mention here that the analyses of our experiment referred to above assume that the charges on the capacitor plates can be viewed as static external classical charges and a number of other assumptions -- see Section \ref{Sect:capac}.) 

Another of our reasons is that, as we will see in Sections \ref{Sect:QED} and \ref{Sect:QEDSchr}, it is also only the $\tilde{\bm\pi}$ framework that leads to a `product picture' for full QED (in which, as well as the electromagnetic field, the charges are regarded as quantum and participate in the dynamics) for which the dynamics is equivalent to that of standard (Coulomb gauge) QED.   We shall explain what a `product picture' is in the second part of this introduction, Section \ref{Sect:Intro2}.  Let us just mention here that the fact that $\tilde{\bm\pi}$ and also the Hamiltonian in the $\tilde{\bm\pi}$ framework, are not self-adjoint will turn out not to be a problem in that full QED context because, as we will show, they \emph{are} self-adjoint on the `product picture physical subspace' which, as we shall see, is what matters.   We will also give arguments, prior to that, in Section \ref{Sect:electrocoh}, as to why the non-self adjointness of $\tilde{\bm\pi}$ and of the Hamiltonian in the $\tilde{\bm\pi}$ framework are also not problems in the context of a static external classically describable charge distribution.
  
Yet another reason why we expect the $\hat{\bm\pi}$ framework to be ruled out relates to the fact which we mentioned above, that, in it, and in the context of a static external charge distribution, Gauss's law only holds in expectation value since this would appear to be difficult to reconcile with the fact that Gauss's law holds in operator form in standard full QED and also be difficult to reconcile with the relativistic invariance of Maxwell's equations.   Furthermore, and as an immediate corollary of the related fact that, in the $\hat{\bm\pi}$ framework, our coherent states are not eigenstates of the Hamiltonian, these states will not be stationary states, even though the charge distributions that give rise to them are static.   This is at least aesthetically objectionable.

Thus, should our proposed experiment turn out to be performable, it would be both a surprise and a puzzle, requiring a rethink of many things,  if it were to rule out the $\tilde{\bm\pi}$ framework rather than the $\hat{\bm\pi}$ framework.   We should perhaps stress that the experiment of course cannot decide betwen standard Coulomb gauge quantum mechanics and our $\tilde{\bm\pi}$ formalism -- or rather the product picture which is based on that which we discuss in Sections \ref{Sect:QED} and \ref{Sect:QEDSchr} --  since, indeed we will show that our product picture is \emph{equivalent} to standard Coulomb gauge QED!  

Returning to the results of \cite{KayNewt}, let us recall that a calculation was performed there that entails, for example, that, defining the 
\emph{spin-1 decoherence exponent} $D_1$ by
\begin{equation}
\label{balloverlap1}
|\langle\Psi_1|\Psi_2\rangle|  = \exp(-D_1),
\end{equation}
if $\Psi_1$ and $\Psi_2$ are the electrostatic coherent states of a single ball of radius $R$ with a uniform charge distribution with total charge $Q$, when it is centred at two different static locations, a distance $a$ apart, then,
$D_1$ is given by 
\begin{equation}
\label{balloverlap2}
D_1=\frac{9Q^2}{4\pi^2\epsilon_0\hbar c}\!\int_0^\infty  {(\sin\kappa-\kappa\cos\kappa)^2\over\kappa^7}
\left ({\kappa\beta-\sin\kappa\beta \over
\kappa\beta}\right)\,d\kappa
\end{equation}
where $\beta=a/R$.   This is smaller by a factor of $4\pi$ than the result stated in \cite{KayNewt} which, unfortunately, appears to have been in error.    Let us note here that, for $a \ll R$, $D_1$ is well-approximated by 
\begin{equation}
\label{balloverlap3}
D_1\approx \frac{3}{2}\frac{Q^2}{16\pi^2\epsilon_0 \hbar c} \frac{a^2}{R^2}= \frac{3\alpha}{8\pi} \frac{Q^2}{{\rm e}^2} \frac{a^2}{R^2}
\end{equation}
where $\alpha$ denotes the fine-structure constant (${\rm e}^2/4\pi\epsilon_0\hbar c\approx 1/137$) and $\rm e$ denotes the charge on the electron.   

Equation (\ref{balloverlap2}) tells us that, if our ball has a surplus of $N$ electrons (or has $N$ holes) uniformly distributed at fixed locations throughout the ball, then, for $a \ll R$, and assuming, as seems reasonable, that, for suitably large $N$, we may, to a good approximation, treat this system as a uniform charge-density ball,
\begin{equation}
\label{balloverlap4}
|\langle\Psi_1|\Psi_2\rangle| \approx \exp\left(-\frac{3N^2\alpha}{8\pi}\frac{a^2}{R^2}\right).
\end{equation}
So, for example, if $N=10^{5}$ (so the magnitude of the charge on the ball is approximately $1.6 \times 10^{-14}$ coulombs -- equivalently, if we measure $R$ in millimeters, if the magnitide of the electrical potential at its surface is around 0.014$/R$ volts)  our formula predicts that, when $a/R$ is around $1/200$, $|\langle\Psi_1|\Psi_2\rangle|$ will be around $0.11$; $|\langle\Psi_1|\Psi_2\rangle|$ will be smaller than that for larger values of $a/R$ and, in order for $|\langle\Psi_1|\Psi_2\rangle|$ to be within one percent of 1, $a/R$ would need to be around $3 \times 10^{-5}$ or smaller than that.   (This is hoped to serve to replace one of the examples on page L94 of \cite{KayNewt} which was incorrect because of the wrong factor of $4\pi$ mentioned above.)

For very large $a/R$, on the other hand, we have (correcting the formula in \cite{KayNewt}) the asymptotic formula
\[
D_1 \approx \frac{\alpha}{\pi}\frac{Q^2}{{\rm e}^2}(\ln(a/R) + 0(1)),
\]
or, in units where $\hbar = c= \epsilon_0 =1$,
\begin{equation}
\label{balloverlap5}
D_1 \approx \frac{Q^2}{4\pi^2}(\ln(a/R) + 0(1)).
\end{equation}
So, in particular, for a single proton, $D_1 \approx \frac{1}{137\pi}\ln(a/R_p)$ where $R_p$ is the radius of the proton (which we could reasonably take to be its Compton wavelength 
$\approx 10^{-15}$ m.).  Thus for example for $a$ equal to 1 metre, $e^{-D_1}$ is around $0.92$ -- significantly smaller than 1.   (This corrects the error, due to the above-mentioned missing factor of $1/4\pi$, in another of the examples in \cite{KayNewt}.)
We observe that it is the ratio $a/R$ which is relevant in all this and the absolute size of the ball is irrelevant.

Let us also remark in passing that we expect the same asymptotic formula to be valid, not just for a uniform charge-density ball, but also for charged bodies with a wide range of other shapes and charge distributions -- each such shape and charge distribution having its own value of `effective radius' $R$.\footnote{\label{ftntAsymp} 
One way to convince oneself that the asymptotic formula (\ref{balloverlap5}), and also the formula (\ref{Asympt}) of Section \ref{Sect:DensOp}, should hold for charged bodies with a wide range of shapes and charge distributions -- and also to check the correctness of the numerical factor $1/4\pi^2$ in (\ref{balloverlap5}) and (\ref{Asympt}) -- is as follows:  First notice that the left hand side of (\ref{Asympt}) (of which (\ref{D1formula}) [re-expressed as in Equation (\ref{altform}) of Footnote \ref{ftntDetails}] is the special case where $q_1=q_2$) is, by (\ref{phichi}), given by the momentum-space integral
\begin{equation}
\label{lhsAsymp}
\int \frac{k}{2}\tilde\phi_1(\bm k)^*(1-e^{i\bm k\bm\cdot\bm a})\tilde\phi_2(\bm k)\, d^3k
\end{equation}
where $\tilde\phi_i(\bm k)$, $i=1,2$ denote the Fourier transforms (see Footnote \ref{ftntFourier}) of $\phi_i(\bm x)$, and $\phi_1(\bm x)$ and $\phi_2(\bm x)$ denote the electrical potentials of the charged objects when their centres are located (say) at the origin.  (Here we just suppose we have defined some suitable notion of `centre' for each of the charged matter distributions involved in (\ref{Asympt}).)  Then notice that, in the case of pointlike charges for which $\rho_i(\bm x) = q_i\delta^{(3)}(\bm x)$, $i = 1,2$, by (\ref{Poisson}) (with $\epsilon_0=1$) and Fourier transformation, we have $\tilde\phi_i(\bm k) = q_i/(2\pi)^{3/2}k^2$.  So, formally, (\ref{lhsAsymp}) becomes
\[
\frac{q_iq_j}{16\pi^2}\int \frac{1}{k^3}(1-e^{i\bm k\bm\cdot\bm a}) \, d^3k,
\]
\begin{equation}
\label{eqwintegral}
= \frac{q_iq_j}{4\pi}\int_0^\infty \frac{1}{k}\left(1-\frac{\sin(ka)}{ka}\right)\, dk.
\end{equation}
The integral in (\ref{eqwintegral}) is of course divergent.  However, its formal derivative with respect to $a$,
\[
\int_0^\infty \frac{\sin(ka) - ka\cos(ka)}{k^2a^2}\, dk,
\]
is convergent and (in view of the fact that $\int_0^\infty \frac{\sin\kappa - \kappa\cos\kappa}{\kappa^2}\, d\kappa= [-\sin\kappa/\kappa]_0^\infty = 1$) is equal to $1/a$.  Hence it is reasonable to assign the value 
$\ln(a/R)$ to the integral in (\ref{eqwintegral}), where $R$ is an unfixable constant.   This further strongly suggests that when one replaces point charges by our charge distributions, the same formula will hold asymptotically for large $a$ but now that, for any given (pair of) charge distribution(s), $R$ will be fixed.    

\smallskip

See also \cite{robust} for a different but relevant consideration written with the linearized gravity case in mind but equally relevant for quantum electrostatics.}   (We should also note here that, while it doesn't affect the usefulness of that paper for the purpose for which we have cited it here\footnote{\label{ftntErr} As far as we can see now, the calculations in \cite{KayNewt} of $D_0$ for the `spin-zero gravity' model there (which is the same as the scalar model discussed in Section \ref{Sect:scalarcoh} here) were carried out correctly.   Also the statement in \cite{KayNewt}) to the effect that (as we put it in Section \ref{Sect:electrocoh} here) the electrostatic (or `spin-1') decoherence exponent, $D_1$, is equal to the spin-0 exponent, $D_0$, when the classical static scalar charge densities, $\sigma_1$ and $\sigma_2$, are equated with $\rho_1$ and $\rho_2$, is correct.  However, unfortunately, the formulae given in \cite{KayNewt} for $D_1$ were a factor of $4\pi$ bigger than the correct formulae (which are given here).   Let us also note that the argument and claim in \cite{KayNewt} that (in the sense explained there) the spin-2 decoherence exponent, $D_2$ is a factor of 6 times bigger than $D_0$ also appears to be in error.   This error also infects \cite{eeee} and \cite{robust}.  It is intended to correct it in the forthcoming paper \cite{QGS}.}, \cite{robust} has inherited a factor of 6 from \cite{KayNewt} which seems to be in error.) 

We next give some more details on what was done on the theory of electrostatic coherent states in \cite{KayNewt} and where the gaps were and how we will fill them in the present paper.

\cite{KayNewt} anticipated the form of the electrostatic (and also gravitostatic) coherent states by pursuing an analogy with a corresponding notion of coherent state describing a static state of a scalar field, $\varphi$, in interaction with an external classical static scalar source, with \textit{scalar charge density} $\sigma$, according to the equation 
\begin{equation}
\label{scalareq}
\ddot\varphi-\nabla^2\varphi=-\sigma.
\end{equation}
However, the arguments given there were based on assumed similarities between such static scalar field configurations and static configurations of an electric field.   But as well as similarities, there are also important differences due to the fact that the static field equation $\nabla^2\varphi=\sigma$ is a special case of the dynamical equation (\ref{scalareq}) (resulting when $\ddot\varphi$ happens to vanish) while the counterpart equation (\ref{Poisson}) for electrostatics is an expression of a constraint (i.e.\ Gauss's law).   In particular \cite{KayNewt} did not explain how the notion of electrostatic coherent state circumvents the above Orthogonality Theorem.   So there may have seemed to be reasons to doubt whether the notion of electrostatic coherent state discussed there was valid -- i.e.\ reasons to doubt whether it was consistent with the established formalism and results of QED.  

In Section \ref{Sect:extsource} here, we clarify all these matters, explain both the similarities and differences between the static scalar and electrostatic situations and introduce our two proposed frameworks (i.e.\ the $\tilde{\bm\pi}$ and $\hat{\bm\pi}$ frameworks mentioned earlier) which are suited to the special nature of the electrostatic case, and which, each in its own way as we discussed already above, circumvent the above Orthogonality Theorem.   The development that helps us to do all this is based on a new formulation of the free electromagnetic field -- equivalent to, but distinct from, the usual Coulomb gauge formulation --  in which the usual Fock space of transverse photons is tensor-producted with a vacuum state for longitudinal photons and regarded as a subspace of an augmented Fock space which includes states of longitudinal as well as transverse photons.  In both frameworks, the electromagnetic field momentum (identified with minus the electric field) is represented as the sum of the usual transverse field momentum operator -- which, as usual, arises as a difference of annihilation and creation operators -- with a longitudinal field-momentum operator.  In the $\hat{\bm\pi}$ framework this also arises as a difference of annihilation and creation operators.  In the $\tilde{\bm\pi}$ framework (and this is the key, and seemingly necessary, innovation which enables the later construction of our product picture) this may be thought of as obtained from the difference of annihilation and creation operators of the $\hat{\bm\pi}$ framework by deleting the creation operator while doubling the annihilation operator (and in consequence, as we discussed above, fails to be self-adjoint).   The electrostatic coherent state describing the electric field due to an external classical charge distribution is the same in both frameworks and is understood as belonging to the augmented Fock space.  Interestingly, in this state, the charged matter and longitudinal photons are entangled with one another.

The main reason why we are interested in defining a notion of `electrostatic coherent state' for a given background classical charge distribution, and in computing inner products, $\langle\Psi_1|\Psi_2\rangle$ between pairs, $\Psi_1$ and $\Psi_2$, of such states, is that these questions are related to the issue of finding a \textit{product picture} for full quantum electrodynamics.   In the next subsection, we will explain what this issue is and then indicate how our coherent states are related to it. 

(We postpone the question of whether or not our electrostatic coherent states, and the inner products between them, have a direct physical interpretation to Sections \ref{Sect:electrocoh} and \ref{Sect:capac}.   As we shall discuss there, the answer depends on whether one adopts the $\hat{\bm\pi}$ or the $\tilde{\bm\pi}$ framework.)

\subsection{\label{Sect:Intro2} The notion of a product picture for QED}

Let us begin by recalling the traditional canonical formulation of full QED, based on Coulomb gauge, $\bm\nabla\bm{\cdot A}=0$ (see e.g.\ \cite{Weinberg}).   From now on we shall adopt units in which $\hbar=1=c$ and also take $\epsilon_0=1$ (see Footnote \ref{ftntUnits}).   The Hamiltonian takes the form  
\begin{equation}
\label{Ham}
H=\int {1\over 2} {\bm\pi^\perp}^2 +{1\over 2}(\bm\nabla \bm\times \bm A)^2-\bm J\cdot\bm A \, d^3x  \ +  \ H^0_{\mathrm{ch\, mat}}  \ +  \ V_{\mathrm {Coulomb}} 
\end{equation}
where
\begin{equation}
\label{VCoulomb}
V_{\mathrm{Coulomb}}={1\over 2} \int {\rho(\bm x)\rho(\bm y)\over 4\pi|\bm x-\bm y|} \, d^3x d^3y
\end{equation}
and $H^0_{\mathrm{ch\, mat}}$ denotes the Hamiltonian for the charged matter -- be it described in terms of fields or particles -- excluding its electromagnetic interactions.  We shall treat, in Section \ref{Sect:QED}, the Dirac field and also, in Section \ref{Sect:QEDSchr},  a model with a collection of non-relativistic charged particles -- modeled as balls, not point particles for reasons which we will explain.  (In the latter case and also e.g.\ in the case of the charged scalar field, further terms, which depend quadratically on $\bf A$, of course arise.) $\bm J$ denotes the electric current and $\rho$ the charge density for whichever model is under consideration, and $\bm\nabla \bm\times {\bm A}$ denotes the magnetic field expressed in terms of the Coulomb gauge vector potential, ${\bm A}$.   (Note that given that $\bm\nabla\bm{\cdot A}=0$, the term $\frac{1}{2}(\bm\nabla \bm\times {\bm A})^2$ is of course the same thing [up to a total divergence] as $\frac{1}{2}\partial^jA^i\partial_jA_i$).  The electromagnetic field momentum operator, ${\bm \pi}^\perp$, satisfies $\bm\nabla{\bm{\cdot\pi}^\perp}=0$.

The specification of the quantum theory is completed with the commutation relations \cite{Weinberg}
\[
[A_i(\bm x), {\pi^\perp}_j(\bm y)]=i\delta_{ij}\delta^{(3)}(\bm x-\bm y) + i{\partial^2\over\partial x^i\partial x^j}
\left ({1\over 4\pi|\bm x-\bm y|}\right )
\]
\begin{equation}
\label{tradCR}
[A_i(\bm x),A_j(\bm y)]=0=[{\pi^\perp}_i(\bm x), {\pi^\perp}_j(\bm y)]
\end{equation}
together with the appropriate anticommutation (or commutation) relations for the charged matter operators, which are also assumed to commute with $\bm A$ and ${\bm\pi}^{\mathrm{perp}}$.

The physical electric field, $\bm E$, is given, in this traditional formulation of the theory, by 
\begin{equation}
\label{Epiphi}
\bm E=-{\bm \pi}^\perp -{\bm\nabla}\phi
\end{equation}
where $\phi$ is related to the charge density operator, $\rho$, of the charged matter by 
(\ref{Poisson}) $\nabla^2\phi=-\rho$ (with the solution (\ref{phirho})).  This ensures that $\bm E$ satisfies Gauss's Law (\ref{Gauss}).

Let us remark in passing here that, with (\ref{phirho}), we may express $V_{\mathrm{Coulomb}}$ alternatively, as ${1\over 2}\int \phi\rho\, d^3x$ and also by 
\begin{equation}
\label{VCoulombphi}
V_{\mathrm{Coulomb}} = \frac{1}{2}\int\bm\nabla\phi{\bm\cdot}\bm\nabla\phi\, d^3x.
\end{equation}

Now a great many simple model quantum theories, involving two interacting systems, have the following basic \textit{product structure}:  One has a system, $a$, described by a free Hamiltonian, $H_a$, acting on a Hilbert space, ${\cal H}_a$, and a system, $b$, described by a free Hamiltonian, $H_b$, acting on a Hilbert space, ${\cal H}_b$, and the full dynamics of the total system is described by a total Hamiltonian,  $H_{\mathrm{total}}$, which can be written
\begin{equation}
\label{HamSum}
H_{\mathrm{total}}=H_a + H_b + H_{\mathrm{interaction}}
\end{equation}
and acts on the tensor-product Hilbert space
\[
{\cal H}_{\mathrm{total}}={\cal H}_a\otimes {\cal H}_b.
\]
Moreover many general notions and results regarding pairs of systems which interact with one another presuppose that they have such a product structure.  For example, once we have such a product structure, and on the further assumption that  $H_{\mathrm{interaction}}$ is suitably small, we may immediately conclude that the possible energy levels of the total system arise approximately as sums of energy levels of system $a$ and system $b$, and the associated energy eigenstates arise approximately as products of ${\cal H}_a$-eigenfunctions and ${\cal H}_b$-eigenfunctions.  This statement about energy levels is, in turn,  a prerequisite for both traditional (see e.g.\ \cite{FeynmanStatMech}) and modern (see e.g.\ \cite{GoldsteinLebowitzEtAl,KayThermality}) explanations (now taking $a$ to stand for the subsystem and $b$ for its environment) of why small subsystems of total systems with approximately fixed total energy will be found to be in approximate Gibbs states -- a result which stands \cite{FeynmanStatMech} at the threshold of statistical mechanics.   Furthermore, once we have such a product structure, it becomes possible, and meaningful, to ask, for any vector, $\bm\Psi$, in ${\cal H}_{\mathrm{total}}$, about the reduced density operator, $\varsigma_a$ of, say, system $a$, i.e.\ about the partial trace of the (pure) density operator $|\bm\Psi\rangle\langle\bm\Psi|$ over the Hilbert space ${\cal H}_b$ (and similarly with $a\leftrightarrow b$) and/or to ask about the extent to which the two interacting systems are \textit{entangled} -- as could, e.g.\ be measured by the von Neumann entropy of $\varsigma_a$ (equal to the von Neumann entropy of $\varsigma_b$, and otherwise known as the $a$-$b$ entanglement entropy).  

The Coulomb gauge Hamiltonian, (\ref{Ham}) does have a product structure in this sense, but only if we identify the system $a$, say, with the transverse degrees of freedom of the electromagnetic field and the system $b$ with the charged matter degrees of freedom.  We find this unsatisfactory and would like to find an alternative, equivalent, formulation of the theory which has a product structure in which the system $a$ can be identified with \textit{all} the degrees of freedom of the electromagnetic field, both transverse and longitudinal, and the system $b$ with the charged matter.   The reason why Coulomb gauge quantization doesn't have such a product structure is because, as we see from (\ref{Epiphi}), in it, the longitudinal modes of the electromagnetic field belong to the charged matter Hilbert space!

As we shall see -- in Section \ref{Sect:QED} when the charged matter is a Dirac field, and in Section \ref{Sect:QEDSchr} when it is a system of nonrelativistic charged balls -- we will achieve that goal with one modification.  Namely the set of physical states will not be the full tensor product, $\cal H_{\mathrm{electromag}}\otimes\cal H_{\mathrm{ch\, mat}}$, of a Hilbert space, ${\cal H}_{\mathrm{electromag}}$, for the electromagnetic field and a Hilbert space, $\cal H_{\mathrm{ch\, mat}}$, for the charged matter, but rather a certain subspace of that tensor-product Hilbert space, which we call the \textit{product picture physical subspace}.   Two points are worth mentioning here.   First, the Hilbert space, ${\cal H}_{\mathrm{electromag}}$,  for the electromagnetic field is to be identified with the   
${\cal F}({\cal H}_{\mathrm{one}})$ of Section \ref{Sect:twoequiv}.  Second, the fact that the product picture physical subspace forms a proper subspace of the full tensor product seems not to be a deficiency, but rather a welcome feature -- it consists, in fact, of states which are entangled between the charged matter and the electromagnetic field in just such a way that, on it, Gauss's law holds as an operator equation.  I.e. we will find (see Equations (\ref{QuantGauss2}), (\ref{QuantGauss3}))  that 
\[
\bm\nabla\bm\cdot\bm E\,\bm\Psi=\rho\,\bm\Psi
\]
for all $\bm\Psi$ belonging to the product picture physical subspace.    

We shall call this reformulation of standard QED the \textit{product picture} of QED.

To explain how the notion of `electrostatic coherent state' of Section \ref{Sect:electrocoh} is related to the issue of providing QED with a product picture in the above sense, which we do in Sections \ref{Sect:QED} and \ref{Sect:QEDSchr}, let us consider a total state that is a static Schr\"odinger cat-like superposition of a single charged glass ball (as discussed in Section \ref{Sect:Intro1}) centred on two different locations.   We might schematically write (cf.\ \cite{KayNewt}) 
\begin{equation}
\label{catschema}
\bm\Psi = c_1|\hbox{ball centred at ${\bm x}_1$}\rangle + c_2|\hbox{ball centred at ${\bm x}_2$}\rangle, 
\end{equation}
where $|c_1|^2 + |c_2|^2 = 1$, 
where $|\hbox{ball centred at ${\bm x}_1$}\rangle$ means the total state of the charged matter of the ball, together with its electromagnetic field, when it is centred at $\bm x_1$ and similarly for $\bm x_2$. 

If we have a product picture, then we expect that, to a good approximation, we could write this as the (entangled) state
\begin{equation}
\label{cat}
\bm\Psi = c_1 |\Psi_1\rangle \otimes |B_1\rangle   + c_2 |\Psi_2\rangle \otimes |B_2\rangle 
\end{equation}
where $|B_1\rangle$ and $|B_2\rangle$ are the quantum states in $\cal H_{\mathrm{ch\, mat}}$ of the charged matter of our charged ball centred on its two locations and $|\Psi_1\rangle$ and $|\Psi_2\rangle$ are the states of their electrostatic fields in $\cal H_{\mathrm{electromag}}$.  (This will only be approximate since there will also be entanglement between the individual charges inside each ball and the electromagnetic field.)  It seems reasonable to assume that these latter states are, in turn, well approximated by the static electromagnetic field states due to our charged ball in the two locations, when it is treated as a classical background charge distribution -- and thus, as we are yet to argue in Section \ref{Sect:extsource}, by the electrostatic coherent states of \cite{KayNewt} as described here in Section \ref{Sect:electrocoh}.

Moreover, if we take the partial trace of (\ref{cat}) over $\cal H_{\mathrm{electromag}}$, which is a meaningful thing to do because we are in a product picture (even though the physical subspace is a proper subspace of the total tensor product, $\cal H_{\mathrm{electromag}}\otimes\cal H_{\mathrm{ch\, mat}}$) we find that the reduced density operator of the matter is given by the formula
\[
\varsigma_{\mathrm{ch \, mat}}\! =\! |c_1|^2|B_1\rangle\langle B_1| +  c_2^*c_1\langle\Psi_2|\Psi_1\rangle |B_1\rangle\langle B_2| + c_1^*c_2\langle\Psi_1|\Psi_2\rangle |B_1\rangle\langle B_2| + |c_2|^2|B_2\rangle\langle B_2|.
\]  
(Thanks to translational invariance (see Footnote \ref{ftntDetails}) $\langle\Psi_2|\Psi_1\rangle$ is actually necessarily real and positive and hence equal to $\langle\Psi_1|\Psi_2\rangle$.) 

Thus our electrostatic coherent states for external classical charge distributions are an ingredient in what we expect to be a good approximation to our Schr\"odinger cat state $\bm\Psi$ of (\ref{catschema}) while the formula for the inner product $\langle\Psi_2|\Psi_1\rangle$ between two such coherent states appears in what we expect to be a good approximation for the reduced density operator $\varsigma_{\mathrm{ch \, mat}}$.

Thus we have shown that -- even in the absence of a complete theory for such states or of knowledge as to how to calculate their inner products -- one would have reason to expect there to be a relation between the notion of `electrostatic coherent state' and the issue of providing QED with a product picture, as we promised to show at the start of this subsection.

And it will in fact become clear in Sections \ref{Sect:QED} and \ref{Sect:QEDSchr} that the theory of electrostatic coherent states, which we shall develop in Section \ref{Sect:electrocoh}, will guide us as to how to obtain a product picture.   This is true for full quantum electrodynamics with Dirac charged matter which we treat in Section \ref{Sect:QED}, but the connection with the work of \cite{KayNewt} is particularly clear for the quantum electrodynamics of a system of many idealized nonrelativistic charged balls which we treat in Section \ref{Sect:QEDSchr}.  Let us mention here that, building on the ideas described above for Schr\"odinger cat-like states, in \cite{KayNewt}, we already anticipated an approximation to some of the results of the product picture in that case, valid in the case of slowly time-evolving wavefunctions and/or in the case of eigenstates of the usual relevant many body Schr\"odinger equation.    In Section \ref{Sect:QEDSchr}, we will rederive those approximations in a systematic way from our new exact product picture and discuss them further around Equation (\ref{NewtForm}) and in Sections \ref{Sect:Hatom} and \ref{Sect:DensOp}, thus both making the present paper self-contained and also retrospectively validating those results in \cite{KayNewt}.

We hope that our product picture of QED in Sections \ref{Sect:QED} and \ref{Sect:QEDSchr} will be seen to be of considerable interest in its own right as an (as far as we are aware) new formulation of standard QED which seems to be more simple, in a number of ways, than the traditional Coulomb gauge formulation.   The product picture is also, in many ways, more physically appealing since, in it, the longitudinal modes of the electromagnetic field are viewed as part and parcel of the full (quantum) electromagnetic field, whereas, in the usual Coulomb gauge treatment, their status is summarized by the traditional admonition \emph{``One should not quantize the longitudinal modes of the electric field''}.  

Indeed Coulomb gauge quantization goes together with two possible views about the ontological status of the longitudinal modes of the electric field, both of which we find unsatisfactory:   \textit{Either} that such modes do not exist at all, the job that they would have done, had they existed, being entirely done by the action-at-a-distance  potential, $V_{\mathrm{Coulomb}}$ of Equation (\ref{VCoulomb}). \textit{Or alternatively} (and this is the view we shall suppose to be taken elsewhere in the paper when we discuss the Coulomb gauge picture) that there \textit{is} a longitudinal electric field, $\bm E^{\mathrm{long}}$, but it is just an epiphenomenon (see Footnote \ref{ftntAnalo}) of the physics of the charged matter sector inasmuch as it is defined to be $-\nabla\phi$ where $\phi$ is given by (\ref{phirho}) -- where (in full QED) $\rho$ is an operator belonging to the charged matter sector.   Let us note that, in the latter case, where the existence of the longitudinal modes is granted, whether or not these modes are quantum in nature depends on how the charged matter is modeled.  In full (Coulomb gauge) QED, the charged matter is quantum (as well as dynamical) in nature and therefore the longitudinal modes inherit that quantum nature, but in the case that the charged matter is modeled as an (external) classical charge distribution, as it is in Section \ref{Sect:extsource} here, then, in consequence, according to Coulomb gauge ideas, so will the longitudinal modes of the electric field be classical in nature.  

The unsatisfactoriness of both of the above two possible Coulomb gauge ontologies is particularly brought into focus e.g.\  by contemplation of the electrostatic field in a capacitor (see Section \ref{Sect:capac}) which we surely would normally think of as a real physical thing.   After all, we would normally take the view that it is the (longitudinal part of the) electric field that can store energy and do work!   As we shall discuss further in Section \ref{Sect:final}, our product picture, on the other hand, is consistent with this `normal' view and, indeed, naturally goes together with an ontology in which the longitudinal modes of the electromagnetic field exist as separate entities and are quantum in nature irrespectively of whether one models charged matter as an (external) c-number source or as a quantum field.

As we shall discuss in Section \ref{Sect:QEDdiscussion}, our product picture resembles, in some ways, previous proposals (see e.g.\ \cite{LMS} and references therein) for quantizing QED in the temporal gauge. But (see the discussion around our `Contradictory Commutator Theorem' in Section \ref{Sect:QEDdiscussion} and the subsequent discussion in Section \ref{Sect:temporal}) it is different from those proposals and appears to be free from their well-known difficulties.  

My own reasons for studying this issue came about as a byproduct of a research program concerning a possible role for quantum gravity in the foundations of quantum mechanics and of thermodynamics.  which I call my  \textit{matter-gravity entanglement hypothesis} and for a description of which I refer to the papers \cite{Kay1,KayNewt,eeee,KayThermality,KayEntropy,KayStringy,KayMore,KayMatGrav,KayRemarks}.   In this work, similar issues arise for quantum gravity to those we address in the present paper for QED. In particular, the issues:  \textit{Is it meaningful to talk about matter-gravity entanglement?}, and, if so, \textit{How does one calculate quantities related to it such as the partial trace over gravity of a total pure density operator of matter-gravity of some given model closed system} (and then how to calculate the von Neumann entropy of the latter, which is equal to the matter-gravity entanglement entropy of the initial total pure density operator).   Indeed, the main purpose of \cite{KayNewt} was to treat what, from the perspective of the present paper, would be regarded as  gravitational analogues of our electrostatic coherent states etc.  And the results from \cite{KayNewt} on which we have focused above have, in \cite{KayNewt}, the subsidiary r\^ole of exhibiting electrodynamic analogues to counterpart quantities in linearized quantum gravity which are, in fact, the main focus of that paper.  In the matter gravity entanglement hypothesis, the reduced density operator of `matter' (i.e.\ all degrees of freedom other than the gravitational ones) has a special status.   One reason for being interested in the reduced density operator of charged matter in QED is that it provides a mathematical analogy to that.   (This is the reason why the process of passing from the total state $\bm\Psi$ to $\varsigma_{\mathrm{ch\, mat}}$ discussed above, is referred to in \cite{KayNewt} as `electromagnetic pseudo decoherence'.)

Work is in progress on a paper \cite{QGS} on the theory of gravitostatic coherent states and on a product picture for linearized quantum gravity which aims to do the same job for linearized gravity -- by complementing and correcting the partial progress on that topic in \cite{KayNewt} -- as the job we do in the present paper for electrodynamics -- which complements and corrects the partial progress on that topic in \cite{KayNewt}.   

It would also be of interest to explore to what extent our results can be generalized to full quantum gravity and to non-abelian gauge theories.

\subsection{\label{Sect:mathnote} A note on the level of mathematical rigour and the relation with mathematically rigorous work} 

We have attempted to be mathematically careful in the spirit e.g.\ of the early chapters of the textbook \cite{Weinberg} of Weinberg and, in some places, e.g.\ in our discussion of Coulomb gauge quantization, we have also adopted the detailed approach of that book.  For the most part, however, we don't deal with issues related to renormalization or the sort of mathematical issues exemplified by the textbook \cite{Haag} of Haag (see also \cite{FewRej}).  Exceptions are Footnote \ref{ftntSubtle} and the above Orthogonality Theorem as well as the Contradictory Commutator Theorem of Section \ref{Sect:QEDdiscussion}.  Regarding the latter two theorems,  we have not spelled out e.g.\ questions relating to domains of self-adjoint operators and sometimes where we write `self-adjoint', symmetric would do.  However the reader who cares about such matters can easily figure out for themselves how to spell them out so as to make these two theorems fully rigorous.

Despite it not being mathematically rigorous, we hope that the present work will be found to be of interest in relation to the mathematically rigorous discussion of the temporal gauge as previously discussed, e.g., in \cite{LMS} and also to rigorous discussions related to the status of Gauss's law in QED.   For a recent brief summary of the latter topic with historical references, see e.g.\ \cite[Sections 1 and 2]{MRS}.  Indeed, we hope it will be seen, more widely, to be of relevance to the general mathematical understanding of QED  in view of the fact that the product picture which we develop in Sections 
\ref{Sect:QED} and \ref{Sect:QEDSchr} provides a formulation of QED, formally equivalent, as we show, to standard QED, in which the Hilbert space is a genuine Hilbert space (i.e.\ with no negative norm states) and in which Gauss's law holds as a genuine operator equation.   I am unaware of any other formalism for standard QED that has both of these properties.  (See the further discussion in Section \ref{Sect:final}.)

\section{\label{Sect:extsource} Coherent states of longitudinal photons in quantum electrostatics}

\subsection{\label{Sect:prelim} Preliminary remarks}

We now turn to study the theory of the quantum electromagnetic field coupled only to a classical external static charge distribution, $\rho$.

In the usual Coulomb gauge quantization (see Section \ref{Sect:Intro2}) the quantum Hamiltonian may be taken to be the same as the Hamiltonian,
\begin{equation}
\label{Hamfree}
H^{\mathrm{EM}}_0=\int {1\over 2} {\bm\pi^\perp}^2 +{1\over 2}(\bm\nabla \bm\times \bm A)^2\, d^3x,
\end{equation}
for the free electromagnetic field i.e.\ just the first two terms of the Hamiltonian (\ref{Ham}) 
and then the only difference from the case of zero external charge is that the electric field, $\bm E$, is identified with $-\bm\pi^\perp - \bm\nabla \phi$, rather than just  $-\bm\pi^\perp$, where $\phi$ is now the c-number scalar field given by (\ref{Poisson}).  One might add the term $V_{\mathrm{Coulomb}}$ of (\ref{VCoulomb}) to the right hand side of (\ref{Hamfree}) but since this is, of course, now a c-number it can't affect any commutation relations.

The question we wish to address is:  \textit{How can the quantum state of the electromagnetic field in the presence of $\rho$ be represented as a vector in a suitable Hilbert space?}  In the traditional representation of $\bm E$, $\bm A$ and $H^{\mathrm{EM}}_0$ on the usual `transverse' Fock space (see Section \ref{Sect:electrocoh} below) and with or without the $V_{\mathrm{Coulomb}}$ term, the only candidate might seem to be the transverse Fock vacuum vector, $\Omega$.  But this would immediately lead to the difficulty that a family of physically distinct electromagnetic fields would then all be represented mathematically by the same state vector.

\subsection{\label{Sect:scalarcoh} A scalar field analogy}

To motivate our proposed resolution to this difficulty, we first digress to consider how one might answer the analogous question for a real (say massless) quantum scalar field, $\varphi$, coupled to a classical external static scalar source with scalar charge density $\sigma$.  Now the free Hamiltonian takes the familiar form
\begin{equation}
\label{Hsc0}
H^{\mathrm{sc}}_0=\int {1\over 2} \pi^2 +{1\over 2}  (\bm\nabla \varphi)^2 \, d^3x
\end{equation}
whilst the Hamiltonian in the presence of the source takes the \textit{different} form
\begin{equation}
\label{Hscrho}
H^{\mathrm{sc}}_\sigma=\int {1\over 2} \pi^2 +{1\over 2}  (\bm\nabla \varphi)^2 + \varphi\sigma \, d^3x.
\end{equation}
In this case (cf.\ \cite{KayNewt}) we claim that the state vector which describes the state of the $\varphi$ field in the presence of the source is 
\begin{equation}
\label{coh}
\Psi=e^{-i\pi(\phi_{\mathrm{sc}})}\Omega
\end{equation}
where $\Omega$ is the usual vacuum vector in the usual Fock space for the free scalar field and $\pi(\phi_{\mathrm{sc}})$ denotes the quantum field-momentum $\pi$ smeared with the classical solution, 
\begin{equation}
\label{phisigma}
\phi_{\mathrm{sc}}(\bm x)=-\int \frac{\sigma(\bm y)}{4\pi|\bm x-\bm y|}\, d^3y, 
\end{equation}
to $\nabla^2\phi_{\mathrm{sc}}=\sigma$ -- i.e.\ $\pi(\phi_{\mathrm{sc}})=\int \pi(\bm x)\phi_{\mathrm{sc}}(\bm x) \, d^3x$.   (There is a mathematical subtlety here\footnote{\label{ftntSubtle} As was pointed out in \cite{KayNewt}, there is a mathematical subtlety here due to the fact that our quantum scalar field is massless.   Due to the $1/x$ tail in $\phi_{\mathrm{sc}}$, the putative one-particle Hilbert space vector $\chi$ of (\ref{chiscalar}) will not be normalizable and so does not strictly belong to the one-particle Hilbert space and so $e^{-i\pi(\phi_{\mathrm{sc}})}$ is not strictly a unitary operator (the integrand in the integral $\int \pi(\bm x)\phi_{\mathrm{sc}}(\bm x) d^3x$ has an infra-red [i.e.\ large distance] divergence in it) and the coherent state $\Psi$ of (\ref{coh}) does not strictly belong to the augmented Fock space.  In terms of notions from the algebraic approach to quantum field theory \cite{Haag} it should really be understood as a state (namely the composition of the vacuum state with the automorphism $\varphi \mapsto \varphi + \phi_{\mathrm{sc}}$ -- see (\ref{scalarshift})) in the sense of a positive linear functional on the relevant field algebra, and a change of Hilbert-space representation should be invoked.  However, we will proceed \textit{as if} these vectors were normalizable since, as long as the sources, $\sigma$, of the scalar field, $\varphi$, that they describe have the same integral $\int \sigma(\bm x)\, d^3x$ (analogous to the total charge in the electrostatic case) the quantity we wish to compute, i.e.\ $|\langle\Psi_1|\Psi_2\rangle|$, given by the formula (\ref{D0formula}), will be finite.   This is because, as long as those integrals are equal, the infra-red divergences in each of the terms, $\chi_1$, $\chi_2$ in  (\ref{D0formula}) will cancel out.  We will assume that finite result to be the physically correct value and we expect that this can be demonstrated rigorously.  

\smallskip

Similar remarks apply to the electrostatic coherent state (\ref{cohem}) and in the linearized gravity case discussed in \cite{KayNewt} and to be discussed further in \cite{QGS}.  Also the integrands in the exponents in the definitions of the formally unitary operators, $U$ of (\ref{Uext}), $U$ of (\ref{U}) in Section \ref{Sect:QED} and its counterpart $U$ in Section \ref{Sect:QEDSchr} all have infra-red divergencies similar to that in $\int \pi(\bm x)\phi_{\mathrm{sc}}(\bm x) d^3x$ while the $U$ of (\ref{U}) in Section \ref{Sect:QED} additionally has ultra-violet divergencies both from the short distance $1/r$ divergence and and also to the field product $\psi^*(x)\psi(x)$ in the $\phi$ of (\ref{phi}).    Thus  the physical subspaces of Sections \ref{Sect:QED} and \ref{Sect:QEDSchr}  are not strictly subspaces of the relevant QED augmented Hilbert spaces.   However, just as we obtained finite results for the inner products of coherent states in Section \ref{Sect:extsource}, we obtain finite results e.g.\ for the reduced density operator of Section \ref{Sect:DensOp}.}.)

By expressing the Fourier transforms\footnote{\label{ftntFourier}  Throughout the paper, we adopt the convention that the Fourier transform $\tilde f(\bm k)$ of a function $f(\bm x)$ on ${\mathbb R}^3$ is $(2\pi)^{-3/2}\int f(\bm x) e^{-i\bm k\bm\cdot\bm x}\, d^3 x$.   We often omit the tilde if it is clear from the context (or from the fact that the value is denoted $\bm k$ rather than $\bm x$) that it is the Fourier transform that is being referred to.}, $\phi(\bm k)$ and $\pi(\bm k)$, of the quantum fields $\phi(\bm x)$ and $\pi(\bm x)$ in the usual way as
\begin{equation}
\label{tradphipi}
\phi(\bm k) = \frac{a(\bm k) + a^+(\bm k)}{\sqrt{2}|k|^{1/2}}, \quad \pi(\bm k)
= - i|k|^{1/2}\left(\frac{a(\bm k) - a^+(\bm k)}{\sqrt 2}\right)
\end{equation}
(actually we shall only need the second of these expressions below)
where $a(\bm k)$ and $a^+(\bm k')$ are the momentum space annihilation and creation operators, satisfying
\begin{equation}
\label{aastar}
[a(\bm k), a^+(\bm k')]=\delta^{(3)}(\bm k + \bm k'),\ [a(\bm k), a(\bm k')] = 0 = [a^+(\bm k), a^+(\bm k')],
\end{equation}
and $a(\bm k)\Omega = 0$, we may see that the exponent, $-i\pi(\phi_{\mathrm{sc}})$, in (\ref{coh}) can be rewritten as $a^+(\chi)-a(\chi)$ where 
\begin{equation}
\label{chiscalar}
\chi(\bm k) = |k|^{1/2}\phi_{\mathrm{sc}}(\bm k)/\sqrt 2
\end{equation}
(where  $\phi_{\mathrm{sc}}(\bm k)$ denotes the Fourier transform of $\phi_{\mathrm{sc}}(\bm x)$) and, as we shall presently demonstrate, it follows from this that (\ref{coh}) can alternatively be written (cf.\ \cite{KayNewt})
\begin{equation}
\label{scalarchiUomega}
\Psi=e^{-\langle\chi|\chi\rangle/2}e^{a^+(\chi)}\Omega 
\end{equation}
where $a(\chi)$ and $a^+(\chi)$ denote $\int a(\bm k)\chi(\bm k)\, d^3 k$ and $\int a^+(\bm k)\chi(\bm k)\, d^3 k$ respectively and where, for general functions, $\chi_1$ and 
$\chi_2$,  $\langle\chi_1|\chi_2\rangle$ is the $L^2$ inner-product $\int \chi_1(\bm k)^*\chi_2(\bm k)\, d^3k$ ($=\int \chi_1(\bm x)^*\chi_2(\bm x)\, d^3x$), where $^*$ denotes complex conjugation.  
 
To demonstrate (\ref{scalarchiUomega}), let us first note that, by (\ref{aastar}), we have
$[a(\chi), a^+(\chi)]=\int \chi(-\bm k)\chi(\bm k)\, d^3k$.   But, in view of the fact that 
$\phi_{\mathrm{sc}}(\bm x)$ is real, $\phi_{\mathrm{sc}}(-\bm k) =                                                                                                                                                                                                                                                                                                                                                                                                      \phi_{\mathrm{sc}}(\bm k)^*$ and therefore $\chi(-\bm k)=\chi(\bm k)^*$.   Thus 
$[a(\chi), a^+(\chi)] = \int \chi(\bm k)^*\chi(\bm k)\, d^3k = \langle\chi|\chi\rangle$.  We then have   
\begin{equation}
\label{UOmegDeriv}
e^{-i\pi(\phi_{\mathrm{sc}})}\Omega = e^{a^+(\chi)-a(\chi)}\Omega = 
e^{-[a(\chi) ,a^+(\chi)]/2}e^{a^+(\chi)}e^{-a(\chi)}\Omega = e^{-\langle\chi|\chi\rangle/2}e^{a^+(\chi)}\Omega.
\end{equation}

The absolute value, $|\langle\Psi_1|\Psi_2\rangle|$, of the inner product between two such coherent states,  $\Psi_1=e^{-\langle\chi_1|\chi_1\rangle/2}e^{a^+(\chi_1)}\Omega$ and $\Psi_2=e^{-\langle\chi_2|\chi_2\rangle/2}e^{a^+(\chi_2)}\Omega$, for two different scalar charge densities, say $\sigma_1$ and $\sigma_2$, is then easily seen\footnote{\label{ftntDetails}  In this footnote, we give some more details on the derivation of Equations (\ref{scalarchiUomega}) and (\ref{D0formula}) and also Equations (\ref{spin1chiUomega}) and (\ref{D1formula}).  The derivation (\ref{UOmegDeriv}) of Equation (\ref{scalarchiUomega}) makes use of the special case, $e^Be^A=e^{A+B-[A,B]/2}$, of the Baker-Campbell-Hausdorff ({\it BCH}) relation for a pair of operators when their commutator is a $c$-number (applied to $a(\psi)$ and $a^+(\phi)$).

\smallskip

As for Equations (\ref{D0formula}), it is easy to see from the commutation relations (\ref{aastar}) for our creation and annihilation operators, using the same BCH formula, or rather $e^Be^A = e^{-[A,B]} e^Ae^B$, that if $\Psi_1$ and $\Psi_2$ are as in (\ref{scalarchiUomega}) (with $\chi$ replaced by $\chi_1$ and $\chi_2$ respectively) then 
$\langle\Psi_1|\Psi_2\rangle = \exp(-\langle\chi_1|\chi_1\rangle/2 - \langle\chi_2|\chi_2\rangle/2+\langle\chi_1|\chi_2\rangle)$.   So $|\langle\Psi_1|\Psi_2\rangle| = \exp(-\|\chi_1-\chi_2\|^2/2)$.

\smallskip

If $\chi_1$ and $\chi_2$ are related by a spatial translation (so we would also say that $\Psi_1$ and $\Psi_2$ are related by that spacelike translation) then, in momentum space, we will have, say $\chi_2(\bm k) = \exp(i\bm k\bm\cdot\bm a)\chi_1(\bm k)$ and hence one sees, by thinking of it as an integral in momentum space, that $\langle\chi_1|\chi_2\rangle$ is real and hence $\langle\Psi_1|\Psi_2\rangle = |\langle\Psi_1|\Psi_2\rangle|$.  Writing the latter as 
$\exp(-D_0)$, we thus confirm (\ref{D0formula}).  We also easily conclude that $D_0$ has the following alternative forms:  
\begin{equation}
\label{altform}
D_0=\|\chi_1-\chi_2\|^2/2 = \langle \chi|(1-\cos(\bm k\bm\cdot\bm a))\chi\rangle =\langle \chi|(1-\exp(i(\bm k\bm\cdot\bm a)))\chi\rangle.
\end{equation}

\smallskip

Similar results to those discussed above hold for electrostatic coherent states and $D_1$.
In particular, to derive (\ref{spin1chiUomega}),  we 
first note (cf.\ the paragraph ending with Equation (\ref{UOmegDeriv})) that the $\bm\chi(\bm k)$ of (\ref{phichi}) is an odd function of $\bm k$ (unlike the $\chi(\bm k)$ of (\ref{chiscalar}) which is even) and thus
\[
[a_i(\chi^i), a^+_j(\chi^j)] = \int \chi^i(-\bm k)\chi^i(\bm k)\, d^3k = - \int {\chi^*}^i(\bm k)\chi^i(\bm k)\, d^3k = -\langle\chi^i|\chi^i\rangle. 
\]
We then have (cf.\ (\ref{UOmegDeriv}))
\[
e^{i\int\hat A^i\partial_i\phi\, d^3x}\Omega = e^{-a^+_i(\chi^i)-a_i(\chi^i)}\Omega = 
e^{[a_i(\chi^i), a^+_j(\chi^j)]/2}e^{-a^+_i(\chi^i)}e^{-a_i(\chi^i)}\Omega 
\]
\[ 
=e^{[a_i(\chi^i), a^+_j(\chi^j)]/2}e^{-a^+_i(\chi^i)}\Omega = e^{-\langle\chi^i|\chi^i\rangle/2}e^{-a^+_i(\chi^i)}\Omega.
\]

\smallskip

Finally, the derivation of (\ref{D1formula}) is similar to that of (\ref{D0formula}) indicated above.  (Essentially all of the results referred to here were already stated in \cite{KayNewt}.)}  to equal 
$\exp(-D_0)$ where the \textit{scalar decoherence exponent}, $D_0$, is given by
\begin{equation}
\label{D0formula}
D_0 = \|\chi_1 - \chi_2\|^2/2
\end{equation}
where $\|\chi_1 - \chi_2\|^2$ is defined to mean $\langle\chi_1 - \chi_2|\chi_1 - \chi_2\rangle$.

As we shall explain in Section \ref{Sect:electrocoh} and as was already said in  \cite{KayNewt}, if we identify the scalar charge, $\sigma$, with the electrostatic charge, $\rho$, then $D_0$ is numerically equal to the $D_1$ discussed in Section \ref{Sect:Intro}.   So, in the case of two spheres of radius $R$ with uniform scalar charge density which are spatial translates of one-another, it is given again by the formula (\ref{balloverlap2}) etc.

Our claim in the sentence containing Equation (\ref{coh}) is justified by the fact that the $\Psi$ of (\ref{coh}) has the properties
\begin{equation}
\label{Psiprops}
\langle\Psi| \varphi\Psi\rangle=\phi_{\mathrm{sc}} \ {\rm and} \ H^{\mathrm{sc}}_\sigma\Psi=V_\sigma\Psi,
\end{equation}
where
\[
V_\sigma=-{1\over 2} \int \int  {\sigma(\bm x)\sigma(\bm y)\over 4\pi|\bm x-\bm y|} \, d^3x d^3y = {1\over 2}\int \phi_{\mathrm{sc}}\sigma\, d^3 x
\]
which may be thought of as the self-energy of the external classical source due to its interaction with the scalar field.

Equation (\ref{Psiprops}) tells us that the expectation value of the quantum scalar field, $\varphi$, in the state $\Psi$ is the classical field $\phi_{\mathrm{sc}}$ and that $\Psi$ is also an eigenstate of $H_\sigma$ with energy $V_\sigma$.

These properties (\ref{Psiprops}) immediately follow once one notes, as may easily be shown, that the unitary operator,
\[
U=e^{-i\pi(\phi_{\mathrm{sc}})},
\]
on the scalar field Fock space,
where $\phi_{\mathrm{sc}}(\bm x)$ is as in (\ref{phisigma}), satisfies
\begin{equation}
\label{scalarshift}
U^{-1}\varphi U=\varphi + \phi_{\mathrm{sc}}
\end{equation}
and hence also
\begin{equation}
\label{scHamshift}
U^{-1}H^{\mathrm{sc}}_\sigma U=H^{\mathrm{sc}}_0 + V_\sigma.
\end{equation}

In view of the properties (\ref{Psiprops}), the state $\Psi$ of equation (\ref{scalarchiUomega}) deserves to be considered a type of coherent state, albeit it is a coherent state corresponding to a static, non-propagating, classical field configuration -- namely that due to our external source.  (The coherent states which are usually considered, e.g.\ in quantum optics, correspond to classical states of radiation -- see e.g.\ \cite{KlauderSudarshan}.)

\subsection{\label{Sect:twoequiv} Two equivalent formulations of the free electromagnetic field}

Inspired by this scalar-field analogy, we seek a solution to our electromagnetic problem with a suitable analogue of the coherent-state vector $\Psi$.   As we shall see, the construction that we arrive at (which comes in two variations or `frameworks' as we will call them) has some interesting similarities to, but also some important differences from, the scalar case.

First we remark that the traditional Hilbert space on which the operators $\bm A$, ${\bm \pi^\perp}$ and $H^{\mathrm{EM}}_0$ act is the \textit{transverse Fock space}.  We pause to recall what is usually meant by this.   One may start with the one-particle Hilbert space, $\cal H_{\mathrm{one}}$, equal to the direct sum of 3 copies of the usual one-particle Hilbert space for a scalar field, which may be taken to be 3 copies of the space of square-integrable complex-valued functions, $L^2(\mathbb R^3)$, on momentum space on which act the usual annihilation and creation operators, $a_i(\bm k)$ and $a^+_i(\bm k)$, $i=1,2,3$, satisfying $[a_i(\bm k), a^+_j(\bm k')]=\delta_{ij}\delta^{(3)}(\bm k+\bm k')$ and $[a_i(\bm k), a_j(\bm k')]=0=[a^+_i(\bm k), a^+_j(\bm k')]$.  One may then define the transverse one-particle Hilbert space,  $\cal H_{\mathrm{one}}^{\mathrm{trans}}$, to be the subspace of  $\cal H_{\mathrm{one}}$ consisting of elements, $\bm\chi = \chi^1\oplus \chi^2 \oplus \chi^3$, of $\cal H_{\mathrm{one}}$ which satisfy the transversality condition $k_i\chi^i(\bm k)=0$.   The transverse Fock space is then simply the Fock space,  $\cal F(\cal H_{\mathrm{one}}^{\mathrm{trans}})$, over (see e.g.\ \cite{RS}) $\cal H_{\mathrm{one}}^{\mathrm{trans}}$ and the operators $\bm\pi^\perp$ and $\bm A$ are then represented on this Fock space by the usual expressions
\begin{equation}
\label{tradApi}
A_i(\bm k)\!=\!\left({a^{\mathrm{trans}}_i(\bm k)+{{a_i^+}^{\mathrm{trans}}}(\bm k)\over 
\sqrt 2|k|^{1/2}}\right),\  {\pi^\perp}_i(\bm k)\!=\!-i|k|^{1/2}\left({a^{\mathrm{trans}}_i
(\bm k)-{{a_i^+}^{\mathrm{trans}}}(\bm k)\over \sqrt 2}\right)
\end{equation}
where $a^{\mathrm{trans}}_i(\bm k)=(\delta_i^j-k_ik^j/k^2)a_j(\bm k)$ and 
${a^+_i}^{\mathrm{trans}}(\bm k)=(\delta_i^j-k_ik^j/k^2)a^+_j(\bm k)$,
while $H^{\mathrm{EM}}_0$  is defined by substituting these expressions into (\ref{Hamfree}) and the usual normal ordering procedure.

Secondly, we notice that a slightly different definition of the same Hilbert space is possible based on the fact that an arbitrary element, $\bm\chi$, of $\cal H_{\mathrm{one}}$ may be uniquely written as the sum of its transverse part, having $i$th component
$\left (\delta^i_j-\frac{k^ik_j}{k^2}\right )\chi^j(\bm k)$, and its longitudinal part, having $i$th component 
${k^ik_j\over k^2}\chi^j(\bm k)$.
Accordingly, $\cal H_{\mathrm{one}}$ arises as the direct sum of our subspace $H_{\mathrm{one}}^{\mathrm{trans}}$ with a longitudinal subspace, 
$\cal H_{\mathrm{one}}^{\mathrm{long}}$.   In symbols,
\begin{equation}
\label{HilbSum}
\cal H_{\mathrm{one}}=\cal H_{\mathrm{one}}^{\mathrm{trans}}\oplus \cal H_{\mathrm{one}}^{\mathrm{long}}.
\end{equation}
We then introduce what we shall call the \textit{augmented Fock space}, 
$\cal F(\cal H_{\mathrm{one}})$, over $\cal H_{\mathrm{one}}$, or, in view of (\ref{HilbSum}), what amounts to the same thing:
\begin{equation}
\label{FofHone}
\cal F(\cal H_{\mathrm{one}})=\cal F(\cal H_{\mathrm{one}}^{\mathrm{trans}})\otimes \cal F(\cal H_{\mathrm{one}}^{\mathrm{long}})
\end{equation}
and take, as our alternative, slightly different, definition for the Hilbert space for the free electromagnetic field, the subspace
\[
{\cal F(\cal H_{\mathrm{one}}^{\mathrm{trans}})}\otimes \Omega^{\mathrm{long}}
\]
of this augmented Fock space which consists of elements of the form 
\[
\Psi^{\mathrm{trans}}\otimes \Omega^{\mathrm{long}}
\]
where $\Omega^{\mathrm{long}}$ is the vacuum vector in $\cal F(\cal H_{\mathrm{one}}^{\mathrm{long}})$.
In other words, \hfil\break
${\cal F(\cal H_{\mathrm{one}}^{\mathrm{trans}})}\otimes \Omega^{\mathrm{long}}$ consists of the result of acting on the vacuum 
\begin{equation}
\label{OmegaOtimes}
\Omega = \Omega^{\mathrm{trans}}\otimes\Omega^{\mathrm{long}}
\end{equation}
(where $\Omega^{\mathrm{trans}}$ is the vacuum vector in $\cal F(\cal H_{\mathrm{one}}^{\mathrm{trans}})$)
of the augmented Fock space with only transverse creation operators.   It is obvious how the traditional operators $\bm A$, ${\bm \pi^\perp}$ and $H^{\mathrm{EM}}_0$ act on ${\cal F(\cal H_{\mathrm{one}}^{\mathrm{trans}})}\otimes \Omega^{\mathrm{long}}$ and it is obvious that this action is equivalent to the action of this same set of operators on the traditional Hilbert space, $\cal F(\cal H_{\mathrm{one}}^{\mathrm{trans}})$ that we mentioned at the start.  The non-vacuum elements of  $\cal F(\cal H_{\mathrm{one}}^{\mathrm{long}})$ correspond to states of longitudinal photons and play no role in the quantization of the free theory with no external charges, but they will play a role, as we will see, in the presence of classical external charges and/or (as we discuss in Section \ref{Sect:QED}) when the electromagnetic field is coupled to another (charged) dynamical quantum field or (as we discuss in Section \ref{Sect:QEDSchr}) system of (charged) quantum particles, in our product picture.

\subsection{\label{Sect:electrocoh} Electrostatic coherent states}

We will indeed next show that a suitable notion of quantum coherent state, describing the static electric field due to a given classical static external charge distribution, is provided by a certain element (see however Footnote \ref{ftntSubtle}) of our augmented Fock space, $\cal F(\cal H_{\mathrm{one}})=\cal F(\cal H_{\mathrm{one}}^{\mathrm{trans}})\otimes \cal F(\cal H_{\mathrm{one}}^{\mathrm{long}})$,
which does not, however, belong to the subspace ${\cal F({\cal H}_{\mathrm{one}}^{\mathrm{trans}})}\otimes \Omega^{\mathrm{long}}$.

To achieve this, we need to introduce suitable extensions of the (usual, transverse) electromagnetic field momentum operator $\bm\pi^\perp$, and of the Hamiltonian, to our augmented Fock space  ${\cal F({\cal H}_{\mathrm{one}}})$.   At first sight, it might seem that the way to extend $\bm\pi^\perp$ would be by defining the operator $\hat{\bm\pi}$ by (cf.\ (\ref{tradApi}))
\[
\hat\pi_i = -i|k|^{1/2}\left({a_i(\bm k)-a_i^+(\bm k)}\over \sqrt 2\right).
\]
In other words, 
\begin{equation}
\label{pihat}
\hat\pi_i={\pi^\perp}_i+\hat\pi^{\mathrm{long}}_i
\end{equation}
where
\begin{equation}
\label{pihatdef}
{\hat\pi_i}^{\mathrm{long}} = -i|k|^{1/2}\left({a_i^{\mathrm{long}}(\bm k)-{a_i^+}^{\mathrm{long}}(\bm k)}\over \sqrt 2\right),  
\end{equation}
where 
\[
a^{\mathrm{long}}_i(\bm k)=\frac{k_ik^j}{k^2}a_j(\bm k).
\]
Then one could extend the Hamiltonian  to  $\hat H^{\mathrm{EM}}_0$, defining the latter to be 
\begin{equation}
\label{Hamfreehat}
\hat H^{\mathrm{EM}}_0=
\int  {1\over 2} \hat{\bm\pi}^2 +{1\over 2}(\bm\nabla \bm\times \bm A)^2 \, d^3x =\int {1\over 2} {\bm\pi^\perp}^2 +{1\over 2}\mbox{$\hat{\bm\pi}^{\mathrm{long}}$}^2+{1\over 2}(\bm\nabla \bm\times \bm A)^2 \, d^3x.
\end{equation} 
However, we shall also consider (and eventually prefer) an alternative framework in which, instead of $\hat{\bm\pi}$ and $\hat H^{\mathrm{EM}}_0$, we take the (non-self adjoint! -- see our third ``likely to be asked question'' below) operators $\tilde{\bm\pi}$ and $\tilde H$ defined by
\begin{equation}
\label{pitilde}
\tilde\pi_i={\pi^\perp}_i+\tilde\pi^{\mathrm{long}}_i
\end{equation}
where
\begin{equation}
\label{pilongtilde}
\tilde\pi^{\mathrm{long}}_i(\bm k)=-2i|k|^{1/2}\left({{a_i^{\mathrm{long}}}(\bm k)\over \sqrt 2}\right) 
\end{equation}
and
\begin{equation}
\label{Hamfreetilde}
\tilde H^{\mathrm{EM}}_0=
\int  {1\over 2} \tilde{\bm\pi}^2 +{1\over 2}(\bm\nabla \bm\times \bm A)^2 \, d^3x =\int {1\over 2} {\bm\pi^\perp}^2 +{1\over 2}\mbox{$\tilde{\bm\pi}^{\mathrm{long}}$}^2+{1\over 2}(\bm\nabla \bm\times \bm A)^2 \, d^3x.
\end{equation} 

We remark that $\hat{\bm\pi}$ is, as usual, the difference of a creation operator term and an annihilation operator term and that $\tilde{\bm\pi}$ may be thought of as obtained from $\hat{\bm\pi}$ by deleting the creation operator term while doubling the annihilation operator term.

It is easy to check that $\hat{\bm\pi}$, $\hat H^{\mathrm{EM}}_0$  and $\bm A$ have the same commutation relations amongst themselves as do $\bm\pi^\perp$, $H^{\mathrm{EM}}_0$ and $\bm A$ amongst themselves and one also easily sees that the same is true of $\tilde{\bm\pi}$, $\tilde H^{\mathrm{EM}}_0$  and $\bm A$.   

Also the right hand side of the quantum Hamilton equations,  $\dot{\bm A}=i[\hat H^{\mathrm{EM}}_0, \bm A]$ and $\dot{\hat{\bm \pi}}=i[\hat H^{\mathrm{EM}}_0, \hat{\bm\pi}]$ for $\bm A$ and $\hat{\bm\pi}$ will be identical with the right hand side of the usual Hamilton equations for $\bm A$ and $\bm\pi^\perp$.    And again, the same is true with $\hat{\bm \pi}$ replaced by $\tilde{\bm \pi}$ and $\hat H^{\mathrm{EM}}_0$ replaced by
$\tilde H^{\mathrm{EM}}_0$.  So, in particular we will have $\dot{\hat{\bm\pi}}=\dot{\bm\pi}^\perp=\dot{\tilde{\bm\pi}}$.  

Despite being nonself-adjoint, there are some properties that $\tilde{\bm\pi}$ and $\tilde H^{\mathrm{EM}}_0$
possess which are \textit{not} possessed by $\hat{\bm\pi}$ and $\hat H^{\mathrm{EM}}_0$ and which will be crucial for what we do in Sections \ref{Sect:QED} and \ref{Sect:QEDSchr}. Namely, along with $\bm A$, $\tilde{\bm\pi}$ and $\tilde H^{\mathrm{EM}}_0$ both map the subspace ${\cal F(\cal H_{\mathrm{one}}^{\mathrm{trans}})}\otimes \Omega^{\mathrm{long}}$ of our augmented Fock space $\cal F(\cal H_{\mathrm{one}})$ to itself and the theory in which the free electromagnetic Hamiltonian,  the vector potential and the electric field, $\bm E$, are identified with the $H^{\mathrm{EM}}_0$ of (\ref{Hamfree}), the $\bm A$ of (\ref{tradApi}), and (minus) the $\bm\pi^\perp$ of (\ref{tradApi}) -- all acting on the transverse Fock space, $\cal F(\cal H_{\mathrm{one}}^{\mathrm{trans}})$ -- is fully equivalent to the theory in which they are identified with the $\tilde H^{\mathrm{EM}}_0$ of (\ref{Hamfreetilde}), the $\bm A$ of (\ref{tradApi}), and (minus) the $\tilde{\bm \pi}$ of (\ref{pitilde}) -- all acting on the 
subspace ${\cal F(\cal H_{\mathrm{one}}^{\mathrm{trans}})}\otimes \Omega^{\mathrm{long}}$ of our augmented Fock space $\cal F(\cal H_{\mathrm{one}})$ (defined above Equation (\ref{FofHone})).    After all, on every element, $\Phi$, in this subspace, we have $\tilde H^{\mathrm{EM}}_0\Phi=H^{\mathrm{EM}}_0\Phi$ and $\tilde{\bm\pi}\Phi=\bm\pi^\perp\Phi$!  In particular, for all $\Phi$ in this subspace, $\bm{\nabla\cdot}\tilde{\bm\pi}\Phi = 0$, equivalently $\bm{\tilde\pi}^{\mathrm{long}}\Omega=0$ -- a special case of which is worth noting:
\begin{equation}
\label{tildepiomega} 
\bm{\nabla\cdot}\tilde{\bm\pi}\,\Omega = 0, \ \hbox{equivalently} \  \bm{\tilde\pi}^{\mathrm{long}}\Omega=0.
\end{equation}
In summary, we have two possible candidates for the extension of the electric field operator, $\bm E$, to our augmented Fock space, $- \hat{\bm\pi}$ and $- \tilde{\bm\pi}$, and corresponding candidates, $\hat H^{\mathrm{EM}}_0$ and $\tilde H^{\mathrm{EM}}_0$, respectively, for the extension of the Hamiltonian.   Depending on whether we choose to use $\hat{\bm\pi}$ and $\hat H^{\mathrm{EM}}_0$ or $\tilde{\bm\pi}$ and $\tilde H^{\mathrm{EM}}_0$, we shall say we are in the $\hat{\bm\pi}$ \textit{framework} or the $\tilde{\bm\pi}$ \textit{framework}.   And we have seen that the two frameworks have some properties in common but there are also significant differences.

In the remainder of this section, we shall give full details for the $\tilde{\bm\pi}$ framework and shall indicate the corresponding results for the $\hat{\bm\pi}$ framework by pointing out similarities and differences as we go along.

Next (and in both frameworks) we introduce the (non-transverse) operator $\hat{\bm A}$ on our augmented Fock space, given simply by
\begin{equation}
\label{Ahat}
\hat A_i(\bm k)=\left({a_i(\bm k)+{a^+_i}(\bm k)\over \sqrt 2|k|^{1/2}}\right).
\end{equation}
As one may easily check, this will have the commutation relations
\begin{equation}
\label{Ahatpitildecomm}
[\hat A_i(\bm x), \tilde \pi_j(\bm y)]=i\delta_{ij}\delta^{(3)}(\bm x-\bm y), 
\end{equation}
with $\tilde{\bm \pi}$, and we'll of course also have
\begin{equation}
\label{Bhatpitildecomm}
[\hat A_i(\bm x), \hat A_j(\bm y)] =0 = [\tilde\pi_i(\bm x), \tilde\pi_j(\bm y)]
\end{equation}
and similar equations to (\ref{Ahatpitildecomm}) and (\ref{Bhatpitildecomm}) will hold with $\tilde{\bm \pi}$ replaced by $\hat{\bm \pi}$ .
We now notice that the unitary operator (see Footnote \ref{ftntSubtle}), 
\begin{equation}
\label{Uext}
U=e^{i\int\hat A^i(\bm x)\partial_i\phi(\bm x)\, d^3x} \quad (= e^{i\int\hat A^{i\,\mathrm{long}}(\bm x)\partial_i\phi(\bm x)\, d^3x})
\end{equation}
satisfies (cf.\ Equations (\ref{scalarshift}), (\ref{scHamshift}))
\begin{equation}
\label{emshift}
U^{-1}\tilde{\bm \pi}U=\tilde{\bm \pi} + \bm\nabla\phi
\end{equation}
and, bearing in mind the formula (\ref{VCoulombphi}), we also have
\begin{equation}
\label{emHamshift}
U^{-1}\tilde H^{\mathrm{EM}}_0U=\tilde H^{\mathrm{EM}}_0+\int \tilde{\bm \pi}\bm{\cdot\nabla}\phi \, 
d^3 x + V_{\mathrm{Coulomb}}, 
\end{equation}
and similar equations to (\ref{emshift}) and (\ref{emHamshift}) hold for $\hat{\bm\pi}$.   

Now however, there are differences between the $\tilde{\bm\pi}$ and the $\hat{\bm\pi}$ framework.   In the case of $\tilde{\bm\pi}$, thanks to the above `crucial properties', we may conclude from (\ref{emHamshift}) that the vector,
\begin{equation}
\label{cohem}
\Psi=U\Omega,
\end{equation}
obtained by acting with $U$ on the vacuum vector, $\Omega$ (\ref{OmegaOtimes}), in our augmented Fock space will satisfy 
\begin{equation}
\label{Psiemprops}
\langle\Psi|\tilde{\bm \pi}\Psi\rangle=\bm\nabla\phi\ \  \hbox{(i.e.}  \ \ \langle\Psi|{\bm E}\Psi\rangle= \bm E^{\mathrm{class}}) \  \hbox{and} \ \tilde H^{\mathrm{EM}}_0\Psi=V_{\mathrm{Coulomb}}\Psi
\end{equation}
(which clearly resemble Equations (\ref{Psiprops}).

In addition, and importantly, it easily follows from operating on both sides of Equation (\ref{emshift}) with $\bm\nabla\cdot$ and recalling Equation (\ref{tildepiomega}) and recalling that\break 
$\nabla^2\phi = - \rho$, that, in the $\tilde{\bm\pi}$ framework, Gauss's law holds at the quantum level in the strong sense that
\begin{equation}
\label{QuantGauss1}
\bm\nabla\bm\cdot\bm E\Psi = -\bm{\nabla\cdot}\tilde{\bm\pi}\Psi \ (= -\bm{\nabla\cdot}\tilde{\bm\pi}^{\mathrm{long}}\Psi) \ = \rho\Psi
\end{equation}
(without the need to take an expectation value).

We remark that (\ref{QuantGauss1}) generalizes from the $\Psi$ of (\ref{cohem}) to any vector\break 
$\Phi\in\cal F(\cal H_{\mathrm{one}})$ of form $\Phi^{\mathrm{trans}}\otimes U\Omega^{\mathrm{long}}$, where $\Phi^{\mathrm{trans}}$ is any vector in $\cal F(\cal H_{\mathrm{one}}^{\mathrm{trans}})$. Each such vector corresponds physically to some electromagnetic radiation (i.e.\ some photons) superposed on our static coherent state.  Also
(\ref{QuantGauss1}) is equivalent to 
\begin{equation}
\label{eigen}
\bm E^{\mathrm{long}}\Psi \  (= - \tilde{\bm\pi}^{\mathrm{long}}\Psi) = (- \bm\nabla\phi)\Psi 
\end{equation}
and, under the assumption of no external classical time-dependent magnetic field, the $(- \bm\nabla\phi)$ here is again the classical electric field, ${\bm E}^{\mathrm{class}}$, of our static charge distribution.   Again, this generalizes to any vector $\Phi^{\mathrm{trans}}\otimes U\Omega^{\mathrm{long}}$ where $\Phi^{\mathrm{trans}} \in \cal F(\cal H_{\mathrm{one}}^{\mathrm{trans}})$.

We see from (\ref{eigen}) that, for a given charge density, $\rho$, our electrostatic coherent state, $\Psi$, is an eigenstate of each of the electric field operators, $\bm E^{\mathrm{long}}(\bm x)$ ($= - \tilde{\bm\pi}^{\mathrm{long}}(\bm x)$), for each point, $\bm x$, with eigenvalue $-(\bm\nabla\phi)(\bm x)$ (= ${\bm E}^{\mathrm{class}}(\bm x)$ when there is no external classical time-dependent magnetic field).

On the other hand, in the $\hat{\bm\pi}$ framework, in place of Equations (\ref{Psiemprops}) we will have the equations
\begin{equation}
\label{Psiemhatprops}
\langle\Psi| \hat{\bm \pi}\Psi\rangle=\bm\nabla\phi \ \ ({\rm and\ so} \  \langle\Psi| {\bm E}\Psi\rangle= \bm E^{\mathrm{class}}) \  \ {\rm and} \ \langle\Psi| \hat H^{\mathrm{EM}}_0\Psi\rangle=V_{\mathrm{Coulomb}}.
\end{equation}
which still resemble (\ref{Psiprops}), albeit the second equation is weaker than the second equation of (\ref{Psiprops}) (and weaker than the second equation of (\ref{Psiemprops})). 
Also, instead of the strong Gauss Law of (\ref{QuantGauss1}), in the $\hat{\bm\pi}$ framework, Gauss's law will only hold in expectation value, i.e.\ we will have
\begin{equation}
\label{QuantGausshat}
\langle\Psi|\bm\nabla\bm\cdot\bm E\Psi\rangle = -\langle\Psi|\bm{\nabla\cdot}\hat{\bm\pi}\Psi\rangle \ (= -\langle\Psi|\bm{\nabla\cdot}\hat{\bm\pi}^{\mathrm{long}}\Psi\rangle) \ = \rho
\end{equation}
or, equivalently
\begin{equation}
\label{weak}
\langle\Psi|\bm E^{\mathrm{long}}\Psi\rangle \  (= - \langle\Psi|\tilde{\bm\pi}^{\mathrm{long}}\Psi\rangle) = - (\bm\nabla\phi) \  
\end{equation}
\[
(= {\bm E}^{\mathrm{class}} \ \hbox{when no time dependent magnetic field}).
\]

Whichever framework we use, in view of the resemblance between (\ref{Psiemprops}) (respectively, (\ref{Psiemhatprops})) and (\ref{Psiprops}),  it is natural to propose that the vector, $\Psi$, of (\ref{cohem}) in our augmented Fock space, should be regarded as the correct description of the quantum state of the electric field due to an external, classically describable, static charge distribution $\rho$.  Note though that the failure of $\Psi$ to be an eigenstate of $\hat H^{\mathrm{EM}}$ would mean that, in the $\hat{\bm\pi}$ formalism, $\Psi$ would not be a stationary state, i.e.\ $\Psi(t)=\exp(-i\hat H^{\mathrm{EM}}t)\Psi$ would not be a phase-multiple of $\Psi$.  So, the state of the quantum electromagnetic field due to a static classical charge distribution source would not itself be static!   This fact alone may seem aesthetically unappealing.   Nevertheless, at any time, $\Psi(t)$, defined as above would still satisfy (\ref{Psiemhatprops}) and (\ref{QuantGausshat})/(\ref{weak}).  So as far as we know, in the absence of an experiment, such as that we discuss in Section \ref{Sect:capac} it doesn't rule it out.  

It is easy to see (cf.\ before Equation (\ref{chiscalar})) from (\ref{Ahat}) that the exponent, 
$i\int\hat A^i(\bm x)\partial_i\phi(\bm x)\, d^3x$ in (\ref{Uext}) can be written
$-a^+_i(\chi^i)-a_i(\chi^i)$ where\break 
$\chi^i$,  $i=1,2,3$, are given by 
\begin{equation}
\label{phichi}
\chi^i(\bm k) = k^i\phi(\bm k)/\sqrt{2}|k|^{1/2} \quad \hbox{($= i{E^{\mathrm{class}}}^i(\bm k)/\sqrt 2|k|^{1/2}$)}.
\end{equation}
And (cf.\ (\ref{scalarchiUomega}) and see Footnote \ref{ftntDetails} for details on the derivation) one then easily has (cf.\ (\ref{scalarchiUomega}))
\begin{equation}
\label{spin1chiUomega}
\Psi = e^{-\langle\bm\chi|\bm\chi\rangle/2}e^{-a^+_i(\chi^i)}\Omega
\end{equation}
where the inner-product $\langle\bm\chi|\bm\chi\rangle$ is taken in $\cal H_{\mathrm{one}}$.  (It is equal to $\langle\chi^i|\chi^i\rangle$ [summed from $i = 1$ to $3$] where the inner products are taken in $L^2({\mathbb R}^3)$.)

Moreover, the absolute value (cf.\ (\ref{D0formula}) and see again Footnote \ref{ftntDetails} for details) of the inner product, $|\langle\Psi_1|\Psi_2\rangle|$, of two such vectors, $\Psi_1$ and $\Psi_2$ -- for two electrostatic potentials, $\phi_1$ and $\phi_2$, for two charge distributions $\rho_1$ and $\rho_2$ --  will equal $\exp(-D_1)$ where the electrostatic (or `spin-1') decoherence exponent, $D_1$, is given by the formula
\begin{equation}
\label{D1formula}
D_1 = \|\bm\chi_1 - \bm\chi_2\|^2/2,
\end{equation}
($= \langle\bm\chi_1 - \bm\chi_2|\bm\chi_1 - \bm\chi_2\rangle/2$).

It is then easy to see, by comparing (\ref{D1formula}) with (\ref{D0formula}), that, as observed in \cite{KayNewt} and mentioned in Section \ref{Sect:scalarcoh}, the electrostatic (or `spin-1') decoherence exponent, $D_1$, is equal to the scalar or `spin-0' decoherence exponent, $D_0$, when the classical static scalar charge densities $\sigma_1$ and $\sigma_2$, are equated with $\rho_1$ and $\rho_2$.   We can also now easily obtain from (\ref{D1formula}) (by correcting what is done in \cite{KayNewt} in the light of Footnote \ref{ftntErr} here) the formula (\ref{balloverlap2}) for the decoherence exponent $D_1$ such that, when $\Psi_1$ and $\Psi_2$ are the two charged ball states discussed in Section \ref{Sect:Intro}, $|\langle\Psi_1|\Psi_2\rangle|=\exp(-D_1)$.

Aside from those similarities, however, let us remind ourselves that there are also notable differences between the scalar and electrostatic cases (as well as between our two frameworks for the latter).   In particular, we recall the contrast between the identity of the usual Coulomb gauge quantum Hamiltonian in the presence of a charge distribution with that in the absence of a charge distribution which we mentioned in the introductory paragraph to this section on the one hand and the difference in the scalar case between the Hamiltonians (\ref{Hsc0}) and (\ref{Hscrho}) in the presence and absence respectively of a scalar charge density on the other hand.  Also the need to introduce an augmented Fock space and the two new operators 
$\tilde{\bm\pi}$ and $\tilde H^{\mathrm{EM}}_0$ in Equations (\ref{pitilde}) and (\ref{Hamfreetilde}) (respectively the $\hat{\bm\pi}$ and $\hat H^{\mathrm{EM}}_0$ of (\ref{pihat}) and (\ref{Hamfreehat})) has no counterpart in the scalar case. Nor do either of our Gauss's law equations, (\ref{QuantGauss1}) or (\ref{QuantGausshat}).   Another notable difference, related to the above points, is that, in the scalar case, the $\Psi$ of Equation (\ref{coh}) can be arbitrarily well approximated by acting on the vacuum vector with products of (suitably smeared) creation operators, while, in the electrostatic case, the $\Psi$ of (\ref{cohem}) cannot be reached by acting on the vacuum vector of our augmented Fock space with products of (suitably smeared) creation operators ${\bm a^+}^{\mathrm{trans}}$.

Next we wish to anticipate, and answer, four likely-to-be-asked questions about all the above:  First, why do we need to extend to our augmented Fock space and add an extra longitudinal piece such as $\tilde{\bm\pi}^{\mathrm{long}}$ (or $\hat{\bm\pi}^{\mathrm{long}}$)  to $\bm{\pi^\perp}$ (and add a corresponding piece to the Hamiltonian) at all; why can't we find a unitary, $U$, on $\cal F(\cal H_{\mathrm{one}}^{\mathrm{trans}})$ (or equivalently on the subspace ${\cal F}({\cal H}_{\mathrm{one}}^{\mathrm{trans}})\otimes\Omega^{\mathrm{long}}$ of our augmented Fock space) such that
\[
U^{-1}\bm \pi^\perp U=\bm\pi^\perp + \bm\nabla\phi?
\]
The answer is that this simply can't be done; any such unitary conjugation on $\cal F(\cal H_{\mathrm{one}}^{\mathrm{trans}})$ would map $\bm\pi^\perp$ to something transverse, while $\bm\nabla\phi$ is longitudinal.   Indeed the obvious candidate for such a unitary, i.e.\ $\exp(i\int A^i(\bm x)\partial_i\phi(\bm x)\, d^3x)$, is in fact just the identity operator, since, in virtue of the Coulomb gauge condition, ${\bm\nabla\bm\cdot \bm A}=0$ (and integrating by parts) $\int A^i(\bm x)\partial_i\phi(\bm x)\, d^3x=0$!   (And if we didn't add a corresponding piece to the Hamiltonian, then the energy of our coherent state would be the same as the energy of the vacuum state.)

Secondly, why have we introduced our $\tilde{\bm\pi}$ framework at all, with its unusual-looking (and non-self adjoint -- see the next likely-to-be asked question) $\tilde{\bm\pi}$ (as we explained above, to be identified with $-\bm E$ on our augmented Fock space in the $\tilde{\bm\pi}$ framework) as in (\ref{pitilde}) with $\tilde{\bm\pi}^{\mathrm{long}}$ defined  in terms only of longitudinal photon \textit{annihilation operators} as in (\ref{pilongtilde}) together with $\tilde H^{\mathrm{EM}}_0$ defined as in (\ref{Hamfreetilde}) rather than contenting ourselves with the $\hat{\bm\pi}$ framework with its (self-adjoint) $\hat{\bm\pi}$ defined as in (\ref{pihatdef}) with the more familiar-looking difference of an annihilation and a creation operator, together with the $\hat H^{\mathrm{EM}}_0$ of (\ref{Hamfreehat})?  

Our answer is that, although the commutation relations amongst the set of operators, $\hat{\bm\pi}$, $\hat H^{\mathrm{EM}}_0$  and $\bm A$ will be the same as those amongst $\tilde{\bm\pi}$, $\tilde H^{\mathrm{EM}}_0$  and $\bm A$, and although the commutation relations between $\hat{\bm\pi}$ and $\hat{\bm A}$ would be the same as those (\ref{Ahatpitildecomm}) between $\tilde{\bm\pi}$ and $\hat{\bm A}$, $\hat{\bm\pi}$ would (obviously) \textit{not} map the subspace ${\cal F(\cal H_{\mathrm{one}}^{\mathrm{trans}})}\otimes \Omega^{\mathrm{long}}$ to itself.  This is, perhaps, not so important for justifying our claim that the $\Psi$ of (\ref{cohem}) deserves to be regarded as the correct mathematical object to identify with the quantum state of the electric field due to our external static charge distribution.  We saw, after all, that we could have replaced $\tilde{\bm\pi}$ by $\hat{\bm\pi}$ in Equations (\ref{emshift}) and (\ref{emHamshift}) and interpreted $\hat{\bm\pi}$ as minus the electric field operator, $\bm E$, and we would still have obtained the only slightly weaker equations (\ref{Psiemhatprops}).  It is true that, since the expectation value this time is in \textit{both} equations, this is a less sharp statement than (\ref{Psiemprops}) or (\ref{Psiprops}).  But it might, arguably, still serve to justify our claim.  But the fact that the subspace ${\cal F(\cal H_{\mathrm{one}}^{\mathrm{trans}})}\otimes \Omega^{\mathrm{long}}$ is mapped to itself by $\tilde{\bm\pi}$ will be crucial for the equivalence of our product picture of full QED with standard full QED (in Coulomb gauge), which we will derive in Section \ref{Sect:QED}  (and similarly crucial for the Maxwell-Schr\"odinger product picture to be discussed in Section \ref{Sect:QEDSchr}).   Moreoever, as we shall discuss next, in Section \ref{Sect:capac}, the fact that the classical energy coincides with the expectation value of the Hamiltonian in the $\hat{\bm\pi}$ framework, rather than being an eigenvalue of the Hamiltonian, as it is in the $\tilde{\bm\pi}$ framework, together with some natural assumptions about the physical interpretation of each of the frameworks, makes it possible, in principle at least, to distinguish between the two frameworks by an experimental test.   (And for reasons that we gave in the introduction (Section \ref{Sect:Intro1}) and will discuss further in Section \ref{Sect:capac}, we expect that the experiment would confirm the correctness of the $\tilde{\bm\pi}$ framework and rule out the $\hat{\bm\pi}$ framework.)

Thirdly, one might ask whether it is not a problem that $\tilde{\bm\pi}$ and $\tilde H^{\mathrm{EM}}_0$ fail to be self-adjoint on our full augmented Fock space $\cal F(\cal H_{\mathrm{one}})$. Our answer is that it isn't a problem as long as we only wish to interpret $\tilde{\bm\pi}$ as an observable (i.e.\ as minus the quantum electric field) when it is restricted to one of the subspaces of form ${\cal F(\cal H_{\mathrm{one}}^{\mathrm{trans}})}\otimes U_\rho\Omega^{\mathrm{long}}\in \cal F(\cal H_{\mathrm{one}})$ where $U_\rho$ now denotes the $U$ of (\ref{Uext}) for some given external charge distribution, $\rho$.   For $\tilde{\bm\pi}$ and $\tilde H^{\mathrm{EM}}_0$ both map each such subspace into itself and, restricted to each such subspace, they \textit{are} self-adjoint.  To see that they each map each such subspace to itself, first notice that, obviously, both $\tilde{\bm\pi}$ and $\tilde H^{\mathrm{EM}}_0$ map the subspace ${\cal F(\cal H_{\mathrm{one}}^{\mathrm{trans}})}\otimes \Omega^{\mathrm{long}}$  into itself because $\tilde{\bm\pi}^{\mathrm{trans}}$ clearly maps any vector, $\psi^{\mathrm{trans}}\otimes\Omega^{\mathrm{long}}$ in that subspace to that subspace, while $\tilde{\bm\pi}^{\mathrm{long}}$ annihilates $\Omega^{\mathrm{long}}$!     Turning to the self-adjointness, $\tilde{\bm\pi}^{\mathrm{trans}}$ is obviously self-adjoint.   Also $\tilde H^{\mathrm{EM}}_0$ is a sum of the manifestly self-adjoint terms
$\tilde H^{\mathrm{EM}}_0 + V_{\mathrm{Coulomb}}$ with the term $\int \tilde{\bm \pi}\bm{\cdot\bm\nabla}\phi \, d^3 x$ which is equal to $\int \tilde{\bm \pi}^{\mathrm{long}}\bm{\cdot\bm\nabla}\phi \, 
d^3 x$.   Thus to complete the proof of self-adjointness of both $\tilde{\bm\pi}$ and $\tilde H^{\mathrm{EM}}_0$, it suffices to prove that $\tilde{\bm \pi}^{\mathrm{long}}$ is self-adjoint on each of our subspaces.  But this follows immediately from Equation (\ref{eigen}) (or rather from the generalization mentioned immediately after Equation (\ref{eigen})) since that shows that, on each of our subspaces, and for each $\bm x$, $\tilde{\bm \pi}^{\mathrm{long}}(\bm x)$ acts as a real-valued multiplication operator (namely $(\bm\nabla\phi)(\bm x)$).

This perhaps seemingly almost trivial self-adjointness result may well still seem to be insufficient and one may continue to wonder whether it is not a problem that $\tilde{\bm\pi}$ fails to be self-adoint on larger subspaces if not on the full augmented Fock space.   But we would argue that it \textit{is} sufficient and, in fact, were $\tilde{\bm\pi}$ to be self-adjoint on larger subspaces or on the whole augmented Fock space -- as is the case for $\hat{\bm\pi}$ -- that would actually be a problem.   

Indeed, for consistency with Gauss's law, for a given charge distribution, $\rho$, we want the longitudinal electric field -- i.e.\ 
$-\tilde{\bm\pi}^{\mathrm{long}}(\bm x)$ -- to be an observable at each point $\bm x$, and for the only possible value of that observable at each such point to be $(\bm\nabla\phi)(\bm x)$.   And this \textit{is} consistent with the fact (see again after Equation (\ref{eigen})) that, for each $\bm x$, $-\tilde{\bm\pi}^{\mathrm{long}}(\bm x)$ is a self-adjoint operator on the subspace ${\cal F}({\cal H}_{\mathrm{one}}^{\mathrm{trans}})\otimes U_\rho\Omega^{\mathrm{long}}\in {\cal F}({\cal H}_{\mathrm{one}})$ if one adopts the usual quantum mechanical interpretation of the eigenvalues of a self adjoint operator as representing the possible values that the observable that it represents may take.  
Were $\tilde{\bm\pi}(\bm x)$ for any value of $\bm x$ to be self-adjoint on a larger subspace, we would expect to have other physically intepretable eigenvectors and eigenvalues which, arguably would be an \textit{embarrass de richesse}.

Likewise, one might feel that one should not be content with the Hamiltonian $\tilde H^{\mathrm{EM}}_0$ being self adjoint only on subspaces of the form ${\cal F(\cal H_{\mathrm{one}}^{\mathrm{trans}})}\otimes U_{\rho}\Omega^{\mathrm{long}}$.   For example, one might worry, should we work in the $\tilde{\bm\pi}$ framework, that if we were to consider two distinct external classical charge distributions, $\rho_1$ and $\rho_2$, the failure of $\tilde{\bm\pi}$ and of $\tilde H^{\mathrm{EM}}_0$ to be self-adjoint on the subspace spanned by
${\cal F(\cal H_{\mathrm{one}}^{\mathrm{trans}})}\otimes U_{\rho_1}\Omega^{\mathrm{long}}$ and ${\cal F(\cal H_{\mathrm{one}}^{\mathrm{trans}})}\otimes U_{\rho_2}\Omega^{\mathrm{long}}$ will make the physical intepretation of our transition amplitudes, $\langle\Psi_1|\Psi_2\rangle$ where $\Psi_1$ is $U_{\rho_1}\Omega$ and $\Psi_2= U_{\rho_2}\Omega$, problematic.  In particular, suppose e.g.\ that $\Psi_1$ happens to be an eigenstate of 
$\tilde H^{\mathrm{EM}}_0$ with eigenvalue (i.e.\ energy) $\epsilon$ (we reserve the symbol  `E' to denote the magnitude of an electric field) then one might worry that we would not be justified in interpreting the squared modulus, $|\langle\Psi_1|\Psi_2\rangle|^2$, of the transition amplitude $\langle\Psi_1|\Psi_2\rangle$ as the probability that the coherent state $\Psi_2$ will turn out to have energy $\epsilon$ when we measure its energy.  But actually we don't necessarily \textit{want} it to be interpretable in this way.   Indeed it is the fact that it would be interpretable in this way in the $\hat{\bm\pi}$ framework but not in the $\tilde{\bm\pi}$ framework that lies behind the fact that the experiment in Section \ref{Sect:capac}, which we mentioned above, would distinguish between the two frameworks -- the point being that $\hat{\bm\pi}$ is self-adjoint on the full augmented Fock space, and therefore it would be warranted in the $\hat{\bm\pi}$ framework on the standard interpretation of self-adjoint operators as observables to interpret $|\langle\Psi_1|\Psi_2\rangle|^2$ in this way.   

Note also that the transition amplitudes, $\langle\Psi_1|\Psi_2\rangle$, will, nevertheless, retain an important physical interpretation, also in our $\tilde{\bm\pi}$ framework, as a basic ingredient in the calculation of the partial trace of the pure density operator, $|\bm\Psi\rangle\langle\bm\Psi|$ of some total state, $\bm\Psi$, of nonrelativistic (many body) Schr\"odinger QED over the electromagnetic field in the product picture, as we first argued in \cite{KayNewt} and as we explained in Section \ref{Sect:Intro2}; and as we will show systematically in Section \ref{Sect:QEDSchr}.  (And such partial traces serve as a useful analogy for partial traces over gravity in linearized non-relativistic quantum gravity in a product picture for that theory, which play a r\^ole in my matter gravity entanglement hypothesis \cite{Kay1,KayNewt,eeee,KayThermality,KayEntropy,KayStringy,KayMore,KayMatGrav,KayRemarks}.)

Let us also point out, in support of contenting ourselves with self-adjointness of $\tilde{\bm\pi}$ and $\tilde H^{\mathrm{EM}}_0$ only on subspaces of the form ${\cal F(\cal H_{\mathrm{one}}^{\mathrm{trans}})}\otimes U_\rho\Omega^{\mathrm{long}}$, that those results are natural counterparts of analogous results for full QED which we prove in Section \ref{Sect:QED} --  namely the self-adjointness of $\tilde{\bm\pi}$ and of the product picture QED Hamiltonian, $H^{\mathrm{PP, Dirac}}_{\mathrm{QED}}$ (and the same is true of $\tilde{\bm\pi}$ and the Hamiltonian $H^{\mathrm{PP, Schr}}_{\mathrm{QED}}$ of Section \ref{Sect:QEDSchr}), on the `product picture physical subspace' of the  (Dirac and Schr\"odinger) `QED augmented Hilbert space' --  the latter self-adjointness results being, as we shall see, all one needs for the unitary equivalence of the latter product picture Hamiltonian with the usual Maxwell-Dirac Coulomb gauge Hamiltonian (respectively Maxwell-Schr\"odinger Coulomb gauge Hamiltonian), and the former self-adjointness result, for $\tilde{\bm\pi}$, being related to the full operator form of Gauss's law ((\ref{QuantGauss2} and (\ref{QuantGauss3})) in our product picture in a closely analogous way to the way our self-adjointness result for $\tilde{\bm\pi}$ here is related to the version of Gauss's law (\ref{QuantGauss1}) appropriate to an external classical charge distribution. 

Fourthly, one might notice that (unlike $\tilde{\bm\pi}$) $\hat{\bm A}$ does not map each subspace of the form ${\cal F(\cal H_{\mathrm{one}}^{\mathrm{trans}})}\otimes U_\rho\Omega^{\mathrm{long}}\in \cal F(\cal H_{\mathrm{one}})$ to itself and worry that this might be a problem.  However this is not a problem because the magnetic field $\bm B = \bm\nabla\bm\times\hat{\bm A}$ does map each such subspace to itself.

This concludes our discussion of our four likely-to-be-asked questions.
  
The coherent state vectors, $\Psi$, defined as in (\ref{cohem}),  are to be identified with the electrostatic coherent states, some of whose properties were derived in \cite{KayNewt}, as we discussed in Section \ref{Sect:Intro1}.   Let us recall that the 
$\tilde{\bm\pi}$ framework for the construction (\ref{cohem}) sidesteps the Orthogonality Theorem of the Section \ref{Sect:Intro1} by virtue of the fact that $\tilde{\bm\pi}$ (identified as explained above with minus the electric field) fails to be self-adjoint on the full augmented Fock space.  On the other hand, on the $\hat{\bm\pi}$ framework, as discussed in our answer to our first likely-to-be-asked question above, the Orthogonality Theorem of the Introduction would still be sidestepped because, even though, in this framework, $\bm E$ (now identified with $-\hat{\bm\pi}$) is self-adjoint, Gauss's law only holds in expectation value as in (\ref{QuantGausshat}). 

For what more is known about the properties of these electrostatic coherent states, we refer to \cite{KayNewt}.  Let us also mention here that the paper \cite{eeee} took some of the results of \cite{KayNewt} as its starting point and we reiterate that the results above (or rather their expected generalization to linearized gravity which it is planned to discuss further in \cite{QGS}) should serve to put both of those papers, \cite{KayNewt} and \cite{eeee}, on a firmer foundation.  In particular, we will explain, in Section \ref{Sect:QEDSchr} here, how some of the results of \cite{KayNewt} (which are used in \cite{eeee}) which were obtained there in a partly heuristic way can be put on a proper foundation -- see the paragraph after equation (\ref{NewtForm}) below.

\subsection{\label{Sect:capac} A further example and a possible experiment to decide between the $\tilde{\bm\pi}$ and $\hat{\bm\pi}$ frameworks}

\begin{figure}

   \centering
    \includegraphics[scale = 0.6, trim = 4cm 17cm 0.0cm 4cm, clip]{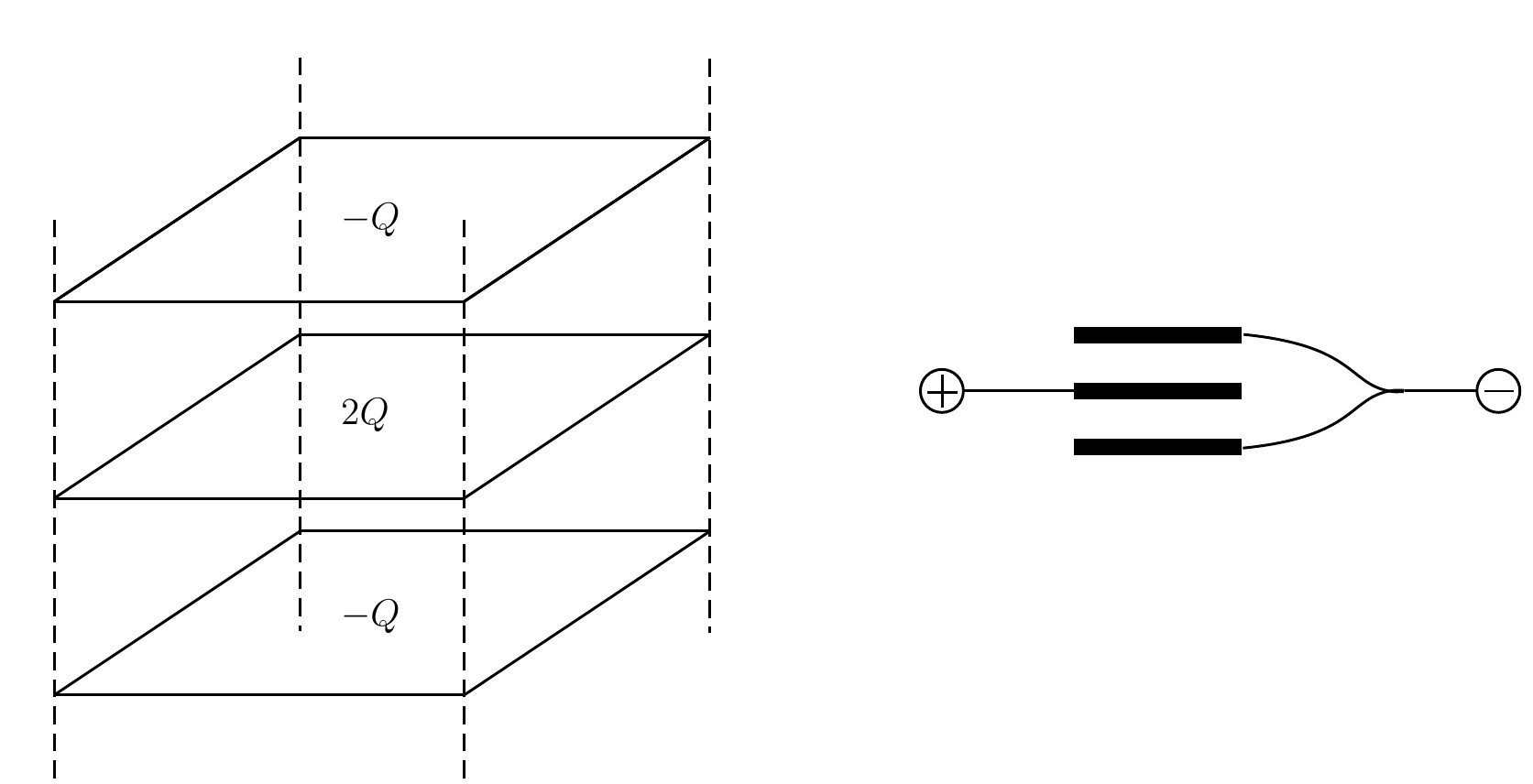}
   \caption{Three parallel electrically charged (conducting) plates.  The diagram on the right shows how they may be connected up so as to be what we call a \textit{double capacitor} -- which can be repeatedly charged up and then discharged through a resistor.}
\label{fig:DblCap}
 \end{figure}

As a further example of a calculation of (the absolute value of) an inner product between two electrostatic coherent states, let us consider $|\langle\Omega|\Psi\rangle|$ where $\Omega$ is the vacuum state of the free electromagnetic field, and $\Psi$ is the state of the electromagnetic field due to the presence of three parallel uniformly charged thin, say square, plates of side $L$ which, taken in order, have total charges $-Q$, $+2Q$, $-Q$ and where the spacing (i.e.\ the perpendicular distance) between neighbouring plates is $a$.  If we coordinatize our three-dimensional space with coordinates $(x,y,z)$, we may think of the middle plate with charge $2Q$ as occupying the region $0< x < L$, $0 < y < L$ of the $z=0$ plane and the upper and lower plates, each with charge $-Q$, as occupying the same regions of the planes $z=\pm a$.   See Figure (\ref{fig:DblCap}).   When the plates are made out of electrical conductors and are connected up to an electrical circuit as indicated at the right of the figure, we will call this a \textit{double capacitor}. (This has nothing to do with `double-layer capacitance', which is something completely unrelated).   And we shall use this terminology even though, for the purposes of calculating  $|\langle\Omega|\Psi\rangle|$, it doesn't matter whether they are conductors or insulators.

For simplicity we shall calculate $|\langle\Omega|\Psi\rangle|$ under the fiction of periodic boundary conditions -- identifying the plane $x=0$ with the plane $x=L$ and the plane $y=0$ with the plane $y=L$.   However we expect that the result will be close to that of actual square plates provided $L \gg a$ -- since edge-effect corrections would then be expected to be small and we shall neglect such edge-effect corrections throughout below.

\begin{figure}
   \centering
    \includegraphics[scale = 0.6, trim = 0cm 18cm -2cm 4cm, clip]{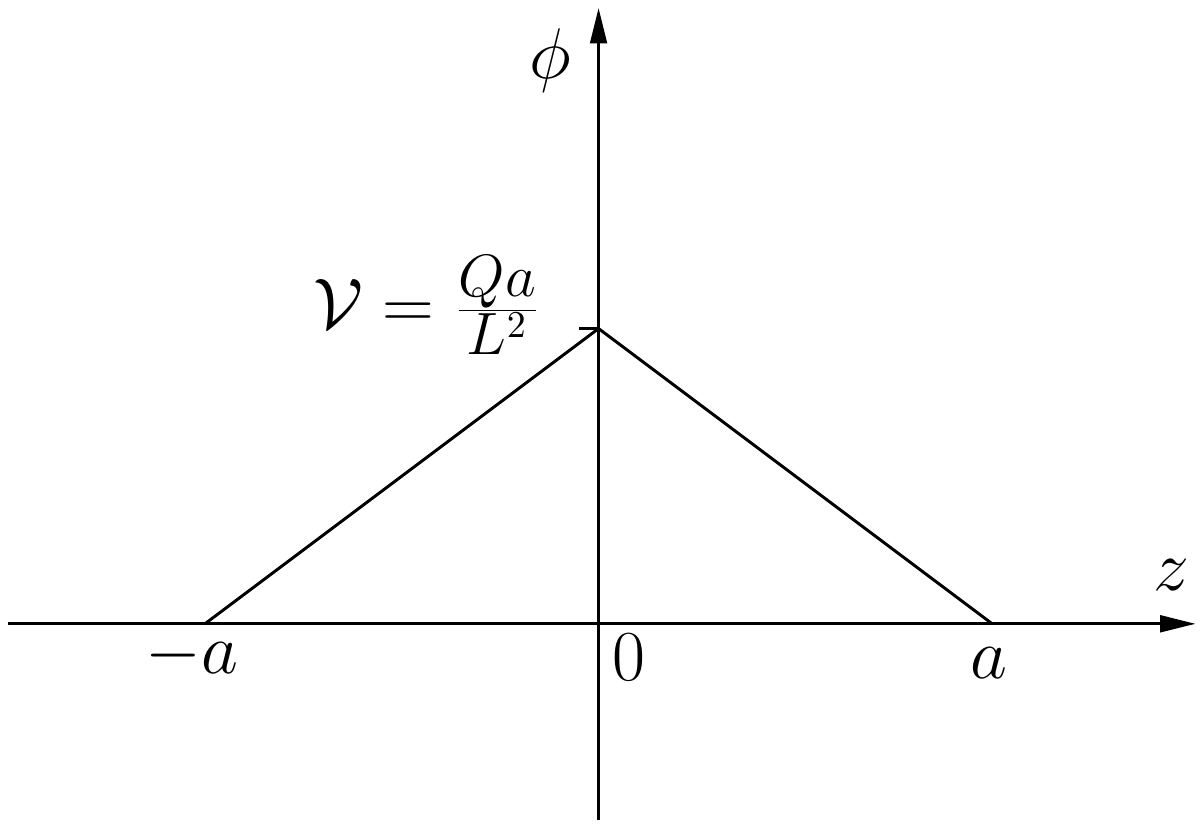}
   \caption{The graph of $\phi(z)$.}
\label{fig:graph}
 \end{figure}

Let us remark that it is only because of our wish to simplify our model by imposing periodic boundary conditions that we have chosen as our example our double capacitor rather than a simple capacitor with two plates (where $\Psi$ would be the quantum state of the electric field due to the presence of equal and opposite charges on two parallel plates occupying, say, the regions $0< x < L$, $0 < y < L$ of the $z=0$ and $z=a$ planes) which might seem more straightforward.  The problem with the simple capacitor is that, in the presence of periodic boundary conditions -- i.e.\ again identifying the plane $x=0$ with the plane $x=L$ and the plane $y=0$ with the plane $y=L$, one would find that $|\langle\Omega|\Psi\rangle|$ will vanish because of an infra red divergence.\footnote{\label {ftntCap}The vanishing of $|\langle\Omega|\Psi\rangle|$ in the simple capacitor model with two plates and periodic boundary conditions is due to an infra-red divergence in the counterpart $\left(Q^2\int_{-\infty}^\infty \frac{\sin^2(ka/2)}{k^3}\, dk/2\pi L^2\right)$ for the simple capacitor to the formula $2Q^2\int_{-\infty}^\infty \frac{\sin^4(ka/2)}{k^3}\, dk/\pi L^2$ (see after Equation (\ref{DblInt})) for the double capacitor that can be traced to the fact that, unlike in the double capacitor, where the classical electrical potential can be taken to vanish above the top plate and below the bottom plate (one could earth them both! -- see also Figure \ref{fig:graph}) in the simple capacitor with those periodic boundary conditions, the classical electrostatic potential must be a non-zero constant either above the top plate or below the bottom plate.  (One can only earth one of the plates!)   This infra red divergence is expected to be an artefact of our periodic boundary conditions since the potential will vanish at infinity for finite physical plates with edges.    Note, though, that we expect that it can be taken as a signal that, in the absence of periodic boundary conditions, edge-effect corrections to $|\langle\Omega|\Psi\rangle|$ will be more severe for a simple capacitor than for a double capacitor.  But we haven't investigated this.} 

Proceeding to analyze our double capacitor model (assuming periodic boundary conditions) the classical electric field, $\bm E^{\mathrm{class}}$, for this system will
be confined to the region between the upper and lower plates and point entirely in the $z$ direction, with the sole nonzero component, $E^{\mathrm{class}}(z)$ given by 
\begin{equation}
\label{DblCapEl}
E^{\mathrm{class}}(z) = 
\begin{cases}
0, & z <-a,\\
-\frac{Q}{L^2}, &-a<z < 0,\\
\frac{Q}{L^2}, &0<z<a,\\
0, & z>a,
\end{cases}
\end{equation}
while the electrical potential, $\phi$ (as usual, arbitrary up to an additive constant) may be taken to be (see Figure (\ref{fig:graph}))
\begin{equation}
\label{DblCapV}
\phi(z) = 
\begin{cases}
0, &z< -a\\
\frac{Q(a+z)}{L^2}, &-a<z < 0, \\
\frac{Q(a-z)}{L^2}, &0<z<a,\\
0, & z>a.
\end{cases}
\end{equation}

So, if we construct a physical capacitor in this way with conducting plates and assume that edge-effect corrections may be ignored, then we may charge them as required e.g.\ by (gently) connecting the top and bottom plates to the negative terminal of a source of emf with a (tunable) electrical potential, $\cal V$, related to the charge, $Q$, by  
\begin{equation}
\label{Volts}
{\cal V}=\frac{Qa}{L^2}
\end{equation}
and connecting the emf source's positive terminal to the middle plate (see Figure (\ref{fig:DblCap})) --  and then gently disconnecting from the emf source.  If we measure $Q$ in coulombs and divide by the SI value for $\epsilon_0$, i.e.\ $8.854\dots \times 10^{-12}$ F m$^{-1}$, (\ref{Volts}) will be the required voltage of the emf source.   

By equations  (\ref{balloverlap1}), (\ref{D1formula}) (with $\bm\chi_1 = \bm 0$ and $\bm\chi_2=\bm\chi$ say) and (\ref{phichi}), suitably modified to take into account our periodic boundary conditions, we will have:
\begin{equation}
\label{OmegaPsi}
|\langle\Omega|\Psi\rangle| = \exp(-D_1)
\end{equation}
where 
\[
D_1 = ||\bm\chi||^2/2
\]
which, in view of (\ref{DblCapEl}), and the obvious appropriate counterpart for our periodic boundary conditions of (\ref{phichi}), is equal to
\begin{equation}
\label{sumint}
\frac{1}{2}\sum_m\sum_n\int |\chi_{mn}(k)|^2 dk
\end{equation}
where the sums and integral are from $-\infty$ to $\infty$ and where
\[
\chi_{mn}(k) = \frac{iE^{\mathrm{class}}_{mn}(k)}{\sqrt 2\left (k^2 + \frac{4\pi^2m^2}{L^2}+\frac{4\pi^2n^2}{L^2}\right)^{1/4}}
\]
where $E^{\mathrm{class}}_{mn}(k)$ is the Fourier part series, part transform
\[
E^{\mathrm{class}}_{mn}(k) = \frac{1}{\sqrt{2\pi}L}\int_{-\infty}^\infty e^{-ikz}\int_0^L\int_0^L E^{\mathrm{class}}(z)e^{-2\pi i(nx + my)/L} \, dx dy dz
\]
of the classical electrostatic field, $E^{\mathrm{class}}(z)$, of (\ref{DblCapEl}). 
We immediately notice that since $E^{\mathrm{class}}$ is constant in the $y$ and $z$ directions, 
${E}^{\mathrm{class}}_{mn}(k)$ takes the form $E^{\mathrm{class}}(k)L\delta_{m0}\delta_{n0}$ where, by (\ref{DblCapEl}), 
\begin{equation}
\label{ordFT}
E^{\mathrm{class}}(k) = \frac{1}{\sqrt{2\pi}}\int_{-\infty}^\infty e^{-ikz} E^{\mathrm{class}}(z) \, dz = -\frac{2\sqrt{2}iQ}{\sqrt{\pi}L^2k}\sin^2{\frac{ka}{2}}
\end{equation}
(i.e.\ the ordinary one-dimensional Fourier transform of $E^{\mathrm{class}}(z)$) whereupon (\ref{sumint}) simplifes to
\begin{equation}
\label{DblInt}
D_1=L^2\int_{-\infty}^\infty \frac{|E^{\mathrm{class}}(k)|^2}{4|k|} dk
\end{equation}
which is easily seen to be $\frac{2Q^2}{\pi L^2}\int_{-\infty}^\infty \frac{\sin^4(ka/2)}{|k|^3}\, dk = 2a^2Q^2\ln 2/4\pi L^2$.

Thus, by (\ref{balloverlap1}), we have (restoring $c$, $\hbar$ and $\epsilon_0$ -- see Footnote \ref{ftntUnits}) that 
\begin{equation}
\label{fineq}
|\langle\Omega|\Psi\rangle|  = 2^{-\frac{Q^2}{2\epsilon_0 \hbar c\pi}\frac{a^2}{L^2}} \ = 2^{-\frac{2\alpha Q^2}{\rm{e}^2}\frac{a^2}{L^2}}
\end{equation}
(where $\alpha$ is the fine structure constant, 
$\approx 1/137$, and e is the charge on the electron) 
\[
= 2^{-\frac{\epsilon_0L^2 {\cal V}^2}{2\hbar c\pi}} 
\]
by (\ref{Volts}).

So for example, to have  $|\langle\Omega|\Psi\rangle|=1/\sqrt{2}$ for our physical double capacitor, and assuming we can ignore edge-effect corrections, we would need the electrical potential, $\cal V$, of our middle plate to be given by
\begin{equation}
\label{voltage}
{\cal V}_{\mathrm{half}} = \sqrt{\frac{\pi\hbar c}{\epsilon_0}} \frac{1}{L} \ \approx \frac{1.06 \times 10^{-5}}{L} \  \hbox{volts if $L$ is measured in centimetres},
\end{equation}
and let us mention in passing,  that, by (\ref{fineq}), the number of surplus electrons on the outer plates (equivalently half the number of holes in the middle plate) will then be $L/\sqrt\alpha a\approx 11.7 L/a$.   So our classical external charge distribution approximation can presumably only be good if $L \gg a$.

It is noteworthy that the formula, (\ref{voltage}), for $\cal V_{\mathrm{half}}$ depends inversely on $L$, the side-length of our plates, and is independent of the spacing, $a$, between them.  

Let us also note that, classically, the energy stored in our double capacitor, is of course equal to $V_{\mathrm{Coulomb}}$ and is given by the formula (\ref{VCoulombphi}) with $\phi$ as in (\ref{DblCapV}) -- i.e.\  by the (almost) familiar formula $Q{\cal V}$ so, by (\ref{Volts}) and restoring $\epsilon_0$, we have  
\begin{equation}
\label{capenerg}
V_{\mathrm{Coulomb}}=\epsilon_0{\cal V}^2L^2/a.
\end{equation}
(It is not quite familiar since it lacks a factor of $1/2$ due to the fact that it is a \textit{double} capacitor!) In other words it is
$\frac{1}{2} C{\cal V}^2$ if, reasonably, we define the capacitance, $C$, of our double capacitor to be $2\epsilon_0 L^2/a$.  (Note though that Equation (\ref{Volts}) then becomes ${\cal V}=2Q/C$.) 
So for the voltage $\cal V_{\mathrm{half}}$ of Equation (\ref{voltage}) for which $|\langle\Omega|\Psi\rangle|=1/\sqrt{2}$, we have that the energy stored in the capacitor, according to classical physics -- let us call it $V_{\mathrm{Coulomb}}^{\mathrm{half}}$ -- is given by
\begin{equation}
\label{VCoulHalf}
V_{\mathrm{Coulomb}}^{\mathrm{half}} = \frac{\pi\hbar c}{a}
\end{equation}
which depends only on the spacing, $a$, of the capacitor plates and not on their size!

So, for example, if  $a$ is chosen to be 5 nm, this will be approximately $2 \times 10^{-17}$\ J. 
This is a very small energy, and if the plate separation $a$ were any larger, it would be even smaller. 

Our calculation of the quantity, $|\langle\Omega|\Psi\rangle|$,  combined with the discussion of our third likely-to-be-asked question in Section \ref{Sect:electrocoh} suggests the following experiment which, we will argue, should, in principle, be able to decide between the $\tilde{\bm\pi}$ and $\hat{\bm\pi}$ frameworks -- and for which, we shall also find, the predictions of the $\tilde{\bm\pi}$ framework coincide with those of the standard Coulomb gauge understanding.  As we shall discuss below, to carry it out would, as far as we can see, be challenging in view of the smallness just noted of $V_{\mathrm{Coulomb}}^{\mathrm{half}}$ for the voltage ${\cal V}_{\mathrm{half}}$ of (\ref{voltage}).  But it is anyway of interest as a \textit{gedanken experiment}. (See Figure \ref{fig:circuit}.)  

\smallskip

\noindent
\textit{Repeatedly charge up a capacitor, and then discharge it through a resistor, each time using the same voltage, ${\cal V}$, for the emf source used to charge the capacitor, tuned so that $|\langle\Omega|\Psi\rangle|=1/\sqrt{2}$. (So in the case of our double capacitor and ignoring edge-effect corrections, $\cal V$ would be given by the ${\cal V}_{\mathrm{half}}$ of (\ref{voltage}).)   Then observe in each run of the experiment whether or not the resistor heats up.}

\smallskip
We note that we could do this experiment with any capacitor as long as it is made out of conductors with vacuum (or maybe air would do) in the spaces between them but we may as well continue to assume it to be done with our double capacitor (with conducting plates) as described above. 

\begin{figure}

   \centering
\label{fig:circuit}
    \includegraphics[scale = 0.6, trim = 2cm 18cm 0cm 4cm, clip]{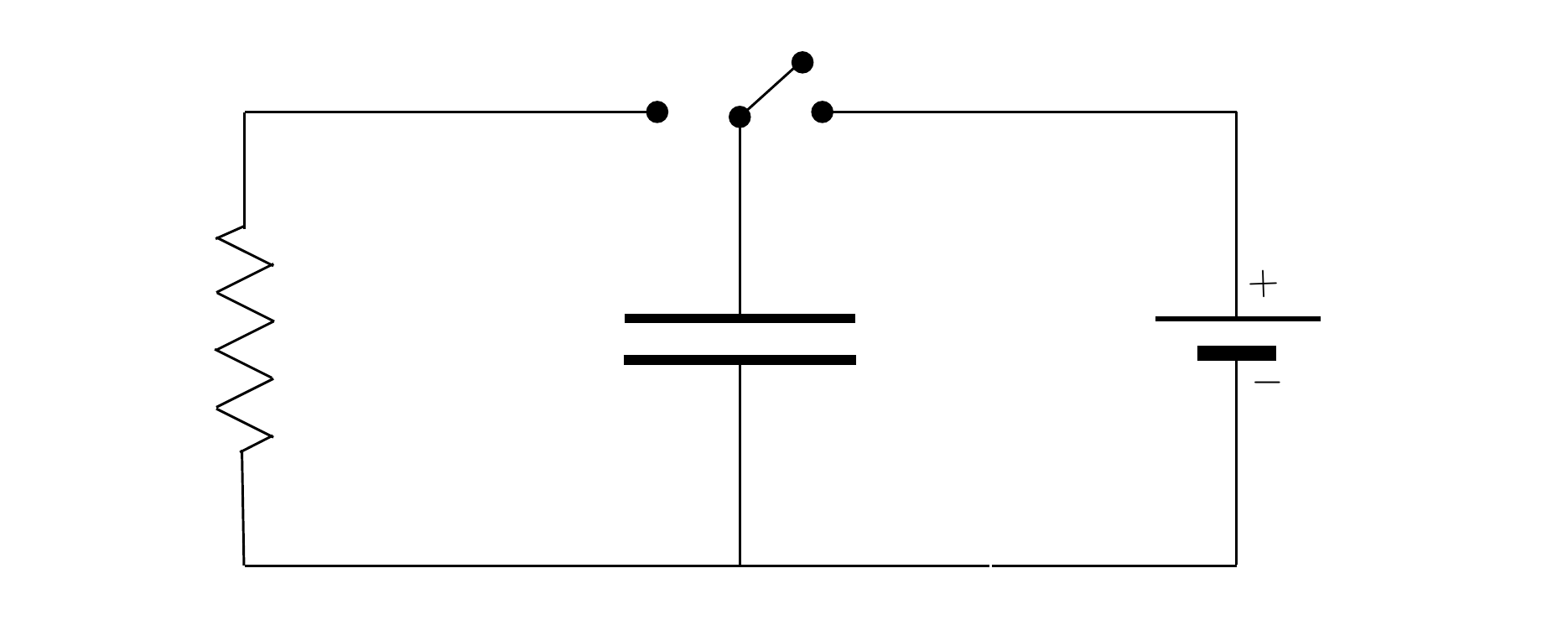}
   \caption{A circuit diagram of our experiment.  The double switch allows the capacitor to be charged up from an emf source and then to discharge through a resistor which heats up, thereby serving as a measuring device for the electrostatic energy that had been stored.}
\label{fig:circuit}
 \end{figure}

We will argue that, both on standard Coulomb gauge QED thinking, and also on our $\tilde{\bm\pi}$ framework, one would predict that, in each run of the experiment, the resistor would heat up by the same amount, equal (leaving aside the caveat we mention below) to the classical energy, $V_{\mathrm{Coulomb}}^{\mathrm{half}}$.  But that on the $\hat{\bm\pi}$ framework, one would predict that in around half of the experimental runs, the resistor would not heat up at all, while in the remaining rougly half of the runs, it would heat up.   In fact it would heat up in such a way that the mean amount by which it heats up over all the runs will approximately equal $V_{\mathrm{Coulomb}}^{\mathrm{half}}$.  So to decide between the $\hat{\bm\pi}$ framework and the 
$\tilde{\bm\pi}$ framework (or standard Coulomb gauge QED) we would not need to make any quantitative measurements but just to observe whether the resistor heats up in only (approximately) half of the runs, or whether it heats up in all of them.

Before justifying this claim, let us remark that, in a fully classical analysis of this experiment, by conservation of energy, the amount by which the resistor heats up will equal $V_{\mathrm{Coulomb}}^{\mathrm{half}}$ since this was the energy stored in the capacitor and energy is conserved.   (This is not quite true since some energy will be radiated away in electromagnetic waves due e.g.\  to the acceleration of the electrons in the circuit that connects the capacitor to the resistance or due to thermal radiation by the heated up resistor etc.   But we expect this to be negligible provided the resistance of the resistor is large enough for the discharge to happen slowly enough and provided the resistor doesn't get too hot etc.\ and, as we shall see below, for realistic parameters, it will happen slowly and the resistor won't get too hot.  This, or rather its obvious quantum counterpart, is the caveat we ignored earlier.) 

In the arguments below in justification of our claim, we will take a few assumptions for granted:   First, that it is valid as far as the physics of the (quantum) electric field in the capacitor is concerned, to a suitable degree of accuracy, to treat the charges on the plates as classical and external.   Secondly, we shall assume that it is valid to treat the second step of each run, where we discharge the capacitor through the resistance and measure by how much it heats up, or, rather, simply measure whether or not it heats up, as an ideal quantum mechanical measurement of the value of the stored electrical energy in the capacitor, or, rather, just of whether that value differs from zero.   We shall not attempt to analyze the physics of the capacitor discharge and the heating up of the resistor, and indeed, we would not expect to be able to analyze this without departing from the approximation of external classical charges.   (One reason for saying this is related to the point we made in Section \ref{Sect:prelim} that the addition of a c-number, i.e.\ $V_{\mathrm{Coulomb}}$, to the Hamiltonian for the free electromagnetic field cannot affect any commutation relations.) In treating the discharge of the capacitor and heating up of the resistor as an ideal quantum mechanical energy measurement, we are implicitly assuming that it takes place at a particular time -- say the time at which one commences the discharge of the capacitor -- i.e.\ the time at which one throws the switch from right to left in the circuit diagram of Figure \ref{fig:circuit}. One might worry that this is an unjustified idealization in the case of the $\hat{\bm\pi}$ framework if the actual discharge of the capacitor, say, takes much longer than the time scales relevant to the time-evolution of the  time-evolving state, $\Psi(t)=\exp(-i\hat H^{\mathrm{EM}}t)\Psi$, since we recall from Section \ref{Sect:electrocoh} that this will be nonstationary.   However, $\Omega$ \emph{will be} an eigenstate of $\hat H$ and hence our condition that the voltage of the capacitor is chosen so that  $\langle\Omega|\Psi\rangle = 1/\sqrt 2$ won't change if we replace $\Psi$ by $\Psi(t)$ for any $t$.   So in an ideal quantum mechanical measurement, at whatever time we observe the energy of the capacitor, it will be zero with probability $1/2$. In view of this, while a fuller analysis of the physics of the discharge process would be worthwhile and should be done, we expect that it will not significantly alter our conclusions.

Finally, we shall omit any discussion of many experimental details such as how an emf source with the voltage $\cal V_{\mathrm{half}}$ is to be provided.   But we shall discuss what might be used as a resistor and the prospects for detecting whether or not it heats up.

Let us first analyse the experiment according to standard Coulomb gauge thinking.   As we discussed in Section \ref{Sect:Intro}, according to a Coulomb gauge understanding  one would either say that, in each run of the experiment, there is no such thing as a longitudinal electric field, and the energy stored in the capacitor before it is discharged is just the (classical) potential energy for the action-at-a-distance force between the (here, assumed classical) charged plates -- namely the classical energy $V_{\mathrm{Coulomb}}^{\mathrm{half}}$, or one would say that there is a longitudinal electric field, but it is a classical field, being given by the right hand side of (\ref{VCoulombphi}) for the classical electrical potential $\phi$ and its energy is again given by $V_{\mathrm{Coulomb}}^{\mathrm{half}}$!  It is true that one would also understand there to be a transverse part of the electromagnetic field which is quantum in nature, but this would be expected to be in its ground state (suitably modified by the presence of the conducting plates) before the discharge of the capacitor and to return to that state after the discharge and thus except for the (we shall assume in line what we wrote above about our caveat) small amount of energy lost to radiation in the form of photons, the quantized transverse field modes will not affect the amount by which the resistor is heated up.   Thus one predicts that the resistor will absorb the classical energy, $V_{\mathrm{Coulomb}}^{\mathrm{half}}$, in each experimental run. 

On the quantum electrostatics ideas developed in Section \ref{Sect:extsource} of the present paper, if we adopt the $\tilde{\bm\pi}$ framework, then although the conceptual framework seems quite different from that of the standard Coulomb gauge discussion above, we will surely arrive at exactly the same prediction because, in that framework, by the last equation of (\ref{Psiemprops}), our electrostatic coherent state, $\Psi$, will be an eigenvector of the electromagnetic field Hamiltonian (taken to be $\tilde H^{\mathrm{EM}}_0$) with eigenvalue equal to the classical energy $V_{\mathrm{Coulomb}}^{\mathrm{half}}$.   Thus, again, we arrive at the conclusion that the energy stored in the capacitor, after charging and before discharge, has this same classical value in every run of the experiment.   Thus we would arrive at the same conclusions as in the standard Coulomb gauge understanding above.  It is also interesting to note that the transverse part of the quantum electric field (taken to be $-\tilde{\bm\pi}^{\mathrm{trans}}$) will act on $\Psi$ in the same way as it acts on the quantum vacuum state (which is, in turn the same as in the standard Coulomb gauge understanding) while the longitudinal part (taken to be $-\tilde{\bm\pi}^{\mathrm{long}}$) will, since Gauss's law holds in the strong form of (\ref{QuantGauss1}), act by multiplying $\Psi$ by the classical electric field $-\bm\nabla\phi$ where $\phi$ is as in (\ref{DblCapV}) -- a conclusion which is surely (should we be able to measure the electric field directly) operationally indistinguishable from the conclusion about the nature of the electric field in the second variant of the above Coulomb gauge analysis.

However, in the $\hat{\bm\pi}$ framework, the operator that represents the electric field is $-\hat{\bm\pi}$ and, as we saw in Section \ref{Sect:electrocoh}, with this representation of the quantum electric field, Gauss's law only holds in the expectation value form (\ref{weak}).  Moreover, and more to the point for our experiment, $\Psi$ is no longer an eigenstate of the electric field Hamiltonian which is now taken to be $\hat H^{\mathrm{EM}}_0$.   All we can say, thanks to the last equation in (\ref{Psiemhatprops}), is that its expectation value in the state $\Psi$ is  $V_{\mathrm{Coulomb}}^{\mathrm{half}}$. On the other hand (see the discussion of the third likely-to-be asked question in Section (\ref{Sect:electrocoh})) since $\hat{\bm\pi}$ is a self-adjoint operator on the full augmented Fock space, it is warranted to interpret the quantity $|\langle\Omega|\Psi\rangle|^2$ (which played no r\^ole in either the Coulomb gauge or the $\tilde{\bm\pi}$ analyses above!) as \textit{the probability that the state of the electric field after the capacitor has been charged up will be the vacuum state, and thereby have zero energy}.   If the voltage has been chosen to be such that $|\langle\Omega|\Psi\rangle|=1/\sqrt{2}$, this probability will be $1/2$.   Thus we predict that in around half of our experimental runs the resistor will not heat up at all.  And it must be that in the remaining runs, the resistor will heat up sufficiently that the mean value over all runs of the experiment of the energy released on discharge of the capacitor will approximately equal the classical energy as we anticipated.

Finally, let us turn to discuss the feasibility of actually carrying out the experiment.  Let us stay with our double capacitor model with plates with side $L$ and spacing between them of $a$ and continue to ignore edge-effect corrections.

We will investigate one possible way of detecting whether or not our resistor heats up.  Namely to choose, as our resistor, a, length, $\ell$ (to be determined) of very thin wire, of radius, $r$, (also to be determined) the idea being that such a resistor might (if its mass is sufficiently low) have such a small heat capacity that its temperature increase, on heating up, may be big enough for its thermal expansion (or the absence thereof) to be noticeable in some way.   (We remark in passing that in practice we might want to replace the resistor with a new one in each run.)  We shall proceed naively as if the physical properties of the material of the wire -- i.e.\ its mass, heat capacity, electrical resistance, $R$, and length increase, $\Delta\ell$, due to thermal expansion -- can be computed as one would for a macroscopic lump of that material from its bulk density, $d$, specific heat, $s$,  electrical resistivity, $\cal R$  and thermal expansion coefficient, $t$.   For no particularly strong reason, we shall illustrate these values for sodium at room temperature and pressure.  See Table \ref{tab1} where we also give the corresponding values for silver.

\begin{table}[ht]
\caption{\textbf{Some physical properties of sodium and of silver} \textit{(at room temperature and pressure)}}

\smallskip

\centering 
\resizebox{\textwidth}{!}{\begin{tabular}{c c c} 
\hline 
Property & Sodium & Silver\\ [0.5ex] 
 \hline\hline
density  $d$ & 970 kg m$^{-3}$ ($= 0.97$ g cm$^{-3}$) & \quad\quad 10500 kg m$^{-3}$ ($= 10.5$ g cm$^{-3}$)\\
specific heat $s$ & 1.23 J g$^{-1}$ K$^{-1}$ & 0.23 J g$^{-1}$ K$^{-1}$\\
resistivity $\mathcal R$ & 43 n$\Omega$m & 16 n$\Omega$m \\
thermal expansion coefficient $t$ & 70 $\times$ 10$^{-6}$ K$^{-1}$ & 18 $\times$ 10$^{-6}$ K$^{-1}$\\
\hline
\end{tabular}}
\label{tab1} 
\end{table}

We easily have that the time scale, $\tau$, for the discharge of our double capacitor through
a resistor (so the time for the resistive wire to heat up by $1-1/e$ of the amount by which it will heat up asymptotically) is $\epsilon_0 L^2R/a$, and, with our naive assumptions, $R$ will be given in terms of the length, $\ell$, and radius, $r$, of the wire by
$R = {\mathcal R}\ell/\pi r^2$ where ${\mathcal R}$ is the bulk resistivity of the material out of which the wire is made, say at room temperature and pressure.  So we have
\begin{equation}
\label{tau}
\tau = \frac{\epsilon_0{\mathcal R}\ell L^2}{\pi ar^2}.
\end{equation}

Also the temperature increase of the wire, $\Delta T$, will, by (\ref{VCoulHalf}) on a similar naive approach, be given by
\begin{equation}
\label{DeltaT}
\Delta T = \frac{\hbar c}{ar^2\ell sd}.
\end{equation}

And consequently, the increase in length, $\Delta\ell$, of the resistive wire, on heating up will be given, naively, by $t\ell\Delta T/3$, where $t$ is the bulk coefficient of expansion of the material of the wire, again at room temperature and pressure.  So we have
\begin{equation}
\label{DeltaEll}
\Delta\ell = \frac{\hbar c t}{3ar^2 sd}.
\end{equation}

Dividing (\ref{DeltaEll}) by (\ref{DeltaT}) we obtain the relation
\[
\Delta\ell = K \frac{\tau}{\ell L^2}
\]
where
\[
K = \frac{\pi\hbar ct}{3\epsilon_0 {\cal R} sd}
\]
For example, if the material of the wire is sodium (see Table \ref{tab1}) this is approximately 
$5\times 10^{-15}$ m$^4$ s$^{-1}$.   So unless the capacitor plates and/or the length of the wire are extremely tiny, and/or the time-scale for the capacitor discharge enormously long, the change in length of our wire will (independently of the capacitor plate spacing, $a$ and the radius, $r$ of the wire) inevitably be tiny.  For example, to have $\tau$ as short as 1 second and $\Delta\ell$ to be no smaller than, say 5 microns, we would therefore need $\ell L^2$ to be no larger than a cubic millimeter, which we could achieve e.g.\ if both the length of our resistive wire, $\ell$, and the side of the capacitor plates, $L$, were 1 mm.   For these values, one then has, from (\ref{DeltaT}), that the temperature increase, $\Delta T$, of the resistive wire would be around $22$ K.  But, from (\ref{tau}) or (\ref{DeltaEll}), we have
\[
ar^2 = \frac{\epsilon_0{\cal R}\ell L^2}{\pi \tau}
\]
which would be $1.2 \times 10^{-28}$ m$^3$.   This could be achieved e.g.\ by taking $a$ (the spacing in the capacitor) as well as $r$ (the radius of the resistive wire) to each be around $5\times 10^{-10}$ m which is only around 10 times the Bohr radius!    I would guess that achieving that would be more difficult than measuring a smaller $\Delta\ell$.  After all, in LIGO,  one can measure displacements of a mirror with laser interferometry which are as small as $10^{-19}$ m -- albeit with very much heavier mirrors than could, say, be suspended from our resistive wire without it breaking, even taking account of the fact that, for smaller $\Delta\ell$, the wire's mass can be somewhat greater -- see below.  So, for example, if a $\Delta\ell$ of $5\times 10^{-12}$ m were to be detectable, then for $\tau =1$ s, 
$\ell L^2$ would have to be $10^{-3}$ m$^3$, which could be achieved, say, with $\ell = 50$ cm and $L \approx 4.5$ cm.   Then $ar^2$ would have to be around $1.2\times 10^{-22}$ m$^3$ whereupon we could, say, take both $a$ and $r$ to be around 50 nm, which is, more comfortably, something like 1000 Bohr radii.   But that is still a very thin wire and a very small plate spacing for our capacitor.  (With plates of side $\approx 4.5$ cm, which is not far off a million times the spacing, at least we presumably won't have to worry about edge-effect corrections.)   The temperature increase would now be the tiny $4.4\times 10^{-7}$ K and the mass of the wire still only around $7.85\times 10^{-5}$ g.  I don't know whether or not it would be preferable to obtain a larger $\Delta\ell$ and/or allow $a$, $L$, $\ell$ and $r$ to be a little larger by contemplating a longer capacitor discharge time (/resistor heating up time), $\tau$, than 1 second or indeed if we should not prefer a shorter time.  

Also one would have to study whether the signal will stand out above other causes of length change due to e.g.\ shot noise in the circuits as they are connected and disconnected, Johnson-Nyquist noise around the frequency $1/\tau$ (so 1 Hz in our examples) environmental vibrations and temperature fluctuations etc.   So perhaps the connections and disconnections of the capacitor would have to be done very gently and perhaps the apparatus would need cooling and/or shielding vibrationally and/or thermally.

What works in our favour though is that, the `signal' we are looking for here is not the amount by which the resistor heats up/changes its length, but just the answer to the yes-no question:   Does the resistor change its length at all by an amount of the order of the calculated $\Delta\ell$ or does it not change its length -- other than due to noise and the other effects just mentioned, which would of course need to be controlled to be, say, an order of magnitude lower than the predicted $\Delta\ell$.

We tentatively conclude that to carry out the experiment successfully, would be extremely challenging, although maybe not impossible.   Of course there may well be other ways of designing a resistor and measuring its increase in energy or perhaps completely different sorts of experiment which would decide between the $\tilde{\bm\pi}$ framework and the $\hat{\bm\pi}$ framework (see also Section \ref{Sect:temporal}.)

In conclusion, let us recall that the ability of our experiment to decide between the $\tilde{\bm\pi}$ and $\hat{\bm\pi}$ frameworks derives from the fact, (\ref{Psiemprops}), that, in the $\tilde{\bm\pi}$ framework, the coherent electrostatic state $\Psi$ (respectively $\Omega$) is an eigenfunction of $\tilde H^{\mathrm{EM}}_0$ with eigenvalue $V_{\mathrm{Coulomb}}^{\mathrm{half}}$ (respectively zero) whereas, in the $\hat{\bm\pi}$ framework, one has only the weaker statement that the expectation value of $\hat H^{\mathrm{EM}}_0$ in the state $\Psi$ is equal to $V_{\mathrm{Coulomb}}$.   (And in the state $\Omega$ is zero.)   And let us recall also that, as is clear from Section \ref{Sect:electrocoh}, these facts are intimately related to the fact that, in the $\tilde{\bm\pi}$ framework, Gauss's law takes the strong form (\ref{QuantGauss1}), whereas, in the $\hat{\bm\pi}$ framework, one has only the weaker statement that Gauss's law  holds in the expectation value sense of (\ref{QuantGausshat}).  So it seems fair to say, speaking loosely, that our experiment can be regarded as a test (at mesoscopic scales -- see Section \ref{Sect:temporal}) of whether Gauss's law holds in strong or weak form.  

\section{\label{Sect:QED} Transformation of Coulomb gauge Maxwell-Dirac QED to a product picture} 

\subsection{\label{Sect:QEDprelim} Preliminaries}

When one adopts the Coulomb gauge, $\bm{\nabla\cdot A}=0$, the usual full Hamiltonian for the quantum electromagnetic field in interaction with a Dirac field, $\psi$, may be written
\begin{equation}
\label{HamDirac}
H_{\mathrm{QED}}^{\mathrm{Dirac}}=
\int {1\over 2} {\bm\pi^\perp}^2 +{1\over 2}({\bm\nabla} {\bm\times} {\bm A})^2 \, d^3x \ 
+ \  H_{\mathrm{Dirac}}  \
\end{equation}
where
\begin{equation}
\label{HDirac}
H_{\mathrm{Dirac}}=
\int \psi^*\gamma^0{\bm \gamma}{\bm \cdot}(-i{\bm\nabla} - {\rm e}{\bm A})\psi+m\psi^*\gamma^0\psi \, d^3x + \ V_{\mathrm{Coulomb}}^{\mathrm{Dirac}}
\end{equation}
where $V_{\mathrm{Coulomb}}^{\mathrm{Dirac}}$ is given by (\ref{VCoulomb}) in the case of the (now quantum) operator
\begin{equation}
\label{rhopsipsi}
\rho(\bm x)={\rm e}\psi(\bm x)^*\psi(\bm x).   
\end{equation}
We recall here that $-{\rm e}$ is the charge on the electron,  also
$\gamma^a$, $a=0, 1, 2, 3$,\break 
are a choice of Dirac matrices, satisfying $\{\gamma^a, \gamma^b\}=2\eta^{ab}$ where\break 
$\eta^{ab}=\mathrm{diag}(1,-1,-1,-1)$, for example the standard Dirac choice, and 
$\psi^*$ denotes the adjoint of $\psi$.
$H_{\mathrm{QED}}^{\mathrm{Dirac}}$ takes the form of (\ref{Ham}) if we identify $H^0_{\mathrm{ch\, mat}}$ with
$\int \psi^*\gamma^0{\bm \gamma}{\bm\cdot}(-i{\bm\nabla}\psi) + m\psi^*\gamma^0\psi \, d^3x$ and
$\bm J$ with ${\rm e}\psi^*\gamma^0{\bm\gamma}\psi$.

The specification of the theory is completed by supplementing the commutation relations (\ref{tradCR}) with the anticommutation relations for the $\psi$ field:
\begin{equation}
\label{tradCAR}
\lbrace\psi(\bm x), \psi(\bm y)\rbrace=0=\lbrace\psi^*(\bm x), \psi^*(\bm y)\rbrace \  \mathrm{and} \  \lbrace\psi(\bm x), \psi^*(\bm y)\rbrace=\delta^{(3)}({\bm x}-{\bm y}),
\end{equation}
together, of course, with the assumption that the commutators of $\psi$ and $\psi^*$ with $\bm A$ and $\bm\pi$ all vanish.   (Above, we have omitted spinor component indices and an identity operator in spinor space on the right hand side.)

$H_{\mathrm{QED}}^{\mathrm{Dirac}}$ will act on the tensor product of the usual transverse Fock space, $\cal F(\cal H_{\mathrm{one}}^{\mathrm{trans}})$, for the electromagnetic field with the usual Hilbert space, ${\cal H}_{\mathrm{Dirac}}$, on which the Dirac field, $\psi$, acts. 

\subsection{\label{Sect:QEDstrategy} Strategy} 

The developments in Section \ref{Sect:extsource} in the $\tilde{\bm\pi}$ framework suggest a strategy for finding a different Hamiltonian on a different Hilbert space, unitarily equivalent to $H_{\mathrm{QED}}^{\mathrm{Dirac}}$ on $\cal F(\cal H_{\mathrm{one}}^{\mathrm{trans}})\otimes{\cal H}_{\mathrm{Dirac}}$, which will be appropriate to a product picture.  

First we observe (cf.\ Section \ref{Sect:twoequiv}) that we may equally regard $H_{\mathrm{QED}}^{\mathrm{Dirac}}$ as a Hamiltonian on the \emph{QED augmented Hilbert space} $\cal F(\cal H_{\mathrm{one}})\otimes {\cal H}_{\mathrm{Dirac}}$ where $\cal F(\cal H_{\mathrm{one}})$ is the augmented Fock space of Section \ref{Sect:twoequiv} and that, when so regarded, this latter Hamiltonian maps the subspace ${\cal F(\cal H_{\mathrm{one}}^{\mathrm{trans}})}\otimes \Omega^{\mathrm{long}}\otimes {\cal H}_{\mathrm{Dirac}}$ to itself and, restricted to this subspace -- which we will call the \textit{Coulomb gauge physical subspace} -- is equivalent, in an obvious way, to $H_{\mathrm{QED}}^{\mathrm{Dirac}}$ on $\cal F(\cal H_{\mathrm{one}}^{\mathrm{trans}})\otimes {\cal H}_{\mathrm{Dirac}}$. 

Secondly, with an eye on Equation (\ref{emHamshift}), we observe that the \textit{different} Hamiltonian $\check H_{\mathrm{QED}}^{\mathrm{Dirac}}$ on $\cal F(\cal H_{\mathrm{one}})\otimes {\cal H}_{\mathrm{Dirac}}$ defined by
\begin{equation}
\label{checkHQED}
\check H_{\mathrm{QED}}^{\mathrm{Dirac}}\!\!=\!\!\!\int\!\!  {1\over 2} \tilde{\bm\pi}^2 +{1\over 2}({\bm\nabla} {\bm\times} {\bm A})^2 +\psi^*\gamma^0{\bm \gamma}{\bm\cdot}(-i{\bm\nabla} - {\rm e}{\bm A})\psi+m\psi^*\gamma^0\psi  + \tilde{\bm \pi}\bm{\cdot\nabla}\phi \, 
d^3 x + V_{\mathrm{Coulomb}}^{\mathrm{Dirac}}
\end{equation}
also maps the Coulomb gauge physical subspace, ${\cal F(\cal H_{\mathrm{one}}^{\mathrm{trans}})}\otimes \Omega^{\mathrm{long}}\otimes {\cal H}_{\mathrm{Dirac}}$, to itself, and, its restriction to this subspace, is \textit{equal} to the restriction of $H_{\mathrm{QED}}^{\mathrm{Dirac}}$ to the same subspace.   (Let us also recall here that the Dirac fields $\psi$ and $\psi^*$ commute with $\tilde{\bm\pi}$.)
In (\ref{checkHQED}),  $\phi$ is the \textit{Dirac electrical potential operator} given by
\begin{equation}
\label{phi}
\phi(\bm x)=\int \frac{{\mathrm{e}}\psi^*(\bm y)\psi(\bm y)}{4\pi|\bm x-\bm y|}\, d^3y.
\end{equation}

To show the equality of $\check H_{\mathrm{QED}}^{\mathrm{Dirac}}$ and $H_{\mathrm{QED}}^{\mathrm{Dirac}}$ on the Coulomb gauge physical subspace, it suffices to notice that 
$\check H_{\mathrm{QED}}^{\mathrm{Dirac}}$ differs  from $H_{\mathrm{QED}}^{\mathrm{Dirac}}$ by the addition of the terms $\int (\frac{1}{2}{\tilde{\bm\pi}}^{\mathrm{long}})^2 + \tilde{\bm \pi}\bm{\cdot\nabla}\phi \, d^3x$ ($=\frac{1}{2}\int ({\tilde{\bm\pi}}^{\mathrm{long}})^2 + \tilde{\bm \pi}^{\mathrm{long}}\bm{\cdot\nabla}\phi \, d^3x$) and that $\tilde{\bm\pi}^{\mathrm{long}}$ annihilates $\Omega^{\mathrm{long}}$!

We remark here that, if we were to replace $\tilde{\bm\pi}$ by $\hat{\bm\pi}$ everywhere, then this crucial step would fail.   See Section \ref{Sect:QEDdiscussion}.

Thirdly, and again motivated by Equation (\ref{emHamshift}), we will introduce the unitary operator defined in (\ref{U}), suggested by the $U$ of Equation (\ref{Uext}) in Section \ref{Sect:electrocoh} and which we shall also call `$U$', but now acting on the QED augmented Hilbert space, $\cal F(\cal H_{\mathrm{one}})\otimes {\cal H}_{\mathrm{Dirac}}$, and obtain our candidate product picture Hamiltonian, $H^{\mathrm{PP, Dirac}}_{\mathrm{QED}}$ on the latter Hilbert space as $U\check H_{\mathrm{QED}}^{\mathrm{Dirac}}U^{-1}$ and our candidate \textit{product picture physical subspace} to be  $U{\cal F(\cal H_{\mathrm{one}}^{\mathrm{trans}})}\otimes \Omega^{\mathrm{long}}\otimes{\cal H}_{\mathrm{Dirac}}$, which we shall feel free to sometimes write 
${\cal F(\cal H_{\mathrm{one}}^{\mathrm{trans}})}\otimes U\Omega^{\mathrm{long}}\otimes{\cal H}_{\mathrm{Dirac}}$ to remind ourselves that $U$ acts nontrivially only on the second and third parts of this triple tensor product.   Clearly, $H^{\mathrm{PP, Dirac}}_{\mathrm{QED}}$ must map that subspace to itself (because $\check H_{\mathrm{QED}}^{\mathrm{Dirac}}$ maps the Coulomb gauge physical subspace, ${\cal F(\cal H_{\mathrm{one}}^{\mathrm{trans}})}\otimes \Omega^{\mathrm{long}}\otimes{\cal H}_{\mathrm{Dirac}}$, to itself) and, when restricted to that product picture physical subspace, it must then be equivalent to $\check H_{\mathrm{QED}}^{\mathrm{Dirac}}$, and hence to $H_{\mathrm{QED}}^{\mathrm{Dirac}}$, on the Coulomb gauge physical subspace.

An important point that needs to be dealt with is that, on the full QED augmented Hilbert space, $\cal F(\cal H_{\mathrm{one}})\otimes {\cal H}_{\mathrm{Dirac}}$, neither $\check H_{\mathrm{QED}}^{\mathrm{Dirac}}$ nor (in consequence, since it is unitarily related) $H^{\mathrm{PP, Dirac}}_{\mathrm{QED}}$ (and nor $\tilde{\bm\pi}$) will be self-adjoint.   However the restriction of $\check H_{\mathrm{QED}}^{\mathrm{Dirac}}$ to the Coulomb gauge physical subspace ${\cal F(\cal H_{\mathrm{one}}^{\mathrm{trans}})}\otimes \Omega^{\mathrm{long}}\otimes{\cal H}_{\mathrm{Dirac}}$ is of course self-adjoint, since, as we pointed out in the opening paragraph of this subsection, restricted to that subspace, it may be identified with the usual Dirac Hamiltonian $H_{\mathrm{QED}}^{\mathrm{Dirac}}$ on $\cal F(\cal H_{\mathrm{one}}^{\mathrm{trans}})\otimes {\cal H}_{\mathrm{Dirac}}$.  In consequence, since it is related to that by a unitary transformation, the restriction of $H^{\mathrm{PP, Dirac}}_{\mathrm{QED}}$ to the product picture physical subspace, $U{\cal F(\cal H_{\mathrm{one}}^{\mathrm{trans}})}\otimes \Omega^{\mathrm{long}}\otimes{\cal H}_{\mathrm{Dirac}}$, will also be self-adjoint.   

We will prove that $\tilde{\bm\pi}$ maps the product picture physical subspace to itself and that its restriction to that subspace is self-adjoint, in the next subsection.

It is perhaps not obvious \textit{a priori}, that this strategy will be successful, i.e.\ that the $H^{\mathrm{PP, Dirac}}$, obtained in this way, will turn out to have the desired form of a product picture Hamiltonian -- i.e.\ that it will arise (cf.\ Section \ref{Sect:Intro2}) as a sum of an electromagnetic Hamiltonian and a Dirac Hamiltonian and an interaction term.  But, we shall see in the next subsection that it turns out to equal the expression in (\ref{prodHam}) which does indeed have that desired product picture form.

\subsection{\label{Sect:QEDcalculation} The calculation of the product picture Hamiltonian, the product picture version of Gauss's law, and the Coulomb gauge to product picture dictionary}

First we notice (cf.\ Equation (\ref{VCoulombphi})) that 
\begin{equation}
\label{VCoulphi}
V_{\mathrm{Coulomb}}^{\mathrm{Dirac}} = \frac{1}{2}\int\bm\nabla\phi{\bm\cdot}\bm\nabla\phi\, d^3x,
\end{equation}
where $\phi$ is given by (\ref{phi}) and hence
\[
\int  {1\over 2} \tilde{\bm\pi}^2   + \tilde{\bm \pi}\bm{\cdot\nabla}\phi \, 
d^3 x + V_{\mathrm{Coulomb}}^{\mathrm{Dirac}}=\int  {1\over 2} (\tilde{\bm\pi} + \nabla\phi)^2   
d^3 x. 
\]
Thus (\ref{checkHQED}) may be written (cf.\ (\ref{emHamshift}))
\begin{equation}
\label{checkHQEDalt}
\check H_{\mathrm{QED}}^{\mathrm{Dirac}}=\int  {1\over 2} (\tilde{\bm\pi} + \nabla\phi)^2 +{1\over 2}({\bm\nabla} {\bm\times} {\bm A})^2 +\psi^*\gamma^0{\bm \gamma}{\bm\cdot}(-i{\bm\nabla} - {\rm e}{\bm A})\psi+m\psi^*\gamma^0\psi  \, 
d^3 x.
\end{equation}

As explained in Section \ref{Sect:QEDstrategy}, we wish to compute $U\check H_{\mathrm{QED}}^{\mathrm{Dirac}} U^{-1}$ where (cf.\ (\ref{Uext})) $U$ now means
\begin{equation}
\label{U}
U=\exp\left(i\int \hat A^i(\bm x)\partial_i\phi(\bm x)\, d^3x\right ),
\end{equation}
where $\phi$ is given by (\ref{phi}),
regarded as a (unitary) operator from our QED augmented Hilbert space, $\cal F(\cal H_{\mathrm{one}})\otimes {\cal H}_{\mathrm{Dirac}}$, to itself.  (But see Footnote \ref{ftntSubtle})

To do this, we first notice that conjugating $\bm A$ with $U$ (i.e.\ taking $U{\bm A}U^{-1}$) leaves $\bm A$ unchanged, and the same is true for $\phi$ and hence, by (\ref{VCoulphi}), also for 
$V_{\mathrm{Coulomb}}^{\mathrm{Dirac}}$. 
On the other hand, by (\ref{Ahatpitildecomm}), we have (cf.\ (\ref{emshift})) that
\begin{equation}
\label{meshift}
U\tilde{\bm\pi} U^{-1}=\tilde{\bm\pi}-\bm\nabla\phi,
\end{equation}
and thus, when conjugated with $U$, the term  ${1\over 2} (\tilde{\bm\pi} + \nabla\phi)^2$ in  (\ref{checkHQEDalt}), becomes simply ${1\over 2} \tilde{\bm\pi}^2$.
Also, on recalling (\ref{rhopsipsi}) that $\rho(\bm x)=e\psi^*(\bm x)\psi(\bm x)$ and using the equation $\nabla^2\phi=-\rho$ and integrating by parts, we see that $U$ can alternatively be written 
\begin{equation}
\label{Ualt}
U=\exp\left(i\int \left(\frac{\partial^i}{\nabla^2}\hat A_i\right)\!(\bm x)\rho(\bm x) \, d^3x\right )
\end{equation}
where $\frac{\partial^i}{\nabla^2}\hat A_i(\bm x)$ means the inverse Fourier transform of $-ik^i/k^2\hat A_i(\bm k)$.  Using this, we easily find that
\begin{equation}
\label{upsiu}
U\psi(\bm x) U^{-1} = \breve\psi(\bm x), \quad \hbox{where}\quad \breve\psi(\bm x) = e^{ - \left (ie \frac{\partial_i}{\nabla^2}\hat A^i\right )(\bm x)}\psi(\bm x),
\end{equation}
and similarly $U\psi^*(\bm x) U^{-1}=\breve\psi^*(\bm x)=e^{\left (ie \frac{\partial_i}{\nabla^2}\hat A^i\right )(\bm x)}\psi^*(\bm x)$.    

Thus we conclude that, on $\cal F(\cal H_{\mathrm{one}})\otimes {\cal H}_{\mathrm{Dirac}}$,

\begin{equation}
\label{equivDir}
U\check H_{\mathrm{QED}}^{\mathrm{Dirac}}U^{-1}=H^{\mathrm{PP, Dirac}}_{\mathrm{QED}}
\end{equation}
where
\begin{equation}
\label{prodHambrev}
H^{\mathrm{PP, Dirac}}_{\mathrm{QED}} =\int \frac{1}{2}\tilde{\bm\pi}^2 +{1\over 2}({\bm\nabla} {\bm\times} {\bm A})^2 + \breve\psi^*\gamma^0{\bm \gamma}{\bm\cdot}(-i{\bm\nabla} - {\rm e}{\bm A})\breve\psi+m\breve\psi^*\gamma^0\breve\psi  \, d^3x.
\end{equation}

We may also derive an alternative expression for $H^{\mathrm{PP, Dirac}}_{\mathrm{QED}}$.   First we notice that  
\begin{equation}
\label{mincouphat}
(({-i\bm\nabla} - {\rm e}{\bm A})\breve\psi)(\bm y) = e^{-i{\rm e} \left(\frac{\partial_i}{\nabla^2}\hat A^i\right )(\bm y)}
\!\left(\!-i({\bm\nabla}\psi)(\bm y)\! -\!\! {\rm e}\!\left [{\bm A}(\bm y) \!+\!{\bm\nabla}\left(\!\frac{\partial_i}{\nabla^2}\hat A^i\!\right )\!\!(\bm y)\!\right ]\!\!\psi(\bm y)\!\right )\!.
\end{equation}
Further, notice (see (\ref{tradApi})) that $\bm A=\hat{\bm A}^{\mathrm{trans}}$, while $\bm\nabla\left(\frac{\partial_i}{\nabla^2}\hat A^i\right )= \hat{\bm A}^{\mathrm{long}}$ and thus, by (\ref{Ahat}), the term in square brackets in (\ref{mincouphat}) is simply $\hat{\bm A}$!   Thus, by (\ref{upsiu}), the third term in the integrand in the right hand side of (\ref{prodHambrev}) can alternatively be written $\psi^*\gamma^0{\bm \gamma}{\bm\cdot}(-i{\bm\nabla} - {\rm e}\hat{\bm A})\psi$.   Also the fourth term is equal to  
$m\psi^*\gamma^0\psi$.   Thus we have
\begin{equation}
\label{prodHam}
H^{\mathrm{PP, Dirac}}_{\mathrm{QED}}=
\int {1\over 2} \tilde{\bm\pi}^2 +{1\over 2}({\bm\nabla} {\bm\times} \hat{\bm A})^2
+ \psi^*\gamma^0{\bm \gamma}{\bm\cdot}(-i{\bm\nabla} - {\rm e}\hat{\bm A})\psi + m\psi^*\gamma^0\psi \, d^3x 
\end{equation}
where we have also used the fact that ${\bm\nabla} {\bm\times} {\bm A}$ is equal to ${\bm\nabla} {\bm\times} \hat{\bm A}$.
  
Thus we have arrived at the candidate product picture Hamiltonian as promised in Section \ref{Sect:QEDstrategy} and see from either of the expressions (\ref{prodHambrev}), (\ref{prodHam}), that it does indeed have the desired product picture form.   We will discuss it further below.   But before doing so we will revisit the unitary transformations of the variables, $\tilde{\bm\pi}$, $\hat{\bm A}$, $\psi$ and $\psi^*$ in terms of which $H^{\mathrm{PP, Dirac}}_{\mathrm{QED}}$ is expressed in Equation (\ref{prodHam}).  

Before doing that, we must fill a gap and prove, as promised in Section \ref{Sect:QEDstrategy}, that $\tilde{\bm\pi}$ maps the product picture physical subspace ${\cal F(\cal H_{\mathrm{one}}^{\mathrm{trans}})}\otimes U\Omega^{\mathrm{long}}\otimes {\cal H}_{\mathrm{Dirac}}$ to itself and that, when restricted to that subspace, it is self-adjoint.  (We recall from Section \ref{Sect:QEDstrategy} that it is an immediate consequence of (\ref{equivDir}) that $H^{\mathrm{PP, Dirac}}_{\mathrm{QED}}$ will also map the product picture physical subspace to itself and be self-adjoint when restricted to it.)

To do this, we write $\tilde{\bm\pi}$ as ${\bm\pi}^\perp + \tilde{\bm\pi}^{\mathrm{long}}$.   ${\bm\pi}^\perp$ is self adjoint on the full QED augmented Fock space, ${\cal F(\cal H_{\mathrm{one}})}\otimes {\cal H}_{\mathrm{Dirac}}$, and one can easily see that it maps the product picture physical subspace, $U{\cal F(\cal H_{\mathrm{one}}^{\mathrm{trans}})}\otimes \Omega^{\mathrm{long}}\otimes {\cal H}_{\mathrm{Dirac}}$, to itself since it acts non-trivially only on the ${\cal F(\cal H_{\mathrm{one}}^{\mathrm{trans}})}$ component of ${\cal F(\cal H_{\mathrm{one}})}\otimes {\cal H}_{\mathrm{Dirac}}$\footnote{\label{ftntAnother} When we say that $\tilde{\bm\pi}^\perp$ acts non-trivially only on the ${\cal F}(\cal H_{\mathrm{one}}^{\mathrm{trans}})$ component of ${\cal F}(\cal H_{\mathrm{one}})\otimes {\cal H}_{\mathrm{Dirac}}$, and that $\phi$ acts non-trivially only on the ${\cal H}_{\mathrm{Dirac}}$ component of ${\cal F}(\cal H_{\mathrm{one}})\otimes {\cal H}_{\mathrm{Dirac}}$, we mean that, when the QED augmented Hilbert space, ${\cal F}(\cal H_{\mathrm{one}})\otimes{\cal H}_{\mathrm{Dirac}}$, is written as
$\cal F(\cal H_{\mathrm{one}}^{\mathrm{trans}})\otimes {\cal F}(\cal H_{\mathrm{one}}^{\mathrm{long}})\otimes {\cal H}_{\mathrm{Dirac}}$, then $\tilde{\bm\pi}^\perp$ takes the form $\tilde{\bm\varpi}^{\mathrm{trans}}\otimes\mathrm{id}\otimes\mathrm{id}$ for some operator $\tilde{\bm\varpi}^\perp$ on $\cal F(\cal H_{\mathrm{one}}^{\mathrm{trans}})$ and that $\phi$ takes the form $\mathrm{id}\otimes\mathrm{id}\otimes\varphi$ for some operator $\varphi$ on ${\cal H}_{\mathrm{Dirac}}$ where $\mathrm{id}$ denote identity operators.} and it commutes with $U$.  So it remains to prove that $\tilde{\bm\pi}^{\mathrm{long}}$ maps $U{\cal F(\cal H_{\mathrm{one}}^{\mathrm{trans}})}\otimes \Omega^{\mathrm{long}}\otimes {\cal H}_{\mathrm{Dirac}}$ to itself and is self-adjoint.   To prove this, first note that equation (\ref{meshift}) entails that the restriction of $\tilde{\bm\pi}^{\mathrm{long}} - \bm\nabla\phi$ to that subspace is unitarily related to the restriction of $\tilde{\bm\pi}^{\mathrm{long}}$ to the Coulomb gauge physical subspace
${\cal F(\cal H_{\mathrm{one}}^{\mathrm{trans}})}\otimes \Omega^{\mathrm{long}}\otimes {\cal H}_{\mathrm{Dirac}}$.   But the latter is clearly zero!   Thus, we have
\begin{equation}
\label{pilongphi}
\tilde{\bm\pi}^{\mathrm{long}}\,\bm\Psi = (\bm\nabla\phi)\bm\Psi, \ \forall \  \bm\Psi \in U{\cal F}({\cal H}_{\mathrm{one}}^{\mathrm{trans}})\otimes \Omega^{\mathrm{long}}\otimes {\cal H}_{\mathrm{Dirac}}. 
\end{equation}
The result then follows from the fact that $\phi$ and hence $\bm\nabla\phi$ maps 
$U{\cal F(\cal H_{\mathrm{one}}^{\mathrm{trans}})}\otimes \Omega^{\mathrm{long}}\otimes {\cal H}_{\mathrm{Dirac}}$ to itself and, when so restricted, is self adjoint.   The latter easily follows from the fact that $\phi$ acts non-trivially and self-adjointly only on the ${\cal H}_{\mathrm{Dirac}}$ component of ${\cal F(\cal H_{\mathrm{one}})}\otimes {\cal H}_{\mathrm{Dirac}}$ (see Footnote \ref{ftntAnother}) and the fact that it commutes with $U$.

Let us next recall that, in Coulomb gauge, the electric field strength operator  -- let us call it here ${\bm E}_{\mathrm C}$ -- viewed as the restriction of $-(\tilde{\bm\pi} + \bm\nabla\phi)$ (cf.\ Equation (\ref{Epiphi})) to the Coulomb gauge physical subspace, ${\cal F(\cal H_{\mathrm{one}}^{\mathrm{trans}})}\otimes \Omega^{\mathrm{long}}\otimes {\cal H}_{\mathrm{Dirac}}$, is, by (\ref{meshift}), unitarily equivalent to $-\tilde{\bm\pi}$ on the product picture physical subspace $U{\cal F(\cal H_{\mathrm{one}}^{\mathrm{trans}})}\otimes \Omega^{\mathrm{long}}\otimes {\cal H}_{\mathrm{Dirac}}$.
Thus we conclude that, in the product picture, the electric field strength operator -- let us call it here $\bm E_{\mathrm{PP}}$ --  is simply $-\tilde{\bm\pi}$ restricted to the product picture physical subspace.   This should not be a surprise since, after all, our strategy was motivated by our treatment of the case of an external classical charge distribution in Section \ref{Sect:electrocoh} in which (in the $\tilde{\bm\pi}$ framework) the electric field was represented by $-\tilde{\bm\pi}$.

Next, let us note that an equivalent statement to (\ref{pilongphi}) is that
\begin{equation}
\label{QuantGauss2}
\bm\nabla\bm\cdot\bm E_{\mathrm{PP}}\,\bm\Psi\  (= -\bm\nabla\cdot\tilde{\bm\pi}\,\bm\Psi) = \rho\,\bm\Psi \quad \forall \ \bm\Psi\in U{\cal F(\cal H_{\mathrm{one}}^{\mathrm{trans}})}\otimes \Omega^{\mathrm{long}}\otimes {\cal H}_{\mathrm{Dirac}}
\end{equation}
where we recall $\rho = e\psi^*\psi$.   In other words, in the product picture, Gauss's law holds as an operator equation!

Returning to the passage from Coulomb gauge to the product picture, we have seen above that, in Coulomb gauge, the electric field is given by $E_{\mathrm C}$ which is $ -\tilde{\bm\pi} - \bm\nabla\phi$ restricted to the Coulomb gauge physical subspace, while, in the product picture, it is given by $E_{\mathrm{PP}}$ which is  $-\tilde{\bm\pi}$ restricted to the product picture physical subspace.   Also, the Dirac field is $\psi$ restricted to the Coulomb gauge physical subspace in Coulomb gauge, while, (\ref{upsiu}) tells us that, in the product picture, it is (but see also Section \ref{Sect:twolast}) 
$\breve\psi(\bm x) = e^{ - \left (ie \frac{\partial_i}{\nabla^2}\hat A^i\right )(\bm x)}\psi(\bm x)$.  Etc.   We thus have the dictionary indicated in Table \ref{tab2}, where it is to be understood that all Coulomb gauge quantities are restricted to the Coulomb gauge physical subspace and all product picture physical quantities are restricted to the product picture physical subspace.

As concerns the vector potentials: while $\bm A$ maps the Coulomb gauge physical subspace to itself,  
$\hat{\bm A}$ maps neither the Coulomb gauge physical subspace, nor the product picture physical subspace  to itself.  However, this is not a problem since the physical quantity associated to $\hat{\bm A}$ is the magnetic field, ${\bm B} = \bm\nabla\bm\times\hat{\bm A}$, and that does map each of these subspaces to itself.

\begin{table}[ht]
\caption{\textbf{Coulomb gauge to product picture dictionary.}\quad\textit{Note that all Coulomb gauge quantities are defined on/restricted to/belong to the Coulomb gauge physical subspace, ${\cal F(\cal H_{\mathrm{one}}^{\mathrm{trans}})}\otimes \Omega^{\mathrm{long}}\otimes {\cal H}_{\mathrm{Dirac}}$, of the QED augmented Hilbert space $\cal F(\cal H_{\mathrm{one}})\otimes {\cal H}_{\mathrm{Dirac}} = {\cal F(\cal H_{\mathrm{one}}^{\mathrm{trans}})}\otimes {\cal F(\cal H_{\mathrm{one}}^{\mathrm{long}})}\otimes {\cal H}_{\mathrm{Dirac}}$ (but we write in parenthesis equivalent forms, which may also be understood as acting on -- or, in the case of the vacuum state, belong to -- the usual Coulomb gauge Hilbert space ${\cal F(\cal H_{\mathrm{one}}^{\mathrm{trans}})}\otimes {\cal H}_{\mathrm{Dirac}}$); all product picture quantities are defined on/restricted to/belong to the product picture physical subspace,\break 
${\cal F(\cal H_{\mathrm{one}}^{\mathrm{trans}})}\otimes U(\Omega^{\mathrm{long}}\otimes {\cal H}_{\mathrm{Dirac}})$, of the same QED augmented Hilbert space.}}

\smallskip

\centering 
\resizebox{\textwidth}{!}{\begin{tabular}{c c c}
\hline 
Quantity & Coulomb gauge & product picture 
\\ [0.5ex] 
\hline
\hline
\quad & \quad & \quad \\
Hamiltonian & $\check H_{\mathrm{QED}}^{\mathrm{Dirac}}$   \ (or $H_{\mathrm{QED}}^{\mathrm{Dirac}}$)  & $H^{\mathrm{PP, Dirac}}_{\mathrm{QED}}$\\
electric field & $\bm E_{\mathrm C} =  -(\tilde{\bm\pi}+\bm\nabla\phi)$ \ (or $ -(\bm\pi^\perp + \bm\nabla\phi)$)  &  ${\bm E}_{\mathrm{PP}} = -\tilde{\bm\pi}$\\ 
magnetic field & $\bm\nabla\bm\times\bm A$ \ (or $\bm\nabla\bm\times{\bm\hat A}$) & $\bm\nabla\bm\times\hat{\bm A}$ \\
Dirac field & $\psi$ &  $\breve\psi = e^{ - \left (ie \frac{\partial_i}{\nabla^2}\hat A^i\right )}\psi$ \\
adjoint Dirac field & $\psi^*$ & $\breve\psi^*= e^{\left (ie \frac{\partial_i}{\nabla^2}\hat A^i\right )}\psi^*$ \\
Dirac electrical potential & $\phi$ & $\phi$\\
vacuum state & $\Omega^{\mathrm{trans}}\otimes\Omega^{\mathrm{long}}\otimes \Omega_{\mathrm{Dirac}}$ \ (or $\Omega^{\mathrm{trans}}\otimes \Omega_{\mathrm{Dirac}}$) \quad\quad & \quad\quad $\Omega^{\mathrm{trans}}\otimes U(\Omega^{\mathrm{long}}\otimes \Omega_{\mathrm{Dirac}})$ (entangled)\\
\quad & \quad & \quad \\
\hline
\end{tabular}}
\label{tab2} 
\end{table}

\bigskip

Returning to the discussion of our product picture Hamiltonian (\ref{prodHambrev})/(\ref{prodHam}), by the equivalence with the Coulomb gauge formulation which we have established above, it must be that the usual Coulomb gauge theory, supplemented by the commutation relations (\ref{tradCR}), (\ref{tradCAR}), is entirely equivalent to the theory defined by the product picture  Hamiltonian, $H^{\mathrm{PP, Dirac}}_{\mathrm{QED}}$ on the product picture physical subspace $U{\cal F(\cal H_{\mathrm{one}}^{\mathrm{trans}})}\otimes \Omega^{\mathrm{long}}\otimes {\cal H}_{\mathrm{Dirac}}$ -- supplemented by the commutation relations (\ref{Ahatpitildecomm}), (\ref{Bhatpitildecomm}), (\ref{tradCAR}).    

Let us point out, however, that, if one works with the expression (\ref{prodHambrev}) for
$H^{\mathrm{PP, Dirac}}_{\mathrm{QED}}$ then one needs to bear in mind that (on the full QED augmented Hilbert space) $\tilde{\bm\pi}$
doesn't commute with $\breve\psi$ and $\breve\psi^*$. In fact, by an easy calculation, their commutation relation is 
\begin{equation}
\label{oddcomm}
[\tilde{\bm\pi}(\bm x), \breve\psi(\bm y)] = [\tilde{\bm\pi}^{\mathrm{long}}(\bm x), \breve\psi(\bm y)] =  -{\rm e}\left(\bm\nabla_{\bm x}\left(\frac{1}{4\pi|\bm x - \bm y|}\right)\right)\breve\psi
\end{equation}
(and similarly for $[\tilde{\bm\pi}(\bm x), \breve\psi^*(\bm y)]$ but with e in place of $-$e).  One may either show this directly from the definition (\ref{upsiu}) of $\breve\psi$ and the commutation relations, (\ref{Ahatpitildecomm}), between $\hat A$ and $\tilde{\bm\pi}$, or instead by noting that, if we denote $U\tilde{\bm\pi}U^{-1} = \tilde{\bm\pi} - \bm\nabla\phi$ by $\breve{\bm\pi}$, then by unitary transformation of the commutation relation $[\tilde{\bm\pi}(\bm x), \psi(\bm y)]=0$, we will have that $[\tilde{\bm\pi}(\bm x), \breve\psi(\bm y)]$ is equal to $[(\bm\nabla\phi)(\bm x), \breve\psi(\bm y)]$ and the latter may be calculated from (\ref{phi}) and (\ref{tradCAR}).
So, in the product picture, the electric field operator, $\bm E_{\mathrm{PP}}$ ($=-\tilde{\bm\pi}$ restricted to the product picture physical subspace) and the product picture Dirac field, $\breve\psi$, don't commute --  their commutator being given by (\ref{oddcomm}) with $\bm E_{\mathrm{PP}} = -\tilde{\bm\pi}$.   

There is in fact a well-known simple qualitative argument that, in any quantization of QED in which Gauss's law holds, the charged matter fields and the (longitudinal part of the) electric field cannot commute at spacelike separation. See e.g.\ around Equation (1.3) in \cite[Sections 1 and 2]{MRS} (which we already referred to in Section \ref{Sect:mathnote}) where it is written:

\medskip

\textit{``$\dots$ because the global charge operator is an integral over
the field strength at spacelike infinity, and has a nontrivial commutator with the charged
field, the latter cannot commute with the field strength at spacelike distance $\dots$''}

\medskip

Since, as we have just seen, our product picture provides a quantization of QED in which Gauss's law holds (as an operator equation) it is therefore not a surprise that we found $[\bm E_{\mathrm{PP}}(\bm x), \breve\psi(\bm y)]$ to be ``nonzero at spacelike distance'' and indeed the right hand side of (\ref{oddcomm}) tells us what that nonzero quantity is in the product picture.

We remark that, in the Coulomb gauge formulation (which can be obtained by quantizing classical electrodynamics after solving the classical Gauss's law) by the quantum counterparts of the classical Equations (\ref{Epot}), (\ref{Poisson}), (\ref{phirho}),   the (longitudinal) electric field, $\bm E_{\mathrm C}^{\mathrm{long}}$ is simply defined to be $-\bm\nabla\phi$ ($\phi$ as in (\ref{phi})) and this has the commutation relation 
\begin{equation}
\label{oddcommCoulomb}
[\bm E_{\mathrm C}^{\mathrm{long}}(\bm x), \psi(\bm y)] =  {\rm e}\left(\bm\nabla_{\bm x}\left(\frac{1}{4\pi|\bm x - \bm y|}\right)\right)\psi
\end{equation}
and similarly for  $[\bm E_{\mathrm C}^{\mathrm{long}}(\bm x), \psi^*(\bm y)]$ with e replaced by $-$e (while $\bm E^{\mathrm{trans}}$\break 
[$= -\bm\pi^\perp$] of course commutes with $\psi$ and $\psi^*$).
It is easy to see by conjugating each side of (\ref{oddcommCoulomb}) with $U$ that (\ref{oddcommCoulomb}) (restricted to the Coulomb gauge physical subspace) and (\ref{oddcomm}) (restricted to the product picture physical subspace) translate into one another as one would expect.  So this could be another entry in our dictionary. But the different ways in which they are arrived at reflect the different statuses of Gauss's law in Coulomb gauge and in the product picture. In Coulomb gauge, the `field strength' of the above quote (or rather its longitudinal part, $\bm E^{\mathrm{long}}$, which is what is relevant) is not actually one of the dynamical variables of the theory.   Rather, it is \emph{defined} to be $-\bm\nabla\phi$ and might, therefore, be said to be, as we put it in Section \ref{Sect:Intro2}, an `epiphenomenon' of the theory.  See also the further discussion in Section  \ref{Sect:final}.  On the other hand,  in the product picture, Gauss's law holds, as we saw above, as the operator equation (\ref{QuantGauss2}) and the two ways of calculating the commutator in (\ref{oddcomm}) mentioned above give a mathematical meaning to the words of the above quote in an interestingly different way.

\subsubsection{\label{Sect:twolast} Two last remarks}

Returning to our formulae, (\ref{prodHambrev}) and (\ref{prodHam}), for the product picture Hamiltonian, $H^{\mathrm{PP, Dirac}}_{\mathrm{QED}}$, it would appear to be simpler, and therefore very possibly preferable for some purposes and, in particular, for the purpose of calculating the dynamics, i.e.\ of calculating $\exp(-iH^{\mathrm{PP, Dirac}}_{\mathrm{QED}}t)$ (restricted to the product picture subspace), to adopt the formula (\ref{prodHam}), rather than (\ref{prodHambrev}), since, unlike $\breve\psi$, $\psi$ commutes with $\tilde{\bm\pi}$.   Relatedly, the Hamilton equation
$\dot\psi=i[{H^{\mathrm{PP, Dirac}}_{\mathrm{QED}}}, \psi]$ looks simpler than the corresponding equation for $\breve\psi$.  Note that the possibility of working with $\psi$, rather than $\breve\psi$, in this connection, is not in contradiction with what we wrote above about the inevitability of noncommutation between the charged field and $\bm E_{\mathrm{PP}}$ and the way that this is reflected in the commutation relation (\ref{oddcomm}).  The reason is that, unlike $\breve\psi$ (and unlike $\tilde{\bm\pi}$ and ${H^{\mathrm{PP, Dirac}}_{\mathrm{QED}}}$) $\psi$ does not map the product picture subspace to itself and thus does not deserve the physical interpretation of `the charged field' in the product picture.  Also related to the preceding comments, note that there are two ways in which one might compute $\exp(-iH^{\mathrm{PP, Dirac}}_{\mathrm{QED}}t)$; one could restrict to the product picture physical subspace first and then exponentiate or one could exponentiate (the non-self adjoint!) $H^{\mathrm{PP, Dirac}}_{\mathrm{QED}}$ on the full augmented QED Hilbert space and then restrict the result to the product picture physical subspace.

Finally, let us remark that (as is briefly noted in our last dictionary entry) the usual vacuum state $\Omega^{\mathrm{trans}}\otimes \Omega_{\mathrm{Dirac}}$ of the Coulomb picture (equivalently the $\Omega^{\mathrm{trans}}\otimes \Omega^{\mathrm{long}}\otimes \Omega_{\mathrm{Dirac}}$ of the QED augmented Hilbert space) gets replaced, in the product picture, by the vector, $U\Omega^{\mathrm{trans}}\otimes \Omega^{\mathrm{long}}\otimes \Omega_{\mathrm{Dirac}}$\break 
($=\Omega^{\mathrm{trans}}\otimes U(\Omega^{\mathrm{long}}\otimes \Omega_{\mathrm{Dirac}})$) in which the Dirac field is entangled with longitudinal photons. 

\subsection{\label{Sect:QEDdiscussion} Further discussion, remarks about the relation with the temporal gauge, and comments on what happens in the $\hat{\bm\pi}$ framework}

It is remarkable that the product picture Hamiltonian, $H^{\mathrm{PP, Dirac}}_{\mathrm{QED}}$ of (\ref{prodHam}), takes the same form as $H_{\mathrm{QED}}^{\mathrm{Dirac}}$ except that (i) the term $V_{\mathrm{Coulomb}}^{\mathrm{Dirac}}$ is absent, (ii) $\bm\pi^\perp$ is replaced by $\tilde{\bm\pi}$, (iii) $\bm A$ is replaced by $\hat{\bm A}$,  and, (iv) in place of the commutation relations (\ref{tradCR}), (\ref{tradCAR}) we have the commutation relations  (\ref{Ahatpitildecomm}), (\ref{Bhatpitildecomm}),  (\ref{tradCAR}).   It seems fair to say that $H^{\mathrm{PP, Dirac}}_{\mathrm{QED}}$, and the latter commutation relations, are simpler.   In particular, the absence of the term
$V_{\mathrm{Coulomb}}^{\mathrm{Dirac}}$ is a simplification since this term of the original Coulomb gauge Hamiltonian (when expressed as (\ref{VCoulomb}) with $\rho=\psi^*\psi$) has the unpleasant features of being both quartic in $\psi$ (i.e.\ quadratic in $\psi^*\psi$) and nonlocal.   And the commutation relations no longer involve the nonlocal term, $i{\partial^2\over\partial x^i\partial x^j}\left({1\over 4\pi|\bm x-\bm y|}\right)$ of (\ref{tradCR}).

It may also be noticed that the new set of commutation relations are well-known as the commutation relations of the \textit{temporal gauge} (see e.g.\ \cite{LMS} and references therein) while the Hamiltonian, $H^{\mathrm{PP, Dirac}}_{\mathrm{QED}}$ is, with one change, what is generally known as the \textit{temporal-gauge Hamiltonian} -- the change being that we identify (minus) the longitudinal part of the electric field with  $\tilde{\bm\pi}^{\mathrm{long}}$ -- defined as in (\ref{pilongtilde}) to be (twice) a (non-self-adjoint) annihilation operator on the full QED augmented Hilbert space,
${\cal F(\cal H_{\mathrm{one}})}\otimes {\cal H}_{\mathrm{Dirac}}$, rather than, e.g.\ identifying it with $\hat{\bm\pi}^{\mathrm{long}}$ -- as defined in (\ref{pihatdef}).  As discussed and analyzed in detail in \cite{LMS}, in those earlier approaches, temporal gauge quantization suffered from a number of difficulties which lead, as the authors of \cite{LMS} themselves describe them, to `peculiar'  mathematical realisations of the heuristic formalism. Those peculiar realisations are, in fact, unsuccessful in the sense that they are not equivalent to standard QED.   We will next characterize what seems to be the essence of what is peculiar by isolating here what we shall call the \textit{Contradictory Commutator Theorem}.   After stating and proving this, we will briefly discuss, in Section \ref{Sect:temporal}, a possible alternative version of QED based on our $\hat{\bm\pi}$ framework and then resume our discussion of previous approaches to temporal gauge quantization and then criticise both of those approaches as well as the $\hat{\bm\pi}$ framework version of QED.   We postpone to the last paragraph of this section an explanation of how our product picture evades the \textit{Contradictory Commutator Theorem}.

\medskip

\noindent
{\bf Contradictory Commutator Theorem:} There can be no pair of 3-vector operators $\bm{\mathsf A}$ and $\bm{\mathsf\pi}$  on a Hilbert space $\mathsf H$ such that 

\smallskip

\noindent
(a) $\bm{\mathsf A}$ and $\bm{\mathsf\pi}$ satisfy the canonical commutation relations
\begin{equation}
\label{CCR}
[{\mathsf A}_i(\bm x), {\mathsf\pi}_j(\bm y)]=i\delta_{ij}\delta^{(3)}(\bm x-\bm y), \ [{\mathsf A}_i(\bm x), {\mathsf A}_j(\bm y)] =0= [{\mathsf\pi}_i(\bm x), {\mathsf\pi}_j(\bm y)]
\end{equation}
(i.e.\ the same commutation relations as satisfied by 
$\hat {\bm A}, \tilde{\bm \pi}$ in (\ref{Ahatpitildecomm}), (\ref{Bhatpitildecomm}) and by $\hat {\bm A}, \hat{\bm \pi}$); 

\smallskip
\noindent
(b) $\bm{\mathsf A}$ and $\bm{\mathsf\pi}$ are each self-adjoint; 

\smallskip
\noindent
(c)  For some vector $\bm\Psi\in \mathsf H$ ($\Psi \ne 0$)
\[
\bm{\nabla\cdot{\mathsf\pi}}\,\bm\Psi= - \rho\,\bm\Psi
\]
for some operator-valued function of $\bm x$, $\rho$.

\smallskip
\noindent
(d)  $\rho$ commutes with $\bm{\mathsf{A}}$.

\medskip

\noindent
\textbf{Note} (A) In the case $\rho=0$, this theorem and its proof has essentially the same content as just one of the ingredients in the proof of just one part of one of the propositions in \cite{LMS}, namely the second part of the proof of Part (1 i) of Proposition 3.2 there. (B)  If we identify the electric field, 
$\bm E$, with $-\bm{\mathsf\pi}$, then Condition (c) would follow from assuming that Gauss's law held as an operator equation.   Indeed that would amount to demanding that the equation in Condition (c) held for all vectors ${\bm\Psi}\in \mathsf H$, whereas in Condition (c), it is only assumed to hold for one single nonzero vector.

\smallskip

This theorem is easily proved by observing that (a) easily implies that the quantity
$\langle\bm\Psi| [{\mathsf A}_i(\bm x), \bm{\nabla\cdot{\mathsf\pi}}(\bm y)] \bm\Psi\rangle$ is equal to $-i(\nabla_i\delta^{(3)})(\bm x-\bm y)$, while (b), (c) and (d) imply that the same quantity is zero -- a contradiction!  

\subsubsection{\label{Sect:temporal} $\hat{\bm\pi}$-framework QED and critique of \cite{LMS}}

Another possible approach to quantizing QED, which is different from our product picture, would be to attempt to do something similar to what we did in Sections \ref{Sect:QEDstrategy} and (\ref{Sect:QEDdiscussion}) but based on the $\hat{\bm\pi}$ framework instead of the $\tilde{\bm\pi}$ framework.  Thus we could start by defining a Hamiltonian, say, $\acute H_{\mathrm{QED}}^{\mathrm{Dirac}}$, exactly as
$\check H_{\mathrm{QED}}^{\mathrm{Dirac}}$ is defined in (\ref{checkHQED}) except with $\tilde{\bm\pi}$ replaced by $\hat{\bm\pi}$.   Then one can compute
$U\acute H_{\mathrm{QED}}^{\mathrm{Dirac}}U^{-1}$ -- let us call this
$\hat H^{\mathrm{PP, Dirac}}_{\mathrm{QED}}$.   The calculation of $\hat H^{\mathrm{PP, Dirac}}_{\mathrm{QED}}$ is easily seen to go very similarly to the calculation of $H^{\mathrm{PP, Dirac}}_{\mathrm{QED}}$ in Section \ref{Sect:QEDcalculation} and one finds that $\hat H^{\mathrm{PP, Dirac}}_{\mathrm{QED}}$ is identical to the $H^{\mathrm{PP, Dirac}}_{\mathrm{QED}}$ of (\ref{prodHam}) except that, again, $\tilde{\bm\pi}$ is replaced by $\hat{\bm\pi}$.
However, unlike $\check H_{\mathrm{QED}}^{\mathrm{Dirac}}$, (a) $\acute H_{\mathrm{QED}}^{\mathrm{Dirac}}$ does not map the Coulomb gauge physical subspace, ${\cal F(\cal H_{\mathrm{one}}^{\mathrm{trans}})}\otimes \Omega^{\mathrm{long}}\otimes {\cal H}_{\mathrm{Dirac}}$, to itself; (b) 
when restricted to the Coulomb gauge physical subspace, $\acute H_{\mathrm{QED}}^{\mathrm{Dirac}}$ does \emph{not} coincide with the standard (Maxwell-Dirac) QED Hamiltonian, $H_{\mathrm{QED}}^{\mathrm{Dirac}}$.  In consequence of (a), it seems natural to consider the physical Hilbert space of the new theory whose Hamiltonian is $\hat H^{\mathrm{PP, Dirac}}_{\mathrm{QED}}$ to be the full QED augmented Hilbert space, $\cal F(\cal H_{\mathrm{one}})\otimes {\cal H}_{\mathrm{Dirac}}$.   Also, in view of (a) and (b), this new theory (acting on the full QED augmented Hilbert space) will \emph{not} be equivalent to standard QED.   It will be a different theory which we shall call \emph{$\hat{\bm\pi}$-framework QED}.

Gauss's law will not hold in this different theory, even just in expectation value.   However, one can remedy this if we give up the `natural' assumption that the physical Hilbert space is the entire QED augmented Hilbert space and (in a Schr\"odinger picture viewpoint) privilege one particular time, say $t=0$, and take the view that the physical state space at that time is the product picture physical subspace (a $\tilde{\bm\pi}$-framework notion!) and at any other time, $t$, consists of vectors of the form $\exp(-i\hat H^{\mathrm{PP, Dirac}}_{\mathrm{QED}}t)\bm\Psi$, where $\bm\Psi$ ranges over the product picture physical subspace.  

If we take that, let us call it \textit{unnatural}, view, then Gauss's law will hold in expectation value at all times, in the sense that, for any vector, $\bm\Psi$, in the product picture physical subspace at $t=0$ we will have,\hfil\break
$\langle\exp(-i\hat H^{\mathrm{PP, Dirac}}_{\mathrm{QED}}t)\bm\Psi|\hat{\bm\pi}^{\mathrm{long}}\exp(-i\hat H^{\mathrm{PP, Dirac}}_{\mathrm{QED}}t)\bm\Psi\rangle$\hfil\break 
$= \langle\exp(-i\hat H^{\mathrm{PP, Dirac}}_{\mathrm{QED}}t)\bm\Psi|\bm\nabla\phi\exp(-i\hat H^{\mathrm{PP, Dirac}}_{\mathrm{QED}}t)\bm\Psi\rangle$.   To see this, notice first that $\hat{\bm\pi}^{\mathrm{long}}$ commutes with $\acute H_{\mathrm{QED}}^{\mathrm{Dirac}}$.  And therefore  $\hat{\bm\pi}^{\mathrm{long}} - \bm\nabla\phi =
U\hat{\bm\pi}^{\mathrm{long}}U^{-1}$ commutes with $U\acute H_{\mathrm{QED}}^{\mathrm{Dirac}} U^{-1} = \hat H^{\mathrm{PP, Dirac}}_{\mathrm{QED}}$.
Thus, if we write $\bm\Psi = U\bm\Phi$ where $\bm\Phi$ is in the Coulomb gauge physical subspace, we will have\hfil\break
 $\langle\exp(-i\hat H^{\mathrm{PP, Dirac}}_{\mathrm{QED}}t)\bm\Psi|(\hat{\bm\pi}^{\mathrm{long}}-\bm\nabla\phi)\exp(-i\hat H^{\mathrm{PP, Dirac}}_{\mathrm{QED}}t)\bm\Psi\rangle = \langle\bm\Psi|(\hat{\bm\pi}^{\mathrm{long}}-\bm\nabla\phi)\bm\Psi\rangle = \langle U\bm\Phi|(\hat{\bm\pi}^{\mathrm{long}}-\bm\nabla\phi)U\bm\Phi\rangle = \langle\bm\Phi|\hat{\bm\pi}^{\mathrm{long}}\bm\Phi\rangle$, which obviously vanishes, thereby proving what we set out to show.

This (weak) version of Gauss's law could obviously alternatively be expressed by saying that, as long as we restrict to $\bm\Psi$ in the product picture physical subspace, then, even though this isn't an invariant subspace for $\hat H^{\mathrm{PP, Dirac}}_{\mathrm{QED}}$, we nevertheless have that $\langle\bm\Psi|\hat{\bm\pi}^{\mathrm{long}}(t)\bm\Psi\rangle = \langle\bm\Psi|\bm\nabla\phi(t)\bm\Psi\rangle$ where the quantities $\hat{\bm\pi}^{\mathrm{long}}(t)$ and $\bm\nabla\phi(t)$ are defined in terms of $\hat{\bm\pi}^{\mathrm{long}}$ and $\bm\nabla\phi$ by the Heisenberg time-evolution rule for the Hamiltonian $\hat H^{\mathrm{PP, Dirac}}_{\mathrm{QED}}$.

Gauss's law (in strong operator form) is presumably routinely verified, along with the rest of standard QED, every day, for example in atomic physics experiments and observations.   But  of course, to test an aspect of a theory quantitatively it is always helpful to have an alternative theory against whose predictions the predictions of the standard theory can be compared.  Perhaps this `unnatural' version of 
$\hat{\bm\pi}$-framework QED could partially fill that r\^ole.   One might expect that as far as transverse (propagating) modes of the electromagnetic field are concerned, its predictions will be quite similar to those of the standard theory.  (Related to this, while the theory may not be Lorentz covariant, it may not be in contradiction with the basic principles of special relativity.)  But one expects differences e.g.\ due to $\hat{\bm\pi}$-framework QED predicting anomalous quantum fluctuations in the Coulomb field of an atomic nucleus.   It might be interesting to study this quantitatively and also to analyse the mesoscopic experiment we proposed in Section \ref{Sect:capac} so as to check whether the predictions on our assumptions of classical external charges which we made on the $\hat{\bm\pi}$ framework of Section \ref{Sect:extsource} do approximately agree with the predictions of full `unnatural' $\hat{\bm\pi}$-framework QED.  But we won't attempt to do that here.

Returning to our discussion of pre-existing work on temporal gauge quantization and the 
r\^ole of what we call here the Contradictory Commutator Theorem, the paper \cite{LMS} considers two approaches to quantizing QED in the temporal gauge, the starting point for both being the wish to represent the $\bm{\mathsf A}$ and $\bm{\mathsf\pi}$ of the canonical commutation relations (\emph{CCR}) of (\ref{CCR}) by self-adjoint (vector) operators.  One of these approaches overcomes the Contradictory Commutator Theorem by not satisfying any version of Gauss's law.   I suspect it is the same as our $\hat{\bm\pi}$-framework QED here (in its `natural' version) but I haven't investigated this.  In the other approach of \cite{LMS}, Gauss's law does hold, but the approach is based on a representation of the associated Weyl relations to the CCR (\ref{CCR}) which is mathematically pathological in the sense that it does \emph{not} arise from (or give rise to) a literal representation of the CCR.  This gets around the Contradictory Commutator Theorem but at the cost that, in the resulting version of QED, there are no quantum electromagnetic fields but only quantities that correspond to the exponentials of such fields.    Obviously, neither of the theories arrived at with these approaches (and nor our $\hat{\bm\pi}$-framework QED here) can be equivalent to standard QED since they do not share all the properties of standard QED and so, in this sense, they have to be regarded as unsuccessful attempts at temporal gauge quantizations.  

In contrast, as we have seen, our product picture -- based on our $\tilde{\bm\pi}$ framework -- gets around the Contradictory Commutator Theorem.   To explain how it does this, let us begin by pointing out that there are actually two ways in which one might fear that the Contradictory Commutator Theorem might threaten the possibility of a satisfactory product picture; in the first, one identifies the Hilbert space, $\mathsf H$, of the theorem with our full augmented Hilbert space, in the second one identifies $\mathsf H$ with our product picture physical subspace.  In the first case, our product picture gets around the theorem by providing a representation of the CCR (\ref{CCR}) in which the (vector) operators $\bm{\mathsf A}$ and $\bm{\mathsf\pi}$ are not both self-adjoint.  Specifically the representor, $\tilde{\bm\pi}^{\mathrm{long}}$, of the longitudinal part of $\bm{\mathsf\pi}$ fails to be self-adjoint.  Nevertheless, as we explained in Section \ref{Sect:QEDcalculation}, the restrictions of the thus-represented operators to the product picture physical subspace are all self-adjoint so the apparent threat is seen to be harmless.    In the second case, our product picture gets around the theorem because $\hat{\bm A}$ does not map the product picture physical subspace to itself.  Nevertheless the gauge-invariant quantity, ${\bm B}=\bm\nabla\bm\times\hat{\bm A}$, does map it to itself so, again, the apparent threat to our product picture is seen to be harmless. And, indeed, as we have seen, the product picture is equivalent to standard QED.   Thus, it appears that our product picture deserves to be regarded as a successful implementation of temporal gauge quantization; albeit the way we arrived at it in Section \ref{Sect:QED} was not by quantizing classical QED in temporal gauge, but rather by a transformation, at the quantum level, of standard Coulomb gauge QED.

\section{\label{Sect:QEDSchr} Transformation of Maxwell-Schr\"odinger QED to a product picture}

\subsection{Main ideas and results}

We now consider the Maxwell-Schr\"odinger version of QED where we have a system of $N$ non-relativistic charged particles with positions, ${\bm x}_I$, and momenta,  ${\bm p}_I$, satisfying the commutation relations
\[
[x_{iI}, x_{jJ}] = 0, \quad [p_{iI}, p_{jJ}]=0, \quad [x_{iI}, p_{jJ}]=i\delta_{ij}\delta_{IJ}
\]
and interacting with the quantum electromagnetic field.   Starting from the standard Coulomb gauge formulation of this system (in which the ${\bm x}_I$ and ${\bm p}_I$  are assumed to commute with the electromagnetic operators $\bm \pi$ and $\bm A$) we wish to establish a product picture much as we did for the Maxwell-Dirac system in Section \ref{Sect:QED}.  In so doing, we will arrive at a systematic derivation of a result, (\ref{NewtForm}) here, which we first derived in \cite{KayNewt} in a partly heuristic way.   We shall then illustrate the product picture by discussing the bound states of the nonrelativistic (spinless) Hydrogen atom from a product picture point of view.   We shall also (re)derive the formula, Equation (\ref{innprod})/(\ref{innprodalt}) below, for the partial trace over the longitudinal part of the electromagnetic field of the density matrix which replaces the usual density matrix $\Psi_{\mathrm{Schr}}(\bm x_1, \dots, \bm x_N)\Psi^*_{\mathrm{Schr}}(\bm x_1', \dots, \bm x_N')$  where $\Psi_{\mathrm{Schr}}(\bm x_1, \dots, \bm x_N)$ is a Schr\"odinger wave function for $N$ charged particles when one goes over to the product picture and which takes into account the longitudinal modes of the electromagnetic field that surround each of the charged particles.   This formula  (or rather its linearized gravity counterpart -- see also Footnote \ref{ftntErr}) was previously arrived at in \cite{KayNewt} (as Equation (16) there) on the basis of partly heuristic arguments based on (the gravitational counterparts of) formulae obtained (such as Equation (\ref{balloverlap1}) here) for the inner products between quantum states of the electric field for given classical charge distributions in the way that we discussed in Section \ref{Sect:Intro2}.    

If we were to model our particles as pointlike, then, in place of the Dirac Hamiltonian, $H_{\mathrm{Dirac}}$ (\ref{HDirac}), of the previous section, we would have the familiar many-body Schr\"odinger Hamiltonian
\begin{equation}
\label{SchroHampoint}
H_{\mathrm{Schr\, pointlike}}=\sum_{I=1}^N \frac{(\bm p_I - q_I \bm A(\bm x_I))^2}{2M_I} + V_{\mathrm{Coulomb}}^{\mathrm{Schr\, pointlike}}
\end{equation}
where (in place of the
$V_{\mathrm{Coulomb}}^{\mathrm{Dirac}}$ of Section \ref{Sect:QED}) we now have the familiar Coulomb potential
\begin{equation}
\label{VCoulombpoint}
V_{\mathrm{Coulomb}}^{\mathrm{Schr\, pointlike}}=\frac{1}{2}\sum_{I=1}^N\sum_{J=1}^N \frac{q_Iq_J}{4\pi|{\bm x}_I-{\bm x}_J|}
\end{equation}
where (see Footnote \ref{ftntSubtle}) one omits equal $I$ and $J$ values from the sum.

Here, $M_I$ and $q_I$, $I:1\dots N$, are the masses and charges of the particles respectively.  The ${\bm x}_I$, ${\bm p}_I$ and  $H_{\mathrm{Schr}}$ will all act (as unbounded operators) on the Hilbert space ${\cal H}_{\mathrm{Schr}}$ which (with the usual Schr\"odinger representation) we may take to be $L^2({\mathbb R}^{3N}$) -- or the appropriately symmetrized and/or antisymmetrized subspace of this according to which (if any) of the particles are identical bosons or identical fermions.  

It would seem to be difficult to obtain an alternative product picture along the lines of that obtained for a Dirac field in the previous section starting from such a point-particle model but the difficulty is overcome by modeling each of the particles as a ball with spherically symmetric mass and charge distributions, which, for the $I$th particle, integrate to $m_I$ and $q_I$ respectively and which we shall take to be rigid in the sense that, relative to each ball centre, they are undistorted by acceleration.\footnote{\label{ftntBalls} The reason we model our particles as having extended charge distributions can be traced to the fact that, while the inner product between logitudinal coherent photon states for the Coulomb potential of two different point charges in different locations is formally finite, the inner product of such a state for a given single point charge with itself (i.e.\ the square of that state's norm) is infinite.  The latter infinity is reflected in the nonexistence of the limit $R\rightarrow 0$ in (\ref{balloverlap2}), (\ref{balloverlap3}), (\ref{balloverlap5}) and in the appearance of the denominator $R$ in the terms in the first product in (\ref{innprod}), while the former finiteness is reflected in the fact that the factors of $R$ cancel out in the remaining terms of that product.   This infinity is of course related to the fact that the  Coulomb potential of a pointlike charged particle due to itself (i.e.\ the `self potential' of a point charge) is infinite.   In Coulomb gauge, this doesn't prevent one from having a model with pointlike charged particles.  One simply omits the self-energy terms i.e.\ the terms with $I=J$ in the formula (\ref{VCoulombpoint}) for $V_{\mathrm{Coulomb}}$.  The problem is that, in order to transform a Coulomb gauge Hamiltonian to the product picture, we need the formula for  $V_{\mathrm{Coulomb}}$ to arise as $\frac{1}{2}\int\bm\nabla\phi{\bm\cdot}\bm\nabla\phi\, d^3x$ for some potential function 
$\phi$. But the point-particle $V_{\mathrm{Coulomb}}$ of (\ref{VCoulombpoint}) where the terms with $I =J$ are necessarily omitted from the sum (because they are infinite!) cannot be written in this way.  (Instead, if $\phi$ denotes the sum $\sum_{I=1}^N\phi_I$ where $\phi_I$ is the potential of the Ith ball, then the latter point-particle $V_{\mathrm{Coulomb}}$ arises as the limit as the charge densities approach delta functions of $\int\bm\nabla\phi{\bm\cdot}\bm\nabla\phi\, d^3x - \sum_{I=1}^M \int\bm\nabla\phi_I{\bm\cdot}\bm\nabla\phi_I\, d^3x$.)  Indeed, it might well be negative, as, e.g.\ in the case where $N=2$ and $q_1$ and $q_2$ have opposite signs.   See, in this connection the remark that ``negative inter-ball potentials $\dots$ are possible $\dots$'' towards the end of Section \ref{Sect:Hatom}. See also the further discussion below.}   We shall then interpret ${\bm x}_I$ and ${\bm p}_I$ as the position and canonical momentum of the centre of mass of the $I$th ball.   Letting $\rho_I(\bm y)$ denote the charge density of the $I$th ball relative to its centre, we thus replace $H_{\mathrm{Schr\, pointlike}}$ by
\begin{equation}
\label{SchroHam}
H_{\mathrm{Schr}}=\sum_{I=1}^N \frac{\left(\bm p_I - \int \bm A(\bm x)\rho_I(\bm x - \bm x_I) \, d^3\bm x\right)^2}{2M_I} + V_{\mathrm{Coulomb}}^{\mathrm{Schr}}
\end{equation}
where 
\begin{equation}
\label{VCoulombrho}
V_{\mathrm{Coulomb}}^{\mathrm{Schr}}=\frac{1}{2}\sum_{I=1}^N\sum_{J=1}^N \int\int\frac{\rho_I(\bm x-\bm x_I)\rho_J(\bm y-\bm x_J)}{4\pi|\bm x-\bm y|}\, d^3x d^3y
\end{equation} 
where now the sum is over \emph{all} $I$ and $J$ whether or not they are equal to one another.  

The full Maxwell-Schr\"odinger QED Hamiltonian will then be
\begin{equation}
\label{HamSchroQED}
H_{\mathrm{QED}}^{\mathrm{Schr}}=
\int {1\over 2} {\bm\pi^\perp}^2 +{1\over 2}({\bm\nabla} {\bm\times} {\bm A})^2 \, d^3x  \
+  \ H_{\mathrm{Schr}}  
\end{equation}
on the Hilbert space $\cal F(\cal H_{\mathrm{one}}^{\mathrm{trans}}) \otimes {\cal H}_{\mathrm{Schr}}$ which, similarly to the Maxwell-Dirac case (see Section \ref{Sect:QEDstrategy}), will be equivalent to 
\begin{equation}
\label{SchroQEDHam}
\check H_{\mathrm{QED}}^{\mathrm{Schr}}=\int  {1\over 2} \tilde{\bm\pi}^2 +{1\over 2}({\bm\nabla} {\bm\times} {\bm A})^2 + \tilde{\bm \pi}\bm{\cdot\nabla}\phi \, 
d^3 x + H_{\mathrm{Schr}} 
\end{equation}
on the Coulomb gauge physical subspace ${\cal F}({\cal H}_{\mathrm{one}}^{\mathrm{trans}})\otimes \Omega^{\mathrm{long}} \otimes {\cal H}_{\mathrm{Schr}}$ of the (now Schr\"odinger) QED augmented Hilbert space ${\cal F}(\cal H_{\mathrm{one}})\otimes {\cal H}_{\mathrm{Schr}}$. 

As in the case of $\check H_{\mathrm{QED}}^{\mathrm{Dirac}}$, we now wish to compute
$U\check H_{\mathrm{QED}}^{\mathrm{Schr}}U^{-1}$ where the unitary operator (but see again Footnote \ref{ftntSubtle}) $U$, now on the just-mentioned Schr\"odinger QED augmented Hilbert space, is defined as in (\ref{U}) but with $\phi$ now given by
\begin{equation}
\label{phiSchro}
\phi(\bm x) = \sum_{I=1}^N \int \frac{\rho_I(\bm y - \bm x_I)}{4\pi|\bm x - \bm y|}\, d^3y
\end{equation}
-- which is to be thought of as a multiplication operator on ${\cal H}_{\mathrm{Schr}}$.
To do this, we again note first that conjugating $\bm A$ with $U$, i.e.\ taking $U\bm A U^{-1}$, leaves it unchanged and the same is true for $\phi$ and for $V_{\mathrm{Coulomb}}^{\mathrm{Schr}}$.   Also, we notice that (just as in the Maxwell-Dirac case in Section \ref{Sect:QEDcalculation}) the terms 
$\int  {1\over 2} \tilde{\bm\pi}^2 + \tilde{\bm \pi}\bm{\cdot\nabla}\phi \, 
d^3 x + V_{\mathrm{Coulomb}}^{\mathrm{Schr}}$ become simply $\int  {1\over 2} \tilde{\bm\pi}^2\, d^3 x$.   To see this, we note (again cf.\  Equation (\ref{VCoulombphi}) and cf.\ also (\ref{VCoulphi})) that $V_{\mathrm{Coulomb}}^{\mathrm{Schr}} =
{1\over 2}\int \phi(\bm x)\rho(\bm x)\, d^3x =
\frac{1}{2}\int\bm\nabla\phi{\bm\cdot}\bm\nabla\phi\, d^3x$.  

Above, $\rho$ is given by  
\begin{equation}
\label{rhoSchro}
\rho(\bm x) =  \sum_{I=1}^N \rho_I(\bm x - \bm x_I).
\end{equation}
Using (\ref{Ualt}) with $\rho$ as in (\ref{rhoSchro}) we easily have
\begin{equation}
\label{pItrans}
U\bm p_I U^{-1} = \bm p_I - \int\hat{\bm A}^{\mathrm{long}}(\bm x)\rho_I(\bm x - \bm x_I)\, d^3x
\end{equation}
and thus conjugating $H_{\mathrm{Schr}}$ with $U$ has the effect of replacing\break 
$\int \bm A(\bm x)\rho_I(\bm x - \bm x_I) \, d^3x$ in (\ref{SchroHam}) by \hfil\break 
$\int \left(\bm A(\bm x) + \hat{\bm A}^{\mathrm{long}}(\bm x)\right)\rho_I(\bm x - \bm x_I)\, d^3x = \int \hat{\bm A}(\bm x)\rho_I(\bm x - \bm x_I) \, d^3x$. (There are interesting similarities but also differences between this passage here and the derivation of (\ref{prodHam}) via Equation (\ref{mincouphat})).\hfil\break   
(Note that this latter step also works for a point-particle model; one simply replaces $\rho_I(\bm x-\bm x_I)$ by $q_i\delta^{(3)}(\bm x - \bm x_I)$ and one finds that conjugating with $U$ has the effect of replacing $\bm A$ by $\hat{\bm A}$ in (\ref{SchroHampoint}).)

We conclude that
\begin{equation}
\label{equivSchro} 
U\check H_{\mathrm{QED}}^{\mathrm{Schr}}U^{-1}=H^{\mathrm{PP, Schr}}_{\mathrm{QED}}
\end{equation}
where
\begin{equation}
\label{PPSchroHamQED}
H^{\mathrm{PP\, Schr}}_{\mathrm{QED}} = \int {1\over 2} \tilde{\bm\pi}^2 +{1\over 2}({\bm\nabla} {\bm\times} \hat{\bm A})^2 \, d^3x + \sum_{I=1}^N \frac{\left(\bm p_I - \int \hat{\bm A}(\bm x)\rho_I(\bm x - \bm x_I) \, d^3\bm x\right)^2}{2M_I}.
\end{equation}

So, similarly to in the Maxwell-Dirac case, the Coulomb gauge Hamiltonian,  $H_{\mathrm{QED}}^{\mathrm{Schr}}$, for QED with many extended
Schr\"odinger charged balls will be equivalent to the product picture Hamiltonian, $H^{\mathrm{PP\, Schr}}_{\mathrm{QED}}$, when the latter is restricted to the product picture physical subspace 
${\cal F}({\cal H}_{\mathrm{one}}^{\mathrm{trans}})\otimes U(\Omega^{\mathrm{long}} \otimes {\cal H}_{\mathrm{Schr}})$ of the augmented Schr\"odinger QED Hilbert space ${\cal F(\cal H_{\mathrm{one}})}\otimes {\cal H}_{\mathrm{Schr}}$.  We remark that, just as in the Maxwell-Dirac case, this product picture physical subspace will be an invariant subspace for both $H^{\mathrm{PP\, Schr}}_{\mathrm{QED}}$ and $\tilde\pi$ and, on this subspace (but not elsewhere), both these operators will be self-adjoint.

We notice that, similarly to in the Maxwell-Dirac case, the term $V_{\mathrm{Coulomb}}^{\mathrm{Schr}}$ -- present in the $H_{\mathrm{Schr}}$ in the $H^{\mathrm{Schr}}_{\mathrm{QED}}$ of (\ref{HamSchroQED}) -- is absent from the $H^{\mathrm{PP\, Schr}}_{\mathrm{QED}}$ of (\ref{PPSchroHamQED}).   

Also as in the Maxwell-Dirac case and by a similar argument to our argument for (\ref{QuantGauss2})), Gauss's law will hold as the operator equation 
\begin{equation}
\label{QuantGauss3}
\bm\nabla\bm\cdot\bm E\,\bm\Psi = \rho\,\bm\Psi \quad \forall \ \bm\Psi\in U{\cal F(\cal H_{\mathrm{one}}^{\mathrm{trans}})}\otimes \Omega^{\mathrm{long}}\otimes {\cal H}_{\mathrm{Schr}}
\end{equation}
where $\rho$ is now as in (\ref{rhoSchro}).

We could also compile a Coulomb gauge to product picture dictionary, in which the electromagnetic fields transform similarly to in Table \ref{tab2}, while $\bm x_I$ translates to $\bm x_I$ and $\bm p_I$ translates to the quantity on the right hand side of Equation (\ref{pItrans}). 

One difference with the Maxwell-Dirac case is that there is of course now no vacuum state.  But similarly to in the Maxwell-Dirac case, in all the vectors in the physical Hilbert space, the charged balls will be entangled with the longitudinal photons.

In fact given a (say unentangled) state of the total system in the traditional Coulomb gauge picture consisting of the tensor product of a state of the transverse part of the electromagnetic field together with a many-body Schr\"odinger wave function 
\begin{equation}
\label{CoulEnt}
\bm\Psi_{\mathrm{Coulomb}}^{\mathrm{QED}}=\Psi^{\mathrm{trans}}_{\mathrm{EM}}\otimes \Psi_{\mathrm{Schr}}
\end{equation}
(a more general state would be a linear combination of such states)
the description of the same physical state in our product picture will be given by 
\begin{equation}
\label{PPent}
\bm\Psi^{\mathrm{PP}}_{\mathrm{QED}}=\Psi^{\mathrm{trans}}_{\mathrm{EM}}\otimes U(\Omega^{\mathrm{long}}\otimes\Psi_{\mathrm{Schr}})
\end{equation}
describing (see Section \ref{Sect:final} for more discussion) a situation in which each of the charged balls is surrounded by, and drags around with it, its electrostatic quantum Coulomb field.

To gain more of a feeling for how things look in the product picture, in a simple setting, let us assume now that we have such an unentangled state in the Coulomb gauge picture and that $\Psi^{\mathrm{trans}}$ is actually the vacuum vector $\Omega^{\mathrm{trans}}$ for the transverse modes of the electromagnetic field.   We might for example consider a time-evolving many body wavefunction, $\Psi_{\mathrm{Schr}}(t)$, which satisfies the time-dependent Schr\"odinger equation for the Hamiltonian
$H_{\mathrm{Schr}}$ of (\ref{SchroHam}) (i.e.\ for the `charged matter part' of (\ref{HamSchroQED})).

If such a time-evolving wave function $\Psi_{\mathrm{Schr}}(t)$ corresponds, in a rough classical description, to a system of charged balls which (due to their initial positions and momenta and their mutual attractions and repulsions) are only accelerating slowly during some time period, then we would expect that the electromagnetic radiation it emits could be neglected to some approximation and then 
$\Omega^{\mathrm{trans}}\otimes\Psi_{\mathrm{Schr}}(t)$ would approximate a solution of the time-dependent Schr\"odinger equation for the full QED Hamiltonian  $H_{\mathrm{QED}}^{\mathrm{Schr}}$ of (\ref{HamSchroQED}).  So (with $\Psi^{\mathrm{trans}}_{\mathrm{EM}} = \Omega^{\mathrm{trans}}$) the $\Psi_{\mathrm{Coulomb}}^{\mathrm{QED}}$ of (\ref{CoulEnt}) could be thought of as a snapshot of such a state, $\Omega^{\mathrm{trans}}\otimes\Psi_{\mathrm{Schr}}(t)$, at a particular moment of time.  Alternatively if $\Psi_{\mathrm{Schr}}$ is an (approximate) solution to the time-independent Schr\"odinger equation for $H_{\mathrm{Schr}}$  such as e.g.\ (in the case $N=2$) for the non-relativistic (spinless) Hydrogen atom (see Section \ref{Sect:Hatom} below) then we would expect $\Omega^{\mathrm{trans}}\otimes\Psi_{\mathrm{Schr}}$ to be an approximate solution to the time-independent Schr\"odinger equation for $H_{\mathrm{QED}}^{\mathrm{Schr}}$.
 
In either case -- slowly time-evolving wavefunction or eigenstate -- we expect the $\bm\Psi^{\mathrm{PP}}_{\mathrm{QED}}$ of Equation (\ref{PPent}) (for $\Psi_{\mathrm{EM}}^{\mathrm{trans}}=\Omega^{\mathrm{trans}}$) will then of course give the description of the same physical state in the product picture.  But we would like to have a more explicit way of presenting that description.   We may drop the $\Omega^{\mathrm{trans}}\otimes$ (which is a constant feature) and focus on gaining an explicit presentation of $U(\Omega^{\mathrm{long}}\otimes\Psi_{\mathrm{Schr}})$.   To do this, it helps to think of the tensor product of the Hilbert-space ${\cal H}_{\mathrm{Schr}} = L^2({\mathbb R}^{3N})$ (suitably symmetrized and/or antisymmetrized), for the Schr\"odinger wave function with the Hilbert space, $\cal F(\cal H_{\mathrm{one}})$, of the electromagnetic field as $L^2$ of ${\mathbb R}^{3N}$ with values in ${\cal F(\cal H_{\mathrm{one}})}$ (again suitably symmetrized and/or antisymmetrized).   It is then straightforward to see that, regarded as an element of that space, $U(\Omega^{\mathrm{long}}\otimes\Psi_{\mathrm{Schr}})$  (where we recall (\ref{U}) that $U$ is given by the formula, $U=\exp\left(i\int \hat A^i(\bm x)\partial_i\phi(\bm x)\, d^3x\right )$ with $\phi$ interpreted as an \textit{operator} on ${\cal H}_{\mathrm{Schr}}$) is equal to the $\cal F(\cal H_{\mathrm{one}})$-valued $L^2$ function on ${\mathbb R}^N$:
\begin{equation}
\label{NewtForm}
(\bm x
_1, \dots , \bm x_N) \mapsto
\Psi_{\mathrm{Schr}}(\bm x_1, \dots, \bm x_N)|\Psi_{\mathrm{EM}}(\bm x_1, \dots , \bm x_N)\rangle
\end{equation}
where $|\Psi(\bm x_1, \dots , \bm x_N)\rangle$ denotes the electrostatic coherent state
defined as in the $U\Omega$ of (\ref{cohem}) in Section \ref{Sect:electrocoh} with $U$ now given by (\ref{Uext}) for $\phi$ as in (\ref{phiSchro}), but with $\phi$ now interpreted as a \textit{$c$-number} -- namely as the electrostatic potential for our charged balls when their centres of mass are located at the fixed positions $(\bm x_1, \dots , \bm x_N)$.

Equation (\ref{NewtForm}) is the quantum electrostatic counterpart of the (unnumbered) displayed formula immediately prior to \cite[Equation (15)]{KayNewt} which we arrived at in that paper by a partly intuitive physical argument based on the ideas that we explained here in Section \ref{Sect:Intro2}.   By rederiving that formula in the way we have done above, we have completed the fulfilment of our promise, made in Section \ref{Sect:Intro}, of furnishing the work in \cite{KayNewt} (in the Maxwell-Schr\"odinger QED case) with a proper theoretical foundation. 

We remark that we may also regard Equation (\ref{cat}) of Section \ref{Sect:Intro2} as a very special case of (\ref{NewtForm}) -- namely when there is a single charged ball and its Schr\"odinger wave function $\Psi(\bm x)$ consists of just two sharp peaks at the locations $\bm x_1$ and $\bm x_2$ --  say $c_1\alpha(\bm x - \bm x_1) + c_2\alpha(\bm x - \bm x_2)$, where $\alpha$ is some normalized $L^2$ function which approximates a delta function.   Thus we have come full circle back to our initial motivating equation.

To end this section, and also the paper, we aim to indicate how our new product picture reduces under appropriate approximations to a new alternative picture of standard nonrelativistic many-body Schr\"odinger quantum mechanics.   We then (for completeness) (re-)derive the quantum electrodynamics counterpart to the (quantum gravitational) formula \cite[Equation (16)]{KayNewt} for the reduced density matrix of a system of Schr\"odinger charged balls when one traces over the electromagnetic field.

\subsection{\label{Sect:Hatom} The product picture description of the Hydrogen atom}

So far we have looked at the full exact transformation between the description of a state of  Maxwell-Schr\"odinger QED in Coulomb gauge and the description of the same state in our product picture.   In this subsection, we look at what this transformation reduces to when we neglect radiative corrections.  In this way we aim to gain a new perspective on the nonrelativistic quantum mechanics of a system of many charged particles (which, as always, we shall need to model as extended charged balls).   For simplicity we invite the reader to have in mind a particular bound state of a two-body system consisting of one positively and one negatively charged particle  and, indeed, for the sake of familiarity, to think of this as the usual model of a nonrelativistic Hydrogen atom (where one neglects spin) -- i.e.\ to think of $M_1$ and $M_2$ in Equation (\ref{HatomHam}) below as the mass of the proton and of the electron, respectively and $q_1$ and $q_2$ in (\ref{VCoulombpoint}) (or rather the integrals of $\rho_1$ and $\rho_2$ in (\ref{VCoulombrho})) in the case $N=2$ to be their charges -- i.e.\ $\rm e$ and $-{\rm e}$.   But what we do will clearly generalize to general nonrelativistic many-body Schr\"odinger theory.

Before we can discuss what our new product picture description reduces to for such a Hydrogen-atom bound state, we need to take into consideration that the usual description of such a bound state in non-relativistic quantum mechanics is only an approximation for two reasons.   First, we need to neglect the radiative corrections we mentioned above.   This amounts to ignoring, in the full Coulomb gauge Hamiltonian (\ref{HamSchroQED}), the terms $q_I\bm p\bm \cdot\bm A/M_I$ and $q_Iq_J\bm A(\bm x_I)\bm \cdot\bm A(\bm x_J)/4M_IM_J$ which arise when one expands the squared bracket in (\ref{SchroHampoint}), or rather the more complicated corresponding terms in (\ref{SchroHam}).  These terms are of course responsible for radiative transitions between excited states (so in our approximation all bound states will be stable) and for radiative corrections to energy levels (which go like the fine structure constant -- this may be traced to the fact that, when we restore $\hbar$ and $c$ [see Footnote \ref{ftntUnits} and see (\ref{tradApi}) and (\ref{Ahat})] then $\bm A$, when expressed in terms of creation and annihilation operators, goes like $1/\sqrt c$).

Second, writing the Schr\"odinger two-ball wave function, $\Psi(\bm x_1, \bm x_2)$, in terms of relative and centre of mass coordinates and separating variables in the usual way so that (with an obvious notation) $\Psi(\bm x_{\mathrm{rel}}, \bm x_{\mathrm{cm}}) = \psi_{\mathrm{rel}}(\bm x_{\mathrm{rel}})\psi_{\mathrm{cm}}(\bm x_{\mathrm{cm}})$,  then while 
$\psi_{\mathrm{rel}}(\bm x_{\mathrm{rel}})$ may be an eigenstate (for the one-body Schr\"odinger equation with the usual reduced mass) for some energy, say $E$, $\psi_{\mathrm{cm}}(\bm x_{\mathrm{cm}})$ will necessarily be in a wave-packet state and hence $\Psi(\bm x_1, \bm x_2)$ will not be an exact eigenstate of the time-independent two-body Schr\"odinger equation.   However we can, say, take $\psi_{\mathrm{cm}}(\bm x_{\mathrm{cm}})$ to be approximately constant in a large region and then to slowly vanish towards infinity and then $\Psi(\bm x_1, \bm x_2)$ will be an \emph{approximate} eigenstate with the same energy, $E$.  That is we will have
\begin{equation}
\label{HatomHam}
\left(\frac{\bm p_1^2}{2M_1} + \frac{\bm p_2^2}{2M_2} + V_{\mathrm{Coulomb}}^{\mathrm{Schr}}\right)\Psi_{\mathrm{Schr}} \approx E\Psi_{\mathrm{Schr}}
\end{equation}
where we take $N$ in (\ref{VCoulombrho}) to be 2.
At the level of full Coulomb gauge Maxwell Schr\"odinger QED, $\Omega^{\mathrm{trans}}\otimes 
\Psi \in {\cal F}({\cal H}_{\mathrm{one}}^{\mathrm{trans}})\otimes {\cal H}_{\mathrm{Schr}}$ will be an approximate eigenstate with the same energy, $E$, of the $H_{\mathrm{QED}}^{\mathrm{Schr}}$ of (\ref{HamSchroQED}) where we neglect the terms mentioned above and where we normal order the electrodynamic piece.  I.e.\ we will have
\begin{equation}
\label{HQEDHam}
\left(\int |\bm k| {{a_i^*}^{\mathrm{trans}}}(\bm k)a^{\mathrm{trans}}_i(\bm k)\, d^3 k  + \frac{\bm p_1^2}{2M_1} + \frac{\bm p_2^2}{2M_2} + V_{\mathrm{Coulomb}}^{\mathrm{Schr}}\right)\Omega^{\mathrm{trans}}\otimes \Psi_{\mathrm{Schr}}
\end{equation}
\[ \approx E \Omega^{\mathrm{trans}}\otimes \Psi_{\mathrm{Schr}}.
\]
The same state in the product picture is modeled as $\Omega^{\mathrm{trans}}\otimes U(\Omega^{\mathrm{long}}\otimes \Psi_{\mathrm{Schr}}) \in
{\cal F(\cal H_{\mathrm{one}})}\otimes {\cal H}_{\mathrm{Schr}}$ $={\cal F}({\cal H}_{\mathrm{one}}^{\mathrm{trans}})\otimes{\cal F}({\cal H}_{\mathrm{one}}^{\mathrm{long}})\otimes {\cal H}_{\mathrm{Schr}}$ where $U(\Omega^{\mathrm{long}}\otimes\Psi_{\mathrm{Schr}})$ is now an entangled state of the form of (\ref{NewtForm}) for $N=2$.    In view of our proof of the equivalence of the exact product picture with the exact Coulomb picture, it must be that this is an approximate eigenstate of the appropriately defined approximate product picture Hamiltonian
\begin{equation}
\label{PPSchroHamQEDApprox}
H^{\mathrm{PP\, Schr}}_{\mathrm{QED\, Approx}}=\int |\bm k| {{a_i^*}^{\mathrm{trans}}}(\bm k)a^{\mathrm{trans}}_i(\bm k)\, d^3\bm k +  \frac{\bm p_1^2}{2M_1} + \frac{\bm p_2^2}{2M_2} + {1\over 2}\mbox{$\tilde{\bm\pi}^{\mathrm{long}}$}^2 
\end{equation}
with the same approximate eigenvalue $E$.   However, it is instructive to (re-)derive this latter fact directly which we now do with the following succession of equalities and approximate equalities.  Their validity is evident once one recalls that $U$ commutes with $\phi$, that 
$\frac{1}{2}\int\bm\nabla\phi{\bm\cdot}\bm\nabla\phi\, d^3x = V_{\mathrm{Coulomb}}^{\mathrm{Schr}}$, and that $\tilde{\bm\pi}^{\mathrm{long}}$ commutes with $\phi$ and annihilates 
$\Omega^{\mathrm{long}}$, while $U$ also commutes with $\bm p_1$ and $\bm p_2$ up to terms which we neglect.   We have
\[
H^{\mathrm{PP\, Schr}}_{\mathrm{QED\, Approx}}\Omega^{\mathrm{trans}}\otimes U(\Omega^{\mathrm{long}}\otimes \Psi_{\mathrm{Schr}})
\] 
\[
= \left(\frac{\bm p_1^2}{2M_1} + \frac{\bm p_2^2}{2M_2} + {1\over 2}\mbox{$\tilde{\bm\pi}^{\mathrm{long}}$}^2 \right)
\Omega^{\mathrm{trans}}\otimes U(\Omega^{\mathrm{long}}\otimes \Psi_{\mathrm{Schr}})
\]
\[
= \Omega^{\mathrm{trans}}\otimes\left(\left(\frac{\bm p_1^2}{2M_1} + \frac{\bm p_2^2}{2M_2}\right)U + {1\over 2} UU^{-1}\mbox{$\tilde{\bm\pi}^{\mathrm{long}}$}^2U\right)\Omega^{\mathrm{long}}\otimes \Psi_{\mathrm{Schr}}
\]
\[
\approx \Omega^{\mathrm{trans}}\otimes U\left(\frac{\bm p_1^2}{2M_1} + \frac{\bm p_2^2}{2M_2} + {1\over 2}\left(\mbox{$\tilde{\bm\pi}^{\mathrm{long}}$} + \bm\nabla\phi\right)^2\right)\Omega^{\mathrm{long}}\otimes \Psi_{\mathrm{Schr}}
\]
\[
= \Omega^{\mathrm{trans}}\otimes U\left(\frac{\bm p_1^2}{2M_1} + \frac{\bm p_2^2}{2M_2} + {1\over 2}\left(\bm\nabla\phi\right)^2\right)\Omega^{\mathrm{long}}\otimes \Psi_{\mathrm{Schr}}
\]
\[
=\Omega^{\mathrm{trans}}\otimes U(\Omega^{\mathrm{long}}\otimes \left(\frac{\bm p_1^2}{2M_1} + \frac{\bm p_2^2}{2M_2} + V_{\mathrm{Coulomb}}^{\mathrm{Schr}}\right)\Psi_{\mathrm{Schr}}
\]
\[
\approx EU\Omega^{\mathrm{trans}}\otimes\Omega^{\mathrm{long}}\otimes \Psi_{\mathrm{Schr}}.
\]

Note that, in Coulomb gauge, for the sake of doing the transformation to our product picture, we have taken $V_{\mathrm{Coulomb}}^{\mathrm{Schr}}$ to be given by (\ref{VCoulomb}) in the case $N=2$.   But, aside from the fact that we treat the charged particles as extended balls, this includes the self-energies of the balls which is not what one normally does in the standard discussion of the Hydrogen atom.  However, one could of course subtract the self-energy terms from $V_{\mathrm{Coulomb}}^{\mathrm{Schr}}$ provided only one also subtracts these same terms from the energy, $E$, in which case what remains is just the familiar Hydrogen atom Hamiltonian with the familiar Coulomb potential term $q_1q_2/4\pi|{\bm x}_1-{\bm x}_2|$ (modified slightly to take into account that the particles are treated as extended charged balls).   

Note too that, in the product picture, there is no potential term in the Hamiltonian at all;  in this picture, the electrostatic attraction between the electron and the proton is instead seen to arise from the fact that they are entangled with longitudinal photons and that the energy of those longitudinal photons depends on the electron and proton locations.  Note further that, after one has subtracted off the self-energy terms as above,  negative inter-ball potentials and negative energy eigenvalues are possible, even though, when one includes the self-energies, $V_{\mathrm{Coulomb}}^{\mathrm{Schr}}$ is (as one can see by writing it in the form $\frac{1}{2}\int\bm\nabla\phi{\bm\cdot}\bm\nabla\phi\, d^3x$) a manifestly positive quantity.    

Note also that our entangled state, $\Omega^{\mathrm{trans}}\otimes U(\Omega^{\mathrm{long}}\otimes\Psi)$ involves longitudinal photons which are non-dynamical since $U$ ($=\exp(i\int\hat A^i(\bm x)\partial_i\phi(\bm x)\, d^3x)$) is determined by the electrical potential operator, $\phi(\bm x)$, on the Hilbert space, ${\cal H}_{\mathrm{Schr}}$, for the charged balls and that operator is, to borrow a term from engineering, a \textit{slave} to the positions of those balls in the sense that it is related to the position operators, $\bm x_1, \dots, \bm x_N$, through Equation (\ref{phiSchro}).   (Similar remarks apply, with suitable modifications, to the unapproximated Maxwell-Schr\"odinger QED model of this section and to the Maxwell-Dirac QED of Section \ref{Sect:QED}.)

Let us also remark that further insight into how the version of Gauss's law (\ref{QuantGauss3}) of this section holds in our product picture (for the Hydrogen atom or more generally) may be had by noting that an alternative derivation of it is to act on each side of (\ref{NewtForm}) with $\bm\nabla\bm\cdot$ and then to apply the coherent state version, (\ref{QuantGauss1}), of Gauss's law of Section \ref{Sect:electrocoh} in the case that $\Psi$ is equal to the 
$\Psi_{\mathrm{Schr}}(\bm x_1, \dots, \bm x_N)$.

\subsection{\label{Sect:DensOp} The reduced density operator of charged Schr\"odinger matter}

As we discussed in Section \ref{Sect:Intro2}, once we have a product picture, it becomes meaningful, say for a given vector state, $\bm\Psi\in {\cal F}({\cal H}_{\mathrm{one}})\otimes {\cal H}_{\mathrm{Schr}} = {\cal F}({\cal H}_{\mathrm{one}}^{\mathrm{trans}})\otimes {\cal F}({\cal H}_{\mathrm{one}}^{\mathrm{long}})\otimes {\cal H}_{\mathrm{Schr}}$, to ask about 
the degree of entanglement between the charged matter and the electromagnetic field, and this, in turn, may be measured e.g.\ by the charged matter-electromagnetic field entanglement entropy which in turn may be thought of as the von Neumann entropy of the reduced density operator of the charged matter.   Here we content ourselves with computing the latter density operator and we refer to \cite{KayNewt} and \cite{eeee} for a qualitative discussion of its von Neumann entropy (in the analogous case of linearized quantum gravity).

Here, by the reduced density operator of charged matter, we mean the partial trace of $|\bm\Psi\rangle\langle\bm\Psi|$ over ${\cal F}({\cal H}_{\mathrm{one}})$.   We shall confine our attention here to cases where $\bm\Psi$ arises (say, approximately as discussed above)
as $\Omega^{\mathrm{trans}}\otimes U(\Omega^{\mathrm{long}}\otimes\Psi_{\mathrm{Schr}})$ whereupon its reduced density operator will be the tensor product of $|\Omega^{\mathrm{trans}}\rangle\langle\Omega^{\mathrm{trans}}|$ with the partial trace -- let us call it $\varsigma_{\mathrm{ch\, mat}}$, of 
$|U(\Omega^{\mathrm{long}}\otimes\Psi_{\mathrm{Schr}})\rangle\langle U(\Omega^{\mathrm{long}}\otimes\Psi_{\mathrm{Schr}})|$ over ${\cal F}({\cal H}_{\mathrm{one}}^{\mathrm{long}})$.  As already discussed in \cite{KayNewt} (in the linearized gravity case) one easily sees, from (\ref{NewtForm}), that we will have
\begin{equation}
\label{parttr}
\varsigma_{\mathrm{ch\, mat}}(\bm x_1, \dots, \bm x_N; \bm x_1', \dots, \bm x_N')
\end{equation}
\[
=\Psi_{\mathrm{Schr}}(\bm x_1, \dots, \bm x_N){\Psi_{\mathrm{EM}}^*}_{\mathrm{Schr}}(\bm x_1', \dots, \bm x_N' )\langle\Psi_{\mathrm{EM}}(\bm x_1, \dots, \bm x_N)|\Psi(\bm x_1', \dots, \bm x_N')\rangle.
\]
For completeness we conclude by finding an asymptotic formula ((\ref{innprod}/(\ref{innprodalt}) below) for the inner-product in the above expression -- and thereby for $\rho_{\mathrm{ch\, mat}}$ itself.    The counterpart to this formula in the linearized gravity case (up to a numerical error -- see Footnote \ref{ftntErr}) was already obtained -- as \cite[Equation (16)]{KayNewt}.  We shall go into a little more detail than was provided in \cite{KayNewt}.    By a straightforward generalization of (\ref{D1formula}),  $\langle\Psi(\bm x_1, \dots, \bm x_N)|\Psi(\bm x_1', \dots, \bm x_N')\rangle$ will equal $\exp(-D_1)$ where now
\[
D_1 = \|(\bm\chi_1(\bm x_1) + \dots + \bm\chi_N(\bm x_N)) - (\bm\chi_1(\bm x_1') + \dots + \bm\chi_N(\bm x_N'))\|^2/2
\]
where $\bm\chi_I^i(\bm x_I)$ now denotes $k^i\phi_I(\bm x_I)/\sqrt{2k}$ (see after Equation (\ref{D1formula})) where, for any $\bm y$, $\phi_I(\bm y)$ denotes the Coulomb potential of the $I$th ball when its centre is located at $\bm y$.
We may rewrite this as
\[
D_1 = \langle (\bm\chi_1(\bm x_1) - \bm\chi_1(\bm x_1')) + \dots + (\bm\chi_N(\bm x_N) - \bm\chi_N(\bm x_N'))|(\bm\chi_1(\bm x_1) - \bm\chi_1(\bm x_1'))
\]
\[
 + \dots + (\bm\chi_N(\bm x_N) - \bm\chi_N(\bm x_N'))\rangle/2
\]
and then write (twice) this as a sum of $N$ inner products of form $\langle (\bm\chi_I(\bm x_I) - \bm\chi_I(\bm x_I))|(\bm\chi_I(\bm x_I) - \bm\chi_I(\bm x_I))\rangle$, $I = 1 \dots N$ and $N(N-1)$ inner products of form $\langle (\bm\chi_I(\bm x_I) - \bm\chi_I(\bm x_I))|(\bm\chi_J(\bm x_J) - \bm\chi_J(\bm x_J))\rangle$, $I = 1 \dots N$, $J=1 \dots N$, $I \ne J$.

In view of (\ref{balloverlap5}) (in the case our charged balls have uniform density) and the subsequent `note in passing'  and Footnote \ref{ftntAsymp} (in the case of other charge distributions) the former inner products (divided by 2) will have the asymptotic form, 
\begin{equation}
\label{II}
\langle (\bm\chi_I(\bm x_I) - \bm\chi_I(\bm x_I'))|(\bm\chi_I(\bm x_I) - \bm\chi_I(\bm x_I'))\rangle/2 \approx \frac{q_I^2}{4\pi^2}\ln\frac{|\bm  x_I - \bm x_I'|}{R_I}
\end{equation}
where $q_I$ is the total charge and $R_I$ the (effective) radius of the charge distribution of the $I$th ball.

We remark here, in preparation for our calculation of the latter inner products (i.e.\ those where $I \ne J$) that an alternative way of writing the left hand side of (\ref{II}) (and, by the way, a useful first step in deriving the above asymptotic form) is (see Footnotes \ref{ftntDetails} and \ref{ftntAsymp}) as
\begin{equation}
\label{IIalt}
\langle \bm\chi_I(\bm 0)| \left(1-e^{i\bm k\bm\cdot (\bm x_I - \bm x_I')}\right)|\bm\chi_I(\bm 0)\rangle.
\end{equation}

The latter inner products (divided by 2) can similarly be written
\begin{equation}
\label{IJ}
\langle \bm\chi_I(\bm 0)| \left(e^{i\bm k\bm\cdot (\bm x_J - \bm x_I)}  + e^{i\bm k\bm\cdot (\bm x_J' - \bm x_I')}   - e^{i\bm k\bm\cdot(\bm x_J - \bm x_I')} - e^{i\bm k\bm\cdot (\bm x_J' - \bm x_I)}\right)\bm\chi_J(\bm 0)\rangle/2.
\end{equation}
To evaluate/estimate this, note first that, by a small extension of the derivation of the asymptotic form for (\ref{IIalt}) (see Footnote \ref{ftntAsymp}) we may show that
\begin{equation}
\label{Asympt}
\langle \bm\chi_I(\bm 0)| \left(1 -  e^{i\bm k\bm\cdot\bm a}\right)\bm\chi_J(\bm 0)\rangle \approx \frac{q_Iq_J}{4\pi^2}\ln \frac{a}{R_{IJ}}
\end{equation}
for some constant, $R_{IJ}$, with dimensions of length (and order of magnitude around/between the sizes of balls $I$ and $J$).    Using this, we then easily have that 
\[
\langle (\bm\chi_I(\bm x_I) - \bm\chi_I(\bm x_I'))|(\bm\chi_J(\bm x_J) - \bm\chi_J(\bm x_J'))\rangle/2
\approx \frac{q_Iq_J}{8\pi^2}\ln\left(\frac{|\bm x_J - \bm x_I'| |\bm x_J' - \bm x_I|}{|\bm x_J - \bm x_I||\bm x_J' - \bm x_I'|}\right),
\]
where we notice that the quantity $R_{IJ}$ no longer appears. 

We conclude that our inner product has the asymptotic form
\[
\langle\Psi(\bm x_1, \dots, \bm x_N)|\Psi(\bm x_1', \dots, \bm x_N')\rangle = e^{-D_1}
\]
\begin{equation}
\label{innprod}
\approx \prod_K \left(\frac{|\bm x_K - \bm x_K'|}{R_K}\right)^{-q_Iq_J/4\pi^2}\prod\prod_{\!\!\!\!\!\!\!\!\!I\ne J}\left(\frac{|\bm x_J - \bm x_I'| |\bm x_J' - \bm x_I|}{|\bm x_J - \bm x_I||\bm x_J' - \bm x_I'|}\right)^{-q_Iq_J/8\pi^2}
\end{equation}
where the first product is over $K$ from 1 to $N$, and the second over $I$ and $J$ from 1 to $N$ for which $I \ne J$.
This can be rewritten as the unrestricted product
\begin{equation}
\label{innprodalt}
\prod_{I= 1}^N\prod_{J=1}^N\left(\frac{|\bm x_J - \bm x_I'| |\bm x_J' - \bm x_I|}{|\bm x_J - \bm x_I||\bm x_J' - \bm x_I'|}\right)^{-q_Iq_J/8\pi^2}
\end{equation}
provided it is understood that the terms in the denominator for which $I=J$ are replaced by $R_I$.

This formula is asymptotic in the sense that it is expected to be a good approximation as long as $\bm x_I$, $\bm x_J$, $\bm x_I'$, $\bm x_J'$ are such that whenever $I \ne J$, all of the quantities $|\bm x_J - \bm x_I|$, $|\bm x_J' - \bm x_I'|$, $|\bm x_J - \bm x_I'|$ and $|\bm x_J' - \bm x_I|$ are much bigger than either of the radii of either of the $I$th and $J$th balls.

As we mentioned above, the gravitational counterpart of (\ref{innprod})/(\ref{innprodalt}) was previously obtained in \cite{KayNewt} (up to a presently uncertain numerical factor -- see Footnote \ref{ftntErr}) on the basis of partly heuristic arguments.

\section{\label{Sect:final} Some final remarks on the relation between the product picture and the Coulomb gauge picture}

While our product picture is, in view of (\ref{equivDir}) and (\ref{equivSchro}), `equivalent' to the Coulomb gauge picture, in some ways it might be said to be an extension of the latter.   This is because new questions arise and can be answered in the product picture formulation of QED which do not arise in the Coulomb gauge picture such as, for example, the question of what is the reduced density operator when one traces over the electromagnetic field, which we discussed in Section \ref{Sect:DensOp}.   These new questions may be seen to arise from the fact that the full set of observables of the theory in the product picture is enlarged by the inclusion, amongst the observables, of an observable (namely $-\tilde{\bm\pi}^{\mathrm{long}}$) for the longitudinal component, $\bm E^{\mathrm{long}}$, of the electric field.   Indeed we also seemingly increase the number of questions we can ask, and answer, concerning the expectation values of field observables.   Thus, as a trivial example, in the product picture, we obtain the result (Gauss's law in expection value) that the expectation value, in a given QED vector state, $\bm\Psi$, in our physical Hilbert space $U{\cal F(\cal H_{\mathrm{one}}^{\mathrm{trans}})}\otimes \Omega^{\mathrm{long}}\otimes {\cal H}_{\mathrm{ch\, mat}}$,
is equal to the expectation value, in the same state, of $\rho$ where $\rho$ is the charge density operator.   This is obtained -- as a trivial consequence of our stronger operator form of Gauss's law (\ref{QuantGauss1}) or (\ref{QuantGauss2}) or (\ref{QuantGauss3}) -- by calculating the inner product $\langle\bm\Psi|\bm\nabla\bm\cdot \bm E\bm\Psi\rangle$ ($= \langle\bm\Psi|\bm\nabla\bm\cdot \bm E^{\mathrm{long}}\bm\Psi\rangle$).   In the Coulomb gauge picture, on the other hand, Gauss's law is imposed as a constraint at the classical level and then solved before quantizing; as far as the quantum theory is concerned, the question of the equality of $\bm\nabla\bm\cdot\bm E$ and $\rho$, or, equivalently of $\bm E^{\mathrm{long}}$ with $-\bm\nabla\phi$ where $\nabla^2\phi=\rho$ does not arise; $\bm E^{\mathrm{long}}$ is just an optional extra quantity which, should we opt to consider it as existing at all, is simply identified with (i.e.\ defined to be) $-\bm\nabla\phi$ and plays no r\^ole in the theory.   It is for this reason that we described the status, in Coulomb gauge, of the longitudinal part of the electric field, in Section \ref{Sect:Intro2}, as (at most) an `epiphenomenon of the physics of the charged matter sector' and it is also for this reason that we would not consider the quantization of QED in Coulomb gauge as a counterexample to our statement in Section \ref{Sect:mathnote} that, as far as we are aware, the fact that our product picture offers a quantization of QED ``in which the Hilbert space is a genuine Hilbert space and in which Gauss's law holds as a genuine operator equation'' is a new feature not shared by any other quantization scheme.

Let us remark here that the replacement of Coulomb gauge QED by our product picture may be of possible wider interest as providing a concrete example of a situation where there are two theories of something -- in this case the longitudinal modes of the quantum electric field -- which are operationally indistinguishable but in which the ontological status of that thing is different.  Namely, in the Coulomb gauge description, those longitudinal modes are described either as an epiphenomenon of the physics of the charged matter, or, in a more radical viewpoint, as not existing at all, while in the product picture those same modes are understood to be a real physical thing \footnote{\label{ftntAnalo} The word `epiphenomenon' is used to mean several different things.  So let us clarify here that, when we say, in Section \ref{Sect:Intro2}, that, in Coulomb gauge, the longitudinal modes of the electric field are an epiphenomenon of the charged matter, we are referring to the fact that the equation $\bm E^{\mathrm{long}}= -\bm\nabla\phi$ is, in a Coulomb gauge understanding, regarded as a \emph{definition} of the left hand side in terms of the right hand side, and not an equation between two independently defined things.  As far as I understand, this use of the word is similar to the use of the word when one says, e.g., that the mind may be an epiphenomenon of brain activity.

\smallskip

Indeed, an obvious place where our analogy in Section \ref{Sect:final} might be relevant is the mind-body problem, where there is a well known and long standing controversy as to whether, on the one hand, the mind is either an epiphenomenon of brain activity, or in a more radical variation, does not exist at all, or, on the other hand, whether it is a real physical thing in its own right.  Pursuing the analogy, one is tempted to speculate that perhaps it is possible to have a theory of the mind in which one has a choice as to whether to regard it as an epiphenomenon of brain activity or even to regard it as not existing at all, and another theory in which it is a real physical thing in its own right,  and yet the two theories turn out, similarly to in our QED analogy, to be operationally indistinguishable.  And, if we pursue the analogy further, we might expect that, were one were to adopt the theory in which it is a real physical thing, then new questions might arise and be answerable -- just as new questions arise and can be answered in our product picture (such as ``What is the partial trace of a total state of QED over the electromagnetic field'').  

\smallskip

Of course, there are several other places in physics that provide such analogies.  Indeed, classical electrodynamics does.   lts Hamiltonian formulation in Coulomb gauge goes together with the ontology that the longitudinal modes of the electric field are either an epiphenomen of the charged matter variables, or don't exist at all; while, on the other hand the classical Maxwell's equations go together with the ontology that all modes of the electromagnetic field are real.  However, having the product picture in QED would be like having an alternative \textit{Hamiltonian formulation} of the classical Maxwell equations, and as far as I know, there is no such thing in classical electrodynamics.   Also the notions of partial trace and entanglement don't have classical counterparts and these perhaps provide a particularly clear example of how new questions can arise in what, from another point of view, is an equivalent theory.  

\smallskip

Finally, since Maxwell himself said that `charge' and `current' are `epiphenomena' of electromagnetic fields, we should clarify that what he had in mind in saying that appears to be unrelated to the sorts of issues and analogies on which we are focusing here.}.    

Let us also remark that, say in the case of Maxwell-Schr\"odinger theory, one implication of the equation, $\bm E^{\mathrm{long}}=-\bm\nabla\phi$ as it is understood in Coulomb gauge, is, (cf.\ where we introduce this term in Section \ref{Sect:Hatom}) that the longitudinal part of the electric field is a \textit{slave} to the charged matter sector.  In fact the same equation (now equivalent to Gauss's law holding in operator form) still holds in the product picture but there the equation is not a definition of the left hand side as it is in Coulomb gauge, but, rather, it is an equality between two independently defined things:  Namely between $-\tilde{\bm\pi}$ which acts nontrivially on the ${\cal H}_{\mathrm{electromag}}$ ($={\cal F}({\cal H}_{\mathrm{one}})$) side of the tensor product ${\cal F}({\cal H}_{\mathrm{one}})\otimes{\cal H}_{\mathrm{Schr}}$
and $\bm\nabla\phi$  which acts nontrivially on the $\cal H_{\mathrm{ch\, mat}}={\cal H}_{\mathrm{Schr}}$ side -- the equality holding when they are both restricted to the product picture physical subspace.   This equality tells us that, also in the product picture, the longitudinal part of the electric field is a slave to the charged matter, but, unlike in the Coulomb gauge picture, it is not just an epiphenomenon; the longitudinal part of the electric field has, in the product picture, an independent existence in its own right.   (It is only equal to $-\bm\nabla\phi$ when we restrict both to the product picture physical subspace.)  

In fact we may say more.  In a model (which can physically only ever be an approximation) where we take the charged matter to be a static classical, external charge distribution, $\rho$, we argued in Section \ref{Sect:extsource}, that the state of the quantum electric field due to that source is one of the coherent states, $\Psi_\rho$, that we defined in Section \ref{Sect:electrocoh}.   We might say that when we promote the charged matter to be a collection of non-relativistic dynamical quantum balls, then the overall quantum state may be understood as the many-body state of those balls in the many body Hilbert space,  ${\cal H}_{\mathrm{Schr}}$, but where the one-body state of each individual ball is tensor producted with the coherent electrostatic state that would surround it, were it to be static, but which is dragged around by that ball as it moves.  (Cf.\ the sentence after Equation (\ref{PPent}) and the sentence around Equation (\ref{NewtForm}).)   So, and, in hindsight, perhaps unsurprisingly, what was  described as classical (i.e.\ the longitudinal part of the electric field) in the Coulomb gauge picture, is given a quantum description in terms of coherent states in our product picture.   Moreover, it now may seem natural that the longitudinal part of the electric field is described mathematically in terms of annihilation operators as in Equation (\ref{pilongtilde}) since coherent states are eigenstates of annihilation operators.   To take stock, the thing (specifically the longitudinal part of the electric field) that in the Coulomb gauge picture is a slave and an epiphenomenon and classical remains a slave in the product picture, but it is no longer just an epiphenomenon and it is no longer classical but quantum in nature.   We should surely find this satisfactory if, as surely we should, we believe that everything is quantum in nature.    And it is clear, on comparing Section \ref{Sect:QED} with Section \ref{Sect:QEDSchr} that everything that we have said, in the present and previous paragraph, about our product picture of Maxwell-Schr\"odinger theory remains true, with suitable technical adjustments, of our product picture for Maxwell-Dirac QED.

\section*{Acknowledgments}

The author thanks the Leverhulme Foundation for the award of Leverhulme Fellowship RF\&G/9/RFG/2002/0377 for the period October 2002 to June 2003 during which some of this work was done. I thank Michael Kay for valuable comments.

\end{document}